\documentclass[superscriptaddress,groupedaddress,nofootnoteinbib,11pt]{article}
\pdfoutput=1

\usepackage{jheppub}
\usepackage{graphicx}					
\usepackage{amssymb}
\usepackage{multirow}
\usepackage[sort&compress,numbers]{natbib}
\usepackage{color,soul}
\definecolor{darkred}{rgb}{0.9, 0.0, 0.0}
\definecolor{darkgreen}{rgb}{0.0, 0.5, 0.0}
\definecolor{darkblue}{rgb}{0.0, 0.0, 0.55}

\usepackage{doi}
\usepackage{bm}
\usepackage{feynmp-auto}
\usepackage{slashed}
\usepackage{amsmath}
\usepackage{physics}
\usepackage{changepage}
\usepackage[frak=euler]{mathalpha}
\usepackage{calligra}

\allowdisplaybreaks[3]

\newcommand{\be}{\begin{equation}}
\newcommand{\ee}{\end{equation}}
\def\ba{\begin{eqnarray}}
\def\ea{\end{eqnarray}}
\def\nn{\nonumber}
\newcommand{\nl}{\nonumber \\ }

\newcommand{\smaller}[1]{\footnotesize{\text{$#1$} }}

\newcommand{\tcomment}[1]{\noindent {\bf \color{blue} \small [TW: #1]} }

\allowdisplaybreaks

\title{\boldmath Capping the positivity cone: dimension-8 Higgs operators in the SMEFT}

\author[a,b]{Qing Chen}
\author[c,d]{\!\!, Ken Mimasu}
\author[a,e,f]{\!\!, Tong Arthur Wu}
\author[a]{\!\!, Guo-Dong Zhang}
\author[a,b,g]{\!\!, Shuang-Yong Zhou}

\affiliation[a]{Interdisciplinary Center for Theoretical Study, University of Science and Technology of China,
Hefei, Anhui 230026, China}
\affiliation[b]{Peng Huanwu Center for Fundamental Theory, Hefei, Anhui 230026, China}
\affiliation[c]{School of Physics and Astronomy, University of Southampton,
Highfield, Southampton S017 1BJ, UK}
\affiliation[d]{Theoretical Particle Physics and Cosmology Group, Department of Physics, King’s College London, London WC2R 2LS, UK}
\affiliation[e]{School of The Gifted Young, University of Science and Technology of China,
Hefei, Anhui 230026, China}
\affiliation[f]{PITT PACC, Department of Physics and Astronomy, University of Pittsburgh, 3941 O’Hara St.,
Pittsburgh, PA 15260, USA}
\affiliation[g]{Theoretical Physics, Blackett Laboratory, Imperial College, London, SW7 2AZ, UK}

\emailAdd{cqpb@ustc.edu.cn}
\emailAdd{ken.mimasu@kcl.ac.uk}
\emailAdd{tow39@pitt.edu}
\emailAdd{guodongz@mail.ustc.edu.cn}
\emailAdd{zhoushy@ustc.edu.cn}

\preprint{{\small USTC-ICTS/PCFT-23-28, KCL-PH-TH/2023-50}}

\abstract{
SMEFT Wilson coefficients are subject to various positivity bounds in order to be consistent with the fundamental principles of S-matrix. Previous bounds on dimension-8 SMEFT operators have been obtained using the positivity part of UV partial wave unitarity and form a (projective) convex cone. We derive a set of linear UV unitarity conditions that go beyond positivity and are easy to implement in an optimization scheme with dispersion relations in a multi-field EFT. Using Higgs scattering as an example, we demonstrate how to obtain closed bounds in the space of the three relevant dimension-8 coefficients, making use of the UV unitarity conditions as well as so-called null constraints that arise from full crossing symmetry. Specifically, we show that they are bounded by inequalities schematically going like $C<\mathcal{O}\left((4\pi)^2\right)$. We compare the newly obtained upper bounds with the traditional perturbative unitarity bounds from within the EFT, and discuss some phenomenological implications of the two-sided positivity bounds in the context of experimental probes of Vector Boson Scattering.
}

\begin{document}

\begin{adjustwidth}{-5pt}{-17pt}
\maketitle
\end{adjustwidth}

\flushbottom

\section{Introduction and summary}

Effective field theories (EFTs) are useful to describe low energy physics within our experimental reach, before the new, more fundamental high energy theory can be probed. Unlike a UV-complete theory, an EFT containing several fields typically needs to be parameterized by a large number of Wilson coefficients, if a sufficiently high accuracy is desired. The utility of an EFT often hinges on the precise determination of the many Wilson coefficients, which can present a significant challenge for the experimental search for new physics. Recently, it has been increasingly appreciated that the general parameter space of the Wilson coefficients is mostly inconsistent with the fundamental principles of S-matrix theory such as causality/analyticity and unitarity, except for a small subspace carved out by so-called positivity bounds (see, {\it e.g.}, \cite{Adams:2006sv, Distler:2007, Ian:2010, deRham:2017avq, deRham:2017zjm, Bellazzini:2017bkb, Arkani-Hamed:2020blm, Bellazzini:2020cot, Tolley:2020gtv, Caron-Huot:2020cmc, Chiang:2021ziz, Sinha:2020win, Zhang:2020jyn, Li:2021lpe, Bellazzini:2014waa, Bellazzini:2016xrt,  Bern:2021ppb, Alberte:2020jsk, Tokuda:2020mlf, Caron-Huot:2021rmr, Guerrieri:2021tak, Du:2021byy, Alberte:2021dnj, Bellazzini:2021oaj, Chiang:2022ltp, Caron-Huot:2022ugt, Chiang:2022jep, CarrilloGonzalez:2023cbf, Hong:2023zgm} and \cite{deRham:2022hpx} for a review).

Lately, there has been a growing focus on extracting the optimal $s^2$ order multi-field positivity bounds ($s$ being the usual Mandelstam invariant) and applying them to the Standard Model Effective Field Theory (SMEFT)~\cite{Zhang:2018shp, Bi:2019phv, Bellazzini:2018paj, Remmen:2019cyz, Zhang:2020jyn, Yamashita:2020gtt, Trott:2020ebl, Remmen:2020vts, Remmen:2020uze, Gu:2020thj, Fuks:2020ujk, Gu:2020ldn, Bonnefoy:2020yee, Li:2021lpe, Davighi:2021osh, Chala:2021wpj, Zhang:2021eeo, Ghosh:2022qqq, Remmen:2022orj, Li:2022tcz, Li:2022rag, Li:2022aby, Altmannshofer:2023bfk, Davighi:2023acq}. As an application of the principles of EFT to the Standard Model (SM) field content, the SMEFT is considered to be the most phenomenologically relevant EFT in the absence of a discovery of new particles at the LHC and has become increasingly popular in both the theoretical and experimental communities. It indeed contains many degrees of freedom, \emph{i.e.,} independent Wilson coefficients~\cite{Lehman:2015coa,Henning:2015alf}, especially from canonical dimension-8 and onwards. Certain dimension-8 operators or squares of dimension-6 operators contribute to $s^2$ terms in $2\to2$ scattering amplitudes, which can be directly constrained by the leading order multi-field positivity bounds. These bounds significantly reduce the vast parameter space, mitigating the experimental challenge of constraining dimension-8 coefficients. 

For example, in vector boson scattering (VBS), even the non-optimal, elastic positivity bounds can restrict the physical Wilson coefficient space to be about 2\% of the space furnished by the dimension-8 operators that parameterize anomalous Quartic Gauge Couplings (aQGCs) \cite{Zhang:2018shp, Bi:2019phv} (see also \cite{Vecchi:2007na}). That is, only within this small physical subspace can the SMEFT be UV completed according to the fundamental principles of quantum field theory. Expanding the inclusion of SMEFT operators will result in a more restricted positivity region in percentage terms. The squares of the dimension-6 coefficients can be dropped in the positivity bounds because they contribute schematically as $\text{(dim-8)}-\text{(dim-6)}^2>0$, such that neglecting them leads to more conservative, nevertheless valid bounds~\cite{Zhang:2018shp}. On the phenomenological side, it is justified to forgo the dimension-6 operators because they are likely to be relatively better constrained by other means.

Generalized elastic positivity bounds, obtained by considering arbitrary superpositions of scattering states~\cite{Zhang:2018shp, Bi:2019phv}, can further restrict the Wilson coefficient space. For example, for the 10D dimension-8 aGQC subspace involving transverse vector bosons, they reduce the viable parameter space down to about 0.7\% of the total space~\cite{Yamashita:2020gtt}. However, the generalized elastic positivity approach still does not give rise to the strongest $s^2$-order bounds. The optimal bounds can be obtained by viewing the $s^2$ amplitude coefficients as forming a convex cone~\cite{Yamashita:2020gtt,Zhang:2020jyn,Zhang:2021eeo,Bellazzini:2014waa}. Interestingly, the extremal rays of the $s^2$ positivity cone correspond to irreducible representations of the symmetries of the low energy EFT, which are projected down from the UV. Thus, the positivity bounds are intertwined with the inverse problem of reverse-engineering the UV theory in the event that non-zero Wilson coefficients are observed. This viewpoint also allows us to easily find the optimal positivity bounds when the low energy degrees of freedom are endowed with sufficiently many symmetries. If there are fewer symmetries, the $s^2$ positivity cone can be obtained numerically via a semi-definite programming method~\cite{Li:2021lpe}.

The previous $s^2$ positivity cone approach gives us projective bounds, where the coefficients are allowed to extend to infinity by their very nature of being a convex cone. In this paper, we will show that the $s^2$ positivity cone can be capped from above. This is largely made possible by the addition of two new ingredients that we will apply on top of the previous positivity bounds in the SMEFT. 

First, notice that the aforementioned positivity cone approach only uses the positivity part of the  unitarity conditions on the UV partial wave, $a_{\ell}^{ijkl}$; that is, it only uses the fact that $\mathrm{Im} a_{\ell}^{ijkl}$ is a positive semi-definite matrix, viewing $ij$ and $kl$ as two indices of the matrix, because it can be written as $\mathrm{Im} a_{\ell}^{ijkl} = \sum'_{X} a_{\ell}^{ij\rightarrow X} \qty(a_{\ell}^{kl\rightarrow X})^*$, where $i,j,k,l$ labels external particles and $X$ labels possible intermediate states. This positivity property can be conveniently implemented fully in a semi-definite program~\cite{Li:2021lpe}. In this paper, we will additionally make use of the non-positivity parts of UV partial wave unitarity\,\footnote{The name ``positivity bounds'' originates from the fact that the positivity part of unitarity implies positivity of certain combinations of the amplitude coefficients. In this paper, we have gone beyond that in using the non-positivity parts of unitarity, but we still call them positivity bounds for lack of a better name.} such as the upper bounds
\begin{align*}
    \mathrm{Im}a_\ell^{ii ii}\leq2\quad\text{and}\quad\mathrm{Im}a_\ell^{ij ij}\leq\frac{1}{2}
\end{align*}
in the elastic case, or the more non-trivial conditions such as
\begin{align*}
-1\leq\mathrm{Im}a_\ell^{ii jj}\leq1\quad\text{and}\quad
    \big|\mathrm{Im}a^{ijkl}_\ell\big|  \leq  \frac{\mathrm{Im}a^{ijij}_\ell+\mathrm{Im}a^{klkl}_\ell}{2} ,~~~~~ \forall ~i,j,k,l,
\end{align*}
for inelastic channels. We refer the eager reader to Section \ref{sec:unitarityBounds} and Appendix \ref{app:partial wave unitarity} for all of the unitarity conditions we will make use of and how to derive them. The non-positivity part of partial wave unitarity has been used in the primal S-matrix numerical bootstrap (see, {\it e.g.,} \cite{Paulos:2017fhb, Guerrieri:2020bto, Guerrieri:2021ivu, EliasMiro:2022xaa, Haring:2022sdp} and \cite{Kruczenski:2022lot} for a review), which constrains the S-matrix directly without using the dispersion relations. However, use of the non-positivity part is less convenient with the dispersion relation approach. We shall further develop the method of implementing partial wave unitarity away from positivity at the level of the dispersion relations, initiated by \cite{Caron-Huot:2020cmc, Chiang:2022ltp}. More precisely, our approach involves discretizing the dispersive integrals over the potential UV states and their corresponding UV spectral densities. To reduce the numerical complexity, we will restrict ourselves to linear unitarity conditions like the ones shown above. These linear conditions are in principle weaker than the full unitarity, but they are able to capture the most salient features of the ${\rm Im} a^{ijkl}_\ell$ space allowed by the full unitarity. We therefore expect that our results are a good approximation to those allowed by the full unitarity.

Second, the fixed $t$ dispersion relations are only manifestly two-channel crossing symmetric. It has been realized that imposing the remaining crossing symmetries on the two-channel dispersion relations results in an infinite set of null constraints, which can be used to obtain two-sided bounds on the Wilson coefficients~\cite{Tolley:2020gtv, Caron-Huot:2020cmc}. For example, for the case of a single scalar that is weakly coupled below the EFT cutoff, all of the dimensionless Wilson coefficients have been constrained by the fully crossing-symmetric positivity bounds to be parametrically ${\cal O}(1)$ (up to factors of $4\pi$). This suggests that dimensional analysis is not merely a natural guess, but is actually mandated by the fundamental principles of the UV S-matrix and cannot be violated by some intricate design of the UV model. These null constraints are the other essential ingredient to obtain the upper bounds on the $s^2$ coefficients from the dispersion relations.  

\begin{figure}[htb!]
    \centering
    \includegraphics[height=0.25\textheight]{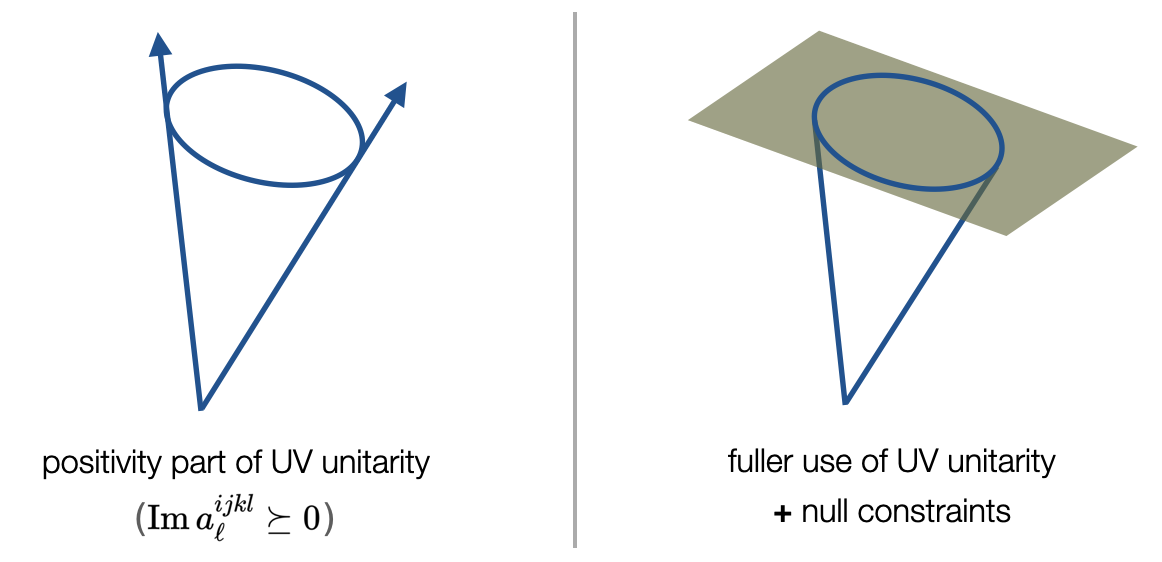}
    \caption{Cartoon view of positivity cone (left) vs capped positivity cone (right). Previous positivity bounds (left) form a convex cone, unbounded away from the vertex, only utilizing the positivity part of the UV unitarity. In this paper, we cap the positivity cone from above (right), using the non-positivity part of the UV unitarity conditions.} \label{fig:twocones}
\end{figure}

For technical simplicity, in this paper we will restrict ourselves to capping the SMEFT Higgs cone of the $s^2$ coefficients, utilizing these new ingredients. Also, we are assuming that the theory is weakly coupled all the way to the EFT cutoff such that we can use the tree level approximation below the cutoff, with the EFT loop contributions suppressed by extra factors of the weak couplings. The central result can be schematically summarized as follows: the previous $s^2$ Higgs positivity cone can extend to infinity away from its vertex (left plot of Figure \ref{fig:twocones}); Now, with the new two-sided bounds, we can constrain the parameters to be within a finite distance from the vertex for a given EFT cutoff, $\Lambda$ (right plot of Figure \ref{fig:twocones}).

We shall first set up the general formalism to bound the $s^2$ coefficients of a multi-scalar field theory. The partial wave expansion of a scalar theory can be performed with Legendre polynomials, while that of a general field theory with spin needs Wigner $d$-matrices, which is technically more involved but does not seem to pose any essential obstacle in applying our general method and is left for future work. We will illustrate how to use the method to bound the single scalar theory (see \cite{Xu:2023lpq} for an application of the upper bound in cosmology) and the theory with two scalars, before applying the method to the Higgs case. 

Numerically, we obtain the two-sided bounds by discretizing the dispersive integrals of the sum rules and the null constraints and truncating the UV partial wave expansion at finite angular momentum. This allows us to reduce the procedure to a linear programming (LP) problem with a large number of decision variables, which nevertheless can be solved relatively easily, thanks to the efficiency of modern LP algorithms. For instance, this can be achieved using the built-in command {\tt LinearProgramming} in mathematica. The numerical efficiency is especially remarkable in the context of a single scalar case. However, in the Higgs case, it does require more computational time due to the increased degrees of freedom, notably the proliferation of decision variables and null constraints. To reduce the running times, we find that we can combine a few sets of spectral densities to significantly reduce the number of decision variables. We plot both 1D (Figure \ref{fig:C123_1D}) and 2D (Figure \ref{fig:4field_2D}) bounds for the three Wilson coefficients for the SMEFT Higgs, from which we can see that the new allowed region only occupies a small portion of the Higgs positivity cone.

We may compare the bounds from the capped Higgs positivity cone to the widely used perturbative unitarity bounds. The upper positivity bounds are similar to perturbative unitarity bounds in the sense that they both make use of the unitarity conditions. However, the fundamental difference between the two kinds of bounds are that, while perturbative unitarity bounds rely on the unitarity bounds within the EFT/below the cutoff, the upper positivity bounds originate from the unitarity conditions of the UV theory. We will see that the upper positivity bounds are often stronger than perturbative unitarity bounds, even when the $\sqrt{s}$ is chosen to be close to the EFT cutoff. We also compare the positivity bounds with the experimental constraints on the SMEFT Higgs coefficients, obtained from measurements of Vector Boson Scattering (VBS) at the LHC. We find that the upper positivity bounds are stronger than the experimental limits when the EFT cutoff is greater than $\sim2$ TeV. 

A caveat about our tree level results is that the sizes of loop contributions are in general dependent on the specific UV theory, a systematic discussion of which is beyond the scope of this work. Since we have not systematically included the loop effects, without any further treatment of the dispersion relations, our results can in general be subject to $\mathcal{O}(1)$ corrections. Nevertheless, our treatment is a good approximation for a UV theory whose dispersion integral is UV-dominated, {\it i.e.,} in cases where the dispersive integral is dominated by the part above the EFT cutoff. Also, we stress that our results are clearly different and complementary to the traditional perturbative unitarity bounds. The unitarity bounds are derived from the breakdown of unitarity within the low-energy EFT, while our bounds are obtained by imposing the unitarity conditions for the unknown UV theory and matching the EFT coefficients to the UV theory via the dispersion relations. The essence of our bounds is the convergence of the dispersive integral at high energies, obtained solely from unitarity and causality of the UV theory. Indeed, the perturbative unitarity bounds, as commonly used in the literature, also assume perturbativity below the cutoff, and are subject to similar loop corrections but are generally neglected. It is in this sense that we compare our positivity bounds with the perturbative unitarity bounds, and we see that the positivity bounds are generally stronger than the perturbative unitarity bounds.

The remainder of the paper is organized as follows. 
Section \ref{sec: mfbounds} introduces and derives the positivity bounds for general multi-field theories, deriving the general null constraints and enumerating the relevant unitarity conditions on the UV partial wave amplitudes beyond positivity. Using examples of the single-scalar and bi-scalar theories, we then establish the LP methodology to obtain new, closed constraints on the coefficients of the associated scattering amplitudes.
Section \ref{sec: higgsupperbounds} applies the methodology to the effective theory for Higgs scattering amplitudes, leading to the upper bounds on the dimension-8 SMEFT coefficients that ``cap'' the $s^2$ positivity cone. Section \ref{sec: pheno} compares our theoretical bounds with existing experimental bounds from VBS measurements at the LHC as well as partial wave unitarity constraints of the low energy EFT. 
Appendix \ref{app:partial wave unitarity} first presents a pedagogical review of the full partial wave unitarity conditions, and then derives a constraining set of linear unitarity  conditions that we use in this paper. In Appendix \ref{app:analytical}, we analytically derive an example upper bound, which helps showcase the inner mechanism of obtaining the upper positivity bounds. Appendix \ref{sec:symHiggs} derives the symmetries of the Higgs amplitudes and connect the amplitude coefficients with the Wilson coefficients. Appendix \ref{app:COV} presents the details about how to reduce the decision variables in the Higgs case. In Appendix \ref{app:convergence}, we provide more details about how to implement the LP numerically and report on a convergence study of our computational method.

\section{Upper bounds in multi-field theories}
\label{sec: mfbounds}

Recently, positivity bounds have established that, at leading order, certain dimension-8 coefficients of the SMEFT should live in a positivity cone~\cite{Zhang:2020jyn, Li:2021lpe}. The vertex of the cone corresponds to all of the coefficients being null, and the cone can extend infinitely away from the vertex, admitting arbitrarily large values of the Wilson coefficients. We shall show in this paper that the positivity cone is actually capped from above, as one would expect on general grounds. In the example of SMEFT Higgs operators that we will explicitly work out, the coefficients are confined to a finite, convex region. In this section, we develop a general formalism to obtain upper bounds on the Wilson coefficients in a multi-field theory. We will only consider scattering amplitudes involving scalar fields, which simplifies the partial wave expansion. For fields with spin, the Wigner small-$d$ matrices should be used instead of Legendre polynomials. 

We will first present the dispersion relations and introduce the two main ingredients for obtaining the upper positivity bounds. The first are so-called \emph{null constraints}, a series of integral equations that constrain the partial wave amplitudes, which we will shortly review based on the formulation of~\cite{Du:2021byy}. The second are the \emph{UV unitarity bounds} away from positivity. The positivity part of the UV unitarity condition gives rise to the aforementioned positivity cone, while the non-positivity part enables us to fully close the allowed region, \emph{i.e} cap the cone. In Section \ref{sec:constWilson}, we formulate the problem of obtaining the positivity bounds (both lower and upper) as a linear program treating the partial wave amplitudes as decision variables. As two illustrating examples, in Section \ref{sec:onephi} and \ref{sec:two_fields}, we apply this method to obtain two-sided bounds for the single scalar theory and the bi-scalar theory. 

\subsection{Null constraints from crossing symmetry}

The null constraints arise when imposing full crossing symmetry~\cite{Tolley:2020gtv, Caron-Huot:2020cmc} on scattering amplitudes.  As we will see, if we start with dispersion relations with only two channels \emph{i.e.} the sum rules are only $su$ crossing symmetric, then the extra $st$ crossing symmetry needs to be imposed to achieve full crossing symmetry. These null constraints turn out to be very potent in restricting the space of Wilson coefficients, which parallels the situation in the conformal bootstrap where full crossing symmetry is fundamental to its recent success~\cite{Poland:2018epd}. Indeed, without the null constraints, naive use of the partial wave unitarity bounds in the sum rules would not constrain the coefficients from above.

We shall first derive the sum rules from the dispersion relations. Let us consider the amplitude, $A_{ijkl}(s,t)$, for a 2-to-2 scattering process $ij\to kl$ in $D=4$ dimensions, where $i,\,j,\,k,\,l=1,2,..., N$ label different low energy states in an effective theory. For simplicity, we will focus on the massless scalar case with Mandelstam variables satisfying $s+t+u=0$. We work in a self-conjugate basis for the scattering states to simplify the crossing relations, which are given by $A_{ijkl}(s,t)=A_{ikjl}(t,s)$ and  $A_{ijkl}(s,t)=A_{ilkj}(u,\,t)$ for $st$ and $su$ crossing respectively\footnote{One can straightforwardly switch to a non-self-conjugate basis by additionally conjugating the crossed states. In the massless limit, crossing symmetry relations are still trivial for fields with spin, and one can simplify the computations by restricting to independent helicity amplitudes.}. By the (complex) analyticity properties of $A_{ijkl}(s,t)$ in the UV theory, we can use Cauchy's integral formula along with the Froissart-Martin bound~\cite{Froissart:1961ux,Martin:1962rt} to relate it to a dispersive integral of its absorptive part, defined by its discontinuity across the real $s$-axis: $\mathrm{Abs}A(s,t) = \frac{1}{2i}\lim_{\epsilon \to 0^+} \left[A(s+i\epsilon,t)-A(s-i\epsilon,t)\right]$. In a time reversal invariant theory, which we focus on in this paper, the absorptive part is simply the imaginary part: ${\rm Abs}A_{ijkl}(s,t)={\rm Im}A_{ijkl}(s,t)$. Additionally making use of $su$ symmetry to infer the discontinuity along the negative $s$-axis in terms of the crossed amplitude, $A_{ijkl}(s+i\epsilon,\,t)=A_{ilkj}(-s-i\epsilon-t,\,t)$, one can express $A_{ijkl}$ as~\cite{Du:2021byy, deRham:2017avq}:
\begin{align}
 A_{ijkl}(s,t)& =  \frac{\lambda_{ijkl}}{-s} + \frac{\lambda_{ijkl}}{-t} + \frac{\lambda_{ijkl}}{-u} +a^{(0)}_{ijkl}(t)+a^{(1)}_{ijkl}(t)s
 \nl
    & +\int _{\Lambda^2}^\infty \frac{d\mu}{\pi(\mu+{t\over 2})^2}\left[\frac{(s+{t\over 2})^2}{\mu-s} \mathrm{Im}\, A_{ijkl}(\mu,\,t)
    +\frac{(u+{t\over 2})^2}{\mu-u} \mathrm{Im}\, A_{ilkj}(\mu,\,t)\right]\,,
    \label{disamp}
\end{align}
where $\lambda_{ijkl}$ are the residues of the poles that are required to be identical by crossing symmetry, and $a^{(0)}_{ijkl}$ and $a^{(1)}_{ijkl}$ are arbitrary functions of $t$ that will be fixed later thanks to $st$ crossing symmetry. $\Lambda$ is the scale associated with the appearance of the lowest modes in the UV theory. It is assumed to be much larger than $s$ and $t$ and can be identified with the cutoff of the EFT. The integration starts from $\Lambda^2$ because we have assumed that below $\Lambda^2$ the theory is perturbative and the tree level amplitude provides a good approximation. For loop amplitudes, a similar dispersive integral from $\Lambda^2$ to infinity can also be obtained by subtracting the low energy part of the dispersive integral \cite{deRham:2017imi, Li:2022aby, Zhang:2021eeo} or equivalently using the arc variables \cite{Bellazzini:2020cot}, the details of which are deferred for future work.

Eq.~\eqref{disamp} is a twice-subtracted dispersion relation for $A_{ijkl}(s,t)$ at fixed $t$, with the arbitrary subtraction point conveniently chosen to at $-{t}/{2}$. Subtracting twice is sufficient to render the terms in the dispersion relations finite, owing to the Froissart-Martin bound. Further subtracting the poles of the amplitude $A_{ijkl}(s,t)$, and defining $v\equiv s +\frac{t}{2}=-u-\frac{t}{2}$ for simplicity yields a dispersion relation for the pole-subtracted amplitude
\begin{align}
    \label{disampv}
   B_{ijkl}(s,t)&\equiv A_{ijkl}(s,t)-\frac{\lambda_{ijkl}}{-s} - \frac{\lambda_{ijkl}}{-t} - \frac{\lambda_{ijkl}}{-u}\\
   \nonumber
    &=\tilde{a}^{(0)}_{ijkl}(t)+a^{(1)}_{ijkl}(t)v
     +\frac{v^2}{\pi}\int_{\Lambda^2}^\infty\! \frac{d\mu}{(\mu+{t\over 2})^2}\left[\frac{\mathrm{Im}\, A_{ijkl}(\mu,\,t)}{\mu-v + {t\over2}} 
    +\frac{ \mathrm{Im}\, A_{ilkj}(\mu,\,t)}{\mu+v - {t\over2}}\right]\,,
\end{align}
where we have also introduced $\tilde{a}^{(0)}_{ijkl}(t)=a^{(0)}_{ijkl}(t)+a^{(1)}_{ijkl}(t){t\over2}$. Since we will be interested in studying the low energy limit of the amplitude modelled by an EFT, we expand $B_{ijkl}$ on the left hand side in terms of small $s$ and $t$,
\begin{align}
\label{Bexpansion}
    B_{ijkl}(s,t) & =  \sum_{n\geq 0}c_{ijkl}^{0,n}t^n + \sum_{n\geq 0}c_{ijkl}^{1,n}v t^n  + \sum_{m\geq 2}\sum_{n\geq 0} c_{ijkl}^{m,n} v^m t^n \,.
\end{align}
 While $c_{ijkl}^{0,n}$ and $c_{ijkl}^{1,n}$ correspond to the expansion coefficients of the soon-to-be determined $\tilde a^{(0)}_{ijkl}(t)$ and $a^{(1)}_{ijkl}(t)$, matching the expansions on both sides of Eq.~\eqref{disampv} for $m\geq 2$ leads to a set of sum rules for the Wilson coefficients
\begin{align}
    c_{ijkl}^{m,n} & = \Bigg\langle  \big[\rho_\ell^{ijkl}(\mu) + (-1)^m \rho_\ell^{ilkj}(\mu) \big]\sum_{p=0}^{n}\frac{L_\ell^p H_{m+1}^{n-p}}{\mu^{m+n+1}} \Bigg\rangle\,,~~~~m\geq 2.
     \label{eq:cijkl}
\end{align}
Here we have used the partial wave expansion for the absorptive part of the amplitude at energy scale $\sqrt{\mu}$,
\begin{align}
\mathrm{Im}\, A_{ijkl}(\mu,\,t)=16\pi \sum_{\ell=0}^\infty(2\ell+1)P_\ell\left(1+\frac{2t}{\mu}\right)\rho_\ell^{ijkl}(\mu)\,,
\end{align}
with $P_\ell$ being the Legendre polynomial and the ``spectral density''\footnote{It is not necessarily positive definite for non-elastic scatterings; see Eq.~(\ref{eq:sb1}).} defined as
\begin{equation}
    \rho^{ijkl}_\ell(\mu)\equiv \mathrm{Im}\, a_\ell^{ijkl}(\mu)\, ,
\end{equation}
and we have introduced shorthand notations
\begin{align}
&~~~~~~~~~~~~\Big\langle\,...\,\Big\rangle \equiv \sum_\ell 16(2\ell+1) \int_{\Lambda^2}^{\infty} d\mu\Big(\,...\,\Big),
\\
L_\ell^n&\equiv\frac{\Gamma(\ell+n+1)}{n!\Gamma(\ell-n+1)\Gamma(n+1)},~~
H_{m+1}^q \equiv\frac{\Gamma(m+q+1)}{(-2)^q\Gamma(q+1)\Gamma(m+1)}.
\end{align}
The sum rules will be the central objects we will use to extract the bounds on the Wilson coefficients.

Due to the crossing symmetry of the amplitude, the $c_{ijkl}^{m,n}$ must satisfy certain relations among themselves, called null constraints~\cite{Du:2021byy, Tolley:2020gtv, Caron-Huot:2020cmc}. Since the dispersion relation (\ref{disampv})/sum rules (\ref{eq:cijkl}), and thus $c_{ijkl}^{m\geq2,n}$ themselves are already $su$ symmetric, we only need to impose $st$ crossing symmetry on the amplitude $B_{ijkl}(s,t)=B_{ikjl}(t,s)$ to achieve full crossing symmetry.
This leads to a series of identities relating coefficients $c_{ijkl}^{m,n}$ with a common $m+n$:
\begin{align}
    n_{ijkl}^{p,q}
    &=\sum_{a=p}^{p+q}\frac{\Gamma(a+1)c_{ijkl}^{a,p+q-a}}{2^{a-p}\Gamma(p+1)\Gamma(a-p+1)}-\sum_{b=q}^{p+q}\frac{\Gamma(b+1)c_{ikjl}^{b,p+q-b}}{2^{b-q}\Gamma(q+1)\Gamma(b-q+1)} =0 \,.
    \label{eq:nc1}
\end{align}
For a fixed order $p+q={f}$, there are $f+1$ identities (given by choosing $p=0,1,...,f$), in which $c_{ijkl}^{0,p+q}$, $c_{ijkl}^{1,p+q-1}$, $c_{ikjl}^{0,p+q}$ and $c_{ikjl}^{1,p+q-1}$ are not useful to us since they do not have a dispersive integral representation like Eq.~(\ref{eq:cijkl}). We therefore eliminate them by first using four of the identities to solve for them as linear sums of $c_{ijkl}^{m\geq2,n}$ and then plugging the solution into the $f-3$ remaining identities.
One additional set of relations between $c_{ijkl}^{1,n}$ can be obtained by applying $su$ crossing symmetry on the amplitude, $B_{ijkl}(s,t)=B_{ilkj}(u,t)$. The fact that $s\leftrightarrow u$ corresponds to $v\to-v$ implies:
\be
\eval{c_{ijkl}^{1,n}+c_{ilkj}^{1,n}}_{c^{1,n}\rightarrow c^{m\geq2,n'}}=0\,,
\label{eq:nc2}
\ee
{in which we again substitute $c^{1,n}$ with $c^{m\geq2,n'}$ using the solution we just obtained.} We are then left with $f-2$ null constraints, which involve only $c_{ijkl}^{m,n}$, $c_{ikjl}^{m,n}$ and $c_{ilkj}^{m,n}$ with $m\geq 2$ and $m+n=f$.
The rest of the $c_{ijkl}^{m\geq2,n}$ terms in the expansion of Eq.~\eqref{eq:cijkl} are $su$ invariant by construction, leaving Eqs.~\eqref{eq:nc1} and~\eqref{eq:nc2} as the full set of null constraints implied by crossing symmetry. For example, the first set of nontrivial null constraints starts at $m+n=f=4$. (see \cite{Du:2021byy} for the higher order results), and they are 
\begin{align}
\begin{split}
\label{eq:nc_lowest}
    c_{{ijkl}}^{{2,2}}-c_{{ikjl}}^{{2,2}}+\frac{3 c_{{ijkl}}^{{3,1}}}{2}-\frac{3 c_{{ikjl}}^{{3,1}}}{2}+\frac{3 c_{{ijkl}}^{{4,0}}}{2}-\frac{3 c_{{ikjl}}^{{4,0}}}{2}=0\,,
    \\
    -c_{{ikjl}}^{{2,2}}-c_{{iklj}}^{{2,2}}+\frac{3 c_{{ijkl}}^{{3,1}}}{4}-\frac{c_{{ikjl}}^{{3,1}}}{2}-\frac{c_{{iklj}}^{{3,1}}}{2}+\frac{3 c_{{ilkj}}^{{3,1}}}{4}+c_{{ijkl}}^{{4,0}}+\frac{c_{{ikjl}}^{{4,0}}}{2}+\frac{c_{{iklj}}^{{4,0}}}{2}+c_{{ilkj}}^{{4,0}}=0\,,    
\end{split}
\end{align}
obtained from $n^{2,2}_{ijkl}$ and Eq.~\eqref{eq:nc2} with $n=3$. One possible way forward is to simply slice out the $c_{{ijkl}}^{m,n}$ coefficient space with the above null constraints, assuming one can extract positivity bounds for sufficiently many $c_{{ijkl}}^{m,n}$ coefficients directly from the sum rules (\ref{eq:cijkl}), via the mathematical moment approach~\cite{Arkani-Hamed:2020blm, Bellazzini:2020cot, Chiang:2021ziz}, for example. These $c_{{ijkl}}^{m,n}$ null constraints simply mean that the independent $c_{{ijkl}}^{m,n}$ coefficients live in a linear subspace of the full $c_{{ijkl}}^{m,n}$ space. However, a more efficient use of the null constraints is,  as we shall see shortly, to turn them into a set of identities for the UV partial wave amplitudes, which makes the name of null constraints more justified. Then, we will find that these reformulated null constraints can be used to facilitate the extraction of positivity bounds for the coefficients, very efficiently for the first few orders. 

Concretely, substituting the sum rules (\ref{eq:cijkl}) into the null constraints (\ref{eq:nc1}-\ref{eq:nc2}), we get a set of constraints on $\rho_{\ell}^{ijkl}$ that take the following general form:
\begin{align}
    & \sum_{\ell}16(2\ell+1)\int_{\Lambda^2}^\infty \frac{\dd\mu}{\mu^{r+4}}  \Bigg[ C_{r,i_r}(\ell) \rho _\ell^{{ijkl}}(\mu) + D_{r,i_r}(\ell) \rho _\ell^{{ijlk}}(\mu) + E_{r,i_r}(\ell) \rho _\ell^{{ikjl}}(\mu) \notag \\
    & \hspace{10em}+ F_{r,i_r}(\ell) \rho _\ell^{{iklj}}(\mu) + G_{r,i_r}(\ell) \rho _\ell^{{iljk}}(\mu) + H_{r,i_r}(\ell) \rho _\ell^{{ilkj}}(\mu)\Bigg] = 0\,, \label{eq:general nc}
\end{align}
where $r\equiv f-3=1,2,...$ is the order of null constraints, which corresponds to $p+q-3$ in Eq.~(\ref{eq:nc1}) or $n-2$ in Eq.~(\ref{eq:nc2}). At the $r$-th order, there are $r+1$ independent constraints for generic $i,j,k,l$, labelled by $i_r=1,...,r+1$.
$C_{r,i_r}(\ell),D_{r,i_r}(\ell),...,H_{r,i_r}(\ell)$ are polynomials in $\ell$ of degree $\leq 2r+2$. For the lowest order constraints given as an example in Eq~\eqref{eq:nc_lowest}, we have:
\begin{align}
\begin{alignedat}{2}\label{eq:nc_lowest_rho}
    C_{1,1}(\ell) &= \frac{\ell^4}{4}+\frac{\ell^3}{2}-\frac{\ell^2}{4}-\frac{\ell}{2},   &
    C_{1,2}(\ell) &= 2,   \\
    D_{1,1}(\ell) &= 0,   &
    D_{1,2}(\ell) &= -\frac{\ell^4}{4}-\frac{\ell^3}{2}+\frac{9 \ell^2}{4}+\frac{5 \ell}{2}-2,\\
    E_{1,1}(\ell) &= -C_{1,1}(\ell),   &
    E_{1,2}(\ell) &= -\frac{\ell^4}{4}-\frac{\ell^3}{2}+\frac{5 \ell^2}{4}+\frac{3 \ell}{2},\\
    F_{1,1}(\ell) &= 0,   &
    F_{1,2}(\ell) &= E_{1,2}(\ell),\\
    G_{1,1}(\ell) &=-\frac{\ell^4}{4}-\frac{\ell^3}{2}+\frac{13 \ell^2}{4}+\frac{7 \ell}{2}-6,\phantom{-  }  &
    G_{1,2}(\ell) &=  D_{1,2}(\ell),\\
    H_{1,1}(\ell) &= -G_{1,1}(\ell),   &
    H_{1,2}(\ell) &=  2\,.
\end{alignedat}
\end{align}
{For a specific choice of $i,j,k,l$, permuting $j,k,l$ can lead to extra constraints. On the other hand, additional degeneracies may exist, which reduces the number of independent ones. For example, if we choose $i,j,k,l$ to be $1,2,3,4$, in addition to the null constraint obtained by setting $i=1,j=2,k=3,l=4$ in Eq.~(\ref{eq:general nc}) in a form of
\begin{equation}
    \sum\int \cdots \Big[ C \rho _\ell^{{1234}} + D \rho _\ell^{{1243}}+ E \rho _\ell^{{1324}} + F \rho _\ell^{{1342}} + G \rho _\ell^{{1423}}+ H \rho _\ell^{{1432}} \Big] = 0\,, \notag
\end{equation}
we can also set $j,k,l$ to be other combinations of $2,3,4$ to obtain more null constraints relating $\rho _\ell^{{1234}}, \rho _\ell^{{1324}}, ...$, such as
\begin{equation}
    \sum\int \cdots \Big[ C \rho _\ell^{{1324}} + D \rho _\ell^{{1342}}+ E \rho _\ell^{{1234}} + F \rho _\ell^{{1243}} + G \rho _\ell^{{1432}}+ H \rho _\ell^{{1423}} \Big] = 0\,, \notag
\end{equation}
among several others. When deriving the null constraints for bi-scalar and four scalars, we need to exhaust all the permutations in Eq.~(\ref{eq:nc1}-\ref{eq:nc2}) to get the full set of null constraints.}

We stress that the $\rho^{ijkl}_\ell(\mu)$'s are unknown in general, since they can only be determined in terms of the scattering amplitudes of the UV theory -- note that in the dispersive integrals $\mu$ goes from $\Lambda^2$ to infinity. Nevertheless, we have shown here that they respect a series of null constraints that we will use later to bound the coefficients of the low energy EFT. These null constraints imply that the contributions from higher spin-$\ell$ components of $\rho^{ijkl}_\ell(\mu)$ are generally highly suppressed, which makes them very efficient at closing in the parameter space of the EFT coefficients. For the particular case of the $s^2$ coefficients, as we will see, they allow us to impose upper bounds on the Wilson coefficients from the sum rules, in conjunction with the upper bounds of the partial wave amplitudes from the unitarity in the UV.

Apart from these null constraints, crossing symmetry also implies symmetries among the spectral densities $\rho^{ijkl}_\ell$ when swapping their initial or final indices. To see this, note that the $tu$ crossing symmetry $A_{ijkl}(\mu,t)=A_{jikl}(\mu,u)=A_{ijlk}(\mu,u)$ implies that $a_\ell^{ijkl}(\mu)=(-1)^\ell a_\ell^{jikl}(\mu)=(-1)^\ell a_\ell^{ijlk}(\mu)$. This is due to the property of the Legendre polynomials $P_\ell(-z)=(-1)^\ell P_\ell(z)$, which leads to
\begin{align}
   A_{jikl}(\mu,u)&= 16\pi \sum_{\ell=0}^\infty \qty(2\ell+1)P_\ell\left(1+\frac{2u}{\mu}\right) a_{\ell}^{jikl}(\mu)
    \\
    &= 16\pi \sum_{\ell=0}^\infty \qty(2\ell+1) 
    (-1)^\ell P_\ell\left(1+\frac{2t}{\mu}\right) a_{\ell}^{jikl}(\mu)\,, 
    \\
    A_{ijkl}(\mu,t)&= 16\pi \sum_{\ell=0}^\infty \qty(2\ell+1) 
     P_\ell\left(1+\frac{2t}{\mu}\right) a_{\ell}^{ijkl}(\mu)\,,
    \label{eq:pwexpand2}
\end{align}
and the corresponding relations
\begin{equation}
\label{eq:rhoellRel}
\rho^{ijkl}_\ell = (-1)^\ell \rho^{jikl}_\ell = (-1)^\ell \rho^{ijlk}_\ell\,.
\end{equation}
In the following examples involving a single scalar, two scalars and the Higgs multiplet, we will use the identities implicitly to eliminate a number of variables. For example, they imply that $\rho^{1111}_\ell=\rho^{1122}_\ell=0$ for odd $\ell$.

\subsection{Unitarity conditions from the UV} 
\label{sec:unitarityBounds}

In order to obtain positivity bounds, we also need some inequality conditions on the spectral densities $\rho^{ijkl}_\ell(\mu)= {\rm Im} a_\ell^{ijkl}(\mu)$ from partial wave unitarity of the UV theory. These conditions lie at the heart of the positivity bounds, which in essence are just the projected-down version of these UV conditions on the IR theory by analyticity. Previous positivity bounds on the SMEFT only utilized the positivity part of these conditions, and now we shall also make use of the non-positivity parts. We will not use the full nonlinear unitarity, but restrict ourselves to the linear conditions, which are easily implemented in the linear programs and are sufficiently strong to close the Higgs positivity cone. In this subsection, we will briefly illustrate how to obtain a couple of these conditions, before listing all of the results we will use in the linear programs, deferring the details of the full derivations to Appendix \ref{app:partial wave unitarity}.

Recall that, for the generic process $ij\to kl$, partial wave unitarity can be written as (see Appendix \ref{app:partial wave unitarity})
\begin{equation}\label{eq:rhoijkl}
    \mathrm{Im} a_{\ell}^{ijkl} = \sum_{mn} \, \eta_{mn} a_{\ell}^{ijmn} \qty(a_{\ell}^{klmn})^* + \sum_{X\neq mn} a_{\ell}^{ij\rightarrow X} \qty(a_{\ell}^{kl\rightarrow X})^*\,, 
\end{equation}
where $a_\ell^{ij\rightarrow X}$ is the partial wave amplitude for the state $ij$ into some intermediate state $X$ and the phase space factor $\eta_{mn}=1/2$ if $m=n$ and $\eta_{mn}=1$ if $m\neq n$. The positivity part of the unitarity conditions refers to the fact that Eq.~(\ref{eq:rhoijkl}) implies that $\mathrm{Im} a_{\ell}^{ijkl}$ is a positive semi-definite matrix, viewing $ij$ and $kl$ as the two indices of the matrix. This allows one to obtain a set of positivity bounds via semi-definite programming \cite{Li:2021lpe, Du:2021byy}. The bounds extracted by this semi-definite programming method are optimal insofar as they only include the positivity part of partial wave unitarity. These ``true'' positivity bounds carve out a convex cone in the coefficient space, starting from its vertex and extending away to infinity. Thus, the ``true'' positivity bounds can be regarded as providing the ``lower bounds'' for the $s^2$ coefficients. In this paper, we will go beyond that by using the non-positivity parts of partial wave unitarity, which allow us to find upper bounds for all of the examples considered in this paper, capping all of the $s^2$ coefficients from above.

Using simple inequality relaxation and Eq.~(\ref{eq:rhoellRel}), we can derive from Eq.~(\ref{eq:rhoijkl}) the following two-sided constraints for the imaginary parts of the partial waves (see Appendix \ref{app:partial wave unitarity}):
\be
    0\leq \mathrm{Im}a_{\ell}^{iiii} \leq 2,~~~~ 
    0\leq \mathrm{Im}a_{\ell}^{ijij} \leq \frac{1}{2},  ~~~~i\neq j .
    ~
\ee
For example, to derive the first inequality above, we can consider the process $ii\to ii$. For this process, the terms on the right hand side of (\ref{eq:rhoijkl}) are all positive, so picking out only $X=ii$ for the intermediate states gives 
 \begin{align}
&\mathrm{Im}a_{\ell}^{ii ii}\geq \frac{1}{2}|a_{\ell}^{ii ii}|^2  = \frac{1}{2}\left(\mathrm{Im}a_{\ell}^{ii ii}\right)^2+ \frac{1}{2}\left(\mathrm{Re}a_{\ell}^{ii ii}\right)^2 ~~~
\Rightarrow ~~~0\leq \mathrm{Im}a_{\ell}^{ii ii} \leq 2.
\end{align}
Similarly, as shown in Appendix \ref{app:partial wave unitarity}, we can also derive two-sided conditions for the inelastic partial waves $\mathrm{Im}a_{\ell}^{iijj}$ and $\mathrm{Im}a_{\ell}^{ijkl}$ ($i\neq j\neq k\neq l$): $-1\leq \mathrm{Im}a_{\ell}^{iijj} \leq 1, ~ -\frac{1}{4}\leq \mathrm{Im}a_{\ell}^{ijkl} \leq \frac{1}{4}$. We see that, in contrast to those of $\mathrm{Im}a_{\ell}^{ii ii}$ and $\mathrm{Im}a_{\ell}^{ij ij}$,  they are not positive semi-definite.

We can actually get stronger conditions on these two coefficients. To see this, let us denote $\mathcal{X}^{ij,kl}$ as the set of all intermediate states that contribute to the scattering process $ij\rightarrow X \rightarrow kl$, and we must have $\mathcal{X}^{ij,kl} \subseteq \mathcal{X}^{ij,ij} \cap \mathcal{X}^{kl,kl}$. This leads to
\begin{align}
    \left|\mathrm{Im}a^{ijkl}_\ell\right|^2 &= \left|\sum_{X \in \mathcal{X}^{ij,kl}}a_{\ell}^{ij\to X} \left(a_{\ell}^{kl\to X}\right)^* \right|^2 
    \nl
    &\leq \sum_{X \in \mathcal{X}^{ij,kl}} \qty|a_{\ell}^{ij\to X}|^2 \sum_{X \in \mathcal{X}^{ij,kl}} \qty|a_{\ell}^{kl\to X}|^2
    \leq \sum_{X \in \mathcal{X}^{ij,ij}} \qty|a_{\ell}^{ij\to X}|^2 \sum_{X \in \mathcal{X}^{kl,kl}} \qty|a_{\ell}^{kl\to X}|^2
    \nl
    &=\mathrm{Im}a^{ijij}_\ell \mathrm{Im}a^{klkl}_\ell
    \leq \left(\frac{\mathrm{Im}a^{ijij}_\ell+\mathrm{Im}a^{klkl}_\ell}{2}\right)^2\,,   \label{eq:proofaijkl} 
\end{align}
where the first inequality arises from the Cauchy-Schwarz inequality and the second is due to the fact that the sets $\mathcal{X}^{ij,ij}$ and $\mathcal{X}^{kl,kl}$ are both larger than their intersection, $\mathcal{X}^{ij,kl}$. Taking the square root of the above inequality and recognizing that $\mathrm{Im}a^{ijij}_\ell\geq0$, we obtain
\be
\label{ImijklUpperB}
\big|\mathrm{Im}a^{ijkl}_\ell\big|  \leq  \frac{\mathrm{Im}a^{ijij}_\ell+\mathrm{Im}a^{klkl}_\ell}{2} 
,~~~~~ \forall ~i,j,k,l
\ee
This condition turns out to be pivotal to cap the Higgs positivity cone. Again using the Cauchy-Schwarz inequality plus some re-arrangements of the partial wave unitarity conditions, as shown in Appendix \ref{app:partial wave unitarity}, we can also derive 
\begin{align}
& \big| \mathrm{Im}a_{\ell}^{ij kl} \big|  \leq \frac{1}{2}-\frac{\mathrm{Im}a_{\ell}^{ij ij} + \mathrm{Im}a_{\ell}^{kl kl}}{2},~~~~
 \big|\mathrm{Im}a_{\ell}^{ii jj}\big|\leq 2-\frac{\mathrm{Im}a^{iiii}_\ell+\mathrm{Im}a^{jjjj}_\ell}{2}\,, \\[2mm]
 & \big|(\mathrm{Im}a^{iijj}_\ell + \mathrm{Im}a^{kkll}_\ell) \pm (\mathrm{Im}a^{iikk}_\ell + \mathrm{Im}a^{jjll}_\ell)\big| \leq 2\,,
\end{align}
where $i\neq j\neq k \neq l$. The first two conditions can be combined with inequality (\ref{ImijklUpperB}) to become
\be
\big|\mathrm{Im}a_{\ell}^{iijj}\big| \leq 1 - \Big|1-\frac{\mathrm{Im}a_{\ell}^{iiii}+\mathrm{Im}a_{\ell}^{jjjj}}{2} \Big| ,
~~~~
\big|\mathrm{Im}a_{\ell}^{ijkl}\big| \leq \frac{1}{4} - \Big|\frac{1}{4}-\frac{\mathrm{Im}a_{\ell}^{ijij}+\mathrm{Im}a_{\ell}^{klkl}}{2} \Big| .
\ee

In summary, we will use the following linear unitarity conditions in the linear programs to constrain the $s^2$ coefficients in the multi-field theory, written in terms of the spectral densities $\rho^{ijkl}_\ell= {\rm Im} a_\ell^{ijkl}$,
\begin{align}
   & 0\leq \rho_{\ell}^{iiii} \leq 2,  \quad \hspace{0.28em} \big|\rho_{\ell}^{iijj}\big| \leq 1 - \Big|1-\frac{\rho_{\ell}^{iiii}+\rho_{\ell}^{jjjj}}{2} \Big|, \, \notag
    \\[1mm]
    & 0\leq \rho_{\ell}^{ijij} \leq \frac{1}{2},  \quad \big|\rho_{\ell}^{ijkl}\big| \leq \frac{1}{4} - \Big|\frac{1}{4}-\frac{\rho_{\ell}^{ijij}+\rho_{\ell}^{klkl}}{2} \Big| ,  ~~~~~ (i\neq j \neq k \neq l)   \label{eq:sb1} 
    \\[3mm]
    & \big|(\rho^{iijj}_\ell + \rho^{kkll}_\ell) \pm (\rho^{iikk}_\ell + \rho^{jjll}_\ell)\big| \leq 2 . \notag
\end{align}

\subsection{Constraining Wilson coefficients}
\label{sec:constWilson}

We will now combine the previous ingredients to constrain the Wilson coefficients in a multi-field EFT. For phenomenologists, the $s^2$ coefficients in the tree-level scattering amplitudes, $c_{ijkl}^{2,0}= {\rm d}^2 A_{ijkl}(s,t)/(2{\rm d} s^2)$, are of greatest interest, as they correspond to a linear combination of dimension-8 Wilson coefficients (and quadratic terms of dimension-6 coefficients). These are the leading coefficients in the EFT expansion of the amplitudes for which positivity bounds can be obtained. Although we will focus on bounding only the $c_{ijkl}^{2,0}$ coefficients in this paper, we stress that the sum rules along with the null constraints and the UV unitarity conditions we have obtained can be used to constrain all of the Wilson coefficients. The positivity part of partial wave unitarity can be fully implemented to constrain the $c_{ijkl}^{2,0}$ coefficients via either the convex cone~\cite{Zhang:2020jyn} or the semi-definite program approach \cite{Li:2021lpe}. 
In this subsection we will formulate the problem of finding the upper bounds of the $c_{ijkl}^{2,0}$ coefficients as a LP in a generic form. 

Essentially, our strategy is to numerically find the upper bounds via the sum rules by brute force. This is made possible thanks to the unitarity conditions and the null constraints. To this end, we first change the integration variable from $\mu$ to $z={\Lambda^2}/{\mu}$, and discretize the $z$ variable as ${n}/{N}$ with $n=1,2,...,N$~\cite{Chiang:2022jep}, approximating $\int_0^1 \dd z$ with $\sum_{n=1}^N\frac{1}{N}$. We can also impose a cutoff $\ell_M$ on the sum over the UV spin $\ell$, since the null constraints require the higher spin spectral densities $\rho_{\ell}^{ijkl}$ to be highly suppressed (see Appendix \ref{app:analytical}). Numerically, we can indeed confirm that the results converge with $\ell_M$ (see Appendix \ref{app:convergence}). Thus, the $c_{ijkl}^{2,0}$ sum rules become
\begin{align}
    c_{ijkl}^{2,0} &= \sum_{\ell}(2\ell+1)\int_{\Lambda^2}^\infty \frac{\dd \mu}{\mu^3} 16(\rho_\ell^{{ijkl}}(\mu) + \rho_\ell^{{ilkj}}(\mu)) 
    \\
    ~~~~~~~~&~~~~~~\Rightarrow ~~
    c^{2,0}_{ijkl} \approx \frac{1}{\Lambda^4} \sum\limits_{\ell=0}^{\ell_M} (2\ell+1) \sum\limits_{n=1}^N \frac{1}{N}\frac{n}{N} 16\left(\rho_{\ell,n}^{ijkl}+\rho_{\ell,n}^{ilkj}\right)  ,
    \label{eq:cijkl20}
\end{align}
and the null constraints are similarly discretized\,\footnote{In the following, we will simply use ``=", instead of ``$\approx$", also for the discretized expressions.}. Thus, $c_{ijkl}^{2,0}$ becomes a linear combination of many unknown variables $\rho^{ijkl}_{\ell,n}\equiv \rho^{ijkl}_\ell({\Lambda^2N}/{n})$, subject to a set of linear inequality and equality constraints. Maximizing $\sum_{ijkl} \alpha^{ijkl} c^{2,0}_{ijkl}$ over all possible variables $\rho^{ijkl}_{\ell,n}$ is a well defined LP problem, 
where $\alpha^{ijkl}$ are constants specifying which direction to optimize in the parameter space furnished by all $c^{2,0}_{ijkl}$\,\footnote{For example, if we want to find the upper bound on an individual coefficient $c^{2,0}_{i_0j_0k_0l_0}$ for given $i_0,j_0,k_0,l_0$, we can set $\alpha^{ijkl}=\delta^i_{i_0} \delta^j_{j_0} \delta^k_{k_0} \delta^l_{l_0}$ in the optimization objective $\sum_{ijkl} \alpha^{ijkl} c^{2,0}_{ijkl}$; the lower bound on $c^{2,0}_{i_0j_0k_0l_0}$, on the other hand, can be obtained by setting $\alpha^{ijkl}=-\delta^i_{i_0}\delta^j_{j_0} \delta^k_{k_0} \delta^l_{l_0}$. Similarly, we can also look to find the bounds on combinations of the coefficients such as the upper bound on $c^{2,0}_{i_0j_0k_0l_0}+ \beta c^{2,0}_{i_1j_1k_1l_1}$, with $\beta$ being constant, by choosing $\alpha^{ijkl}=\delta^i_{i_0} \delta^j_{j_0} \delta^k_{k_0} \delta^l_{l_0} + \beta \delta^i_{i_1}\delta^j_{j_1} \delta^k_{k_1} \delta^l_{l_1}$.}. More explicitly, the programming task can be summarised as follows:
 \\[5mm]
\begin{align} 
    &\mathbf{Decision~variables} \notag \\
    &\hspace{1em} \rho_{\ell,n}^{ijkl} ,~~~ \ell=0,...,\ell_M,~~~n=1,2,...,N ,~~~ (\forall \ i,j,k,l) \label{LPgeneral1} \\
    &\mathbf{Maximize}  \notag \\
    &\hspace{1.2em} 
     \sum_{ijkl} \alpha^{ijkl} c^{2,0}_{ijkl}, ~~~{\rm where}~~ c^{2,0}_{ijkl} = \frac{1}{\Lambda^4} \sum\limits_{\ell=0}^{\ell_M} (2\ell+1) \sum\limits_{n=1}^N \frac{1}{N}\frac{n}{N} 16\left(\rho_{\ell,n}^{ijkl}+\rho_{\ell,n}^{ilkj}\right)\,, 
    \label{eq:c20_objective}\\
    &\mathbf{Subject\ to} \notag  \\
    &  \hspace{0.2em}
    \left. \begin{array}{lr}
    0\leq \rho_{\ell,n}^{iiii} \leq 2, \hspace{0.6em} \big|\rho_{\ell,n}^{iijj}\big| \leq 1 - \Big|1-\frac{\rho_{\ell,n}^{iiii}+\rho_{\ell,n}^{jjjj}}{2} \Big| \\[3mm]   0\leq \rho_{\ell,n}^{ijij} \leq \frac{1}{2}, ~ \big|\rho_{\ell,n}^{ijkl}\big| \leq \frac{1}{4} - \Big|\frac{1}{4}-\frac{\rho_{\ell,n}^{ijij}+\rho_{\ell,n}^{klkl}}{2} \Big| \\[4mm]
    \big|(\rho^{iijj}_{\ell,n} + \rho^{kkll}_{\ell,n}) \pm (\rho^{iikk}_{\ell,n} + \rho^{jjll}_{\ell,n})\big| \leq 2
    \end{array} \right\} ~~  (i\neq j \neq k \neq l);
    \notag \\[2mm]
    &~~ \rho^{ijkl}_{\ell,n} = (-1)^\ell \rho^{jikl}_{\ell,n} = (-1)^\ell \rho^{ijlk}_{\ell,n}, \hspace{6.5em} (\forall \ i,j,k,l) ;\notag\\[2mm]
    &~~ 
    \sum\limits_{\ell=0}^{\ell_M} (2\ell+1) \sum\limits_{n=1}^N
    \frac{1}{N} \left(\frac{n}{N}\right)^{r+2} \Bigg[ C_{r,i_r}(\ell) \rho _{\ell,n}^{{ijkl}} + D_{r,i_r}(\ell) \rho_{\ell,n}^{{ijlk}} + E_{r,i_r}(\ell) \rho_{\ell,n}^{{ikjl}} \notag\\ 
    & \hspace{14em}+ F_{r,i_r}(\ell) \rho_{\ell,n}^{{iklj}} + G_{r,i_r}(\ell) \rho_{\ell,n}^{{iljk}} + H_{r,i_r}(\ell) \rho_{\ell,n}^{{ilkj}}\Bigg] = 0\,,  \label{eq:discrete general nc}
\end{align}
Sometimes, the numerical results will converge faster if we also include a much larger partial wave, which will be labelled as $\ell_\infty$. In the next subsections, we will work through two example theories of a single scalar or a pair of scalars, illustrating how we perform the LP in detail.

The constraints listed in (\ref{eq:discrete general nc}) are generic ones. For a specific model, the symmetries of the theory can give rise to additional constraints, such as extra relations among the amplitude coefficients, that can also be implemented in the LP. They can act similarly to the null constraints in constraining the parameter space, reducing the number of decision variables by relating different $\rho_{\ell,n}^{ijkl}$.

We would like to emphasise here that positivity bounds are a unique, UV probe of the allowed space of IR Wilson coefficients. For a Lagrangian of dimension-8 operators of the form
\be
\mathcal{L}_{\mathrm{dim-8}}=\sum_I C_I^{(8)} \mathcal{O}_I^{(8)}=\sum_I \frac{\bar{C}_I^{(8)}}{\Lambda_{\mathrm{_{EFT}}}^4} \mathcal{O}_I^{(8)}\,,
\ee
the $s^2$ amplitude coefficients, {\it i.e.}, our objective functions $c^{2,0}_{ijkl}$, are simple combinations of $C_I^{(8)}$'s. $C_I^{(8)}$ therefore acquires a bound of the form $C_I^{(8)}\leq a/\Lambda^4$ via  Eq.~\eqref{eq:c20_objective}, where $a$ is a dimensionless number, and the dimensionless coefficient $\bar{C}_I^{(8)}$ is bounded by $C_I^{(8)}\leq a(\Lambda_{\mathrm{_{EFT}}}/\Lambda)^4$. If we identify the cutoff scale of EFT $\Lambda_{\mathrm{_{EFT}}}$ with our lower limit of integration $\Lambda$ in the dispersion relation, which is always true if we assume perturbativity of the UV theory below $\Lambda_{\mathrm{_{EFT}}}$ (see Eq.~(\ref{disamp}) and the discussion below it), then our result directly bounds the dimensionless $\bar{C}_I^{(8)}$ through $\bar{C}_I^{(8)}\leq a$ with no dependence on $\Lambda_{\rm EFT}$ or any scale. This is in contrast to other bounds that might be obtained from experiments or partial wave unitarity in the low energy EFT which clearly suffer from a degeneracy due to the lack of knowledge on both $\bar C_I^{(8)}$ and $\Lambda_{\mathrm{_{EFT}}}$. The former sets limits on the combination $\bar C_I^{(8)}/\Lambda_{\mathrm{_{EFT}}}^{4}\sim c^{2,0}_{ijkl}$, while the latter constrains $\bar C_I^{(8)}s^2/\Lambda_{\mathrm{_{EFT}}}^{4}\sim s^2 c^{2,0}_{ijkl}$, meaning that certain theoretical assumptions are always required to make statements about the couplings or mass scale of the associated UV theory. Being able to bound $\bar C_I^{(8)}$ directly is therefore highly complementary to the information obtained from the other bounds, as we discuss in Section~\ref{sec: pheno} for the case of Higgs operators in the SMEFT.

Before proceeding to some simple examples, we would like to comment on the robustness of our approximation to ignoring loop contributions in the low energy part of the dispersive integral (Eq.~(\ref{disamp})). As we will see shortly, a typical two-sided positivity bound under this approximation is
\begin{equation}
    0\leq \frac{\bar{C}}{(4\pi)^2} \leq {\cal O}(1) \,,
\end{equation}
where $\bar{C}$ is a dimensionless Wilson coefficient. After including the one-loop contribution of the EFT operators, it becomes
\begin{equation}
    \label{eq:loopcorr}
    0\leq \frac{\bar{C}}{(4\pi)^2} - k \frac{\bar{C}^2}{(4\pi)^4} \leq {\cal O}(1) \,,
\end{equation}
where the second term is from the dispersive integral from the threshold to $\Lambda$, and $k$ is a factor depending on the specific UV theory that shapes the dispersive relation. For a UV-dominated dispersive relation, $k$ is usually much less than unity \cite{EliasMiro:2023fqi}. Taking $k=0.1$ and the upper limit to be 1, then Eq.~(\ref{eq:loopcorr}) yields
\begin{equation}
    0\leq \frac{\bar{C}}{(4\pi)^2} \leq 1.13\,,
\end{equation}
which gives a small correction to the tree-level approximation. For a UV theory whose dispersive integral is IR-dominated, the loop contribution could be sizable, a systematic treatment of which is left for future work.

\subsection{Example: single scalar} 
\label{sec:onephi}

We start with the simplest case where the low energy theory includes a single massless, real scalar. As is well known, the associated EFT has a single independent operator at dimension-8, whose Wilson coefficient maps to $c^{2,0}$ at tree-level:
 \begin{align}
 \mathcal{L}^{(8)} = \frac{g_{2}}{2
 }(\partial^\mu\phi\partial_\mu \phi)^2 
\,\quad\Rightarrow\quad c^{2,0}=2g_2 .
 \end{align}
 It reduces the generic scattering situation to $i=j=k=l$ and we therefore omit those indices. Also, as a result of Eq.~(\ref{eq:rhoellRel}), the spectral density $\rho_{\ell,n}$ vanishes for odd $\ell$. Since the $c^{2,0}$ coefficient is a positive sum of spectral densities which are themselves positive for the single scalar case, the lower bound is trivially zero and corresponds to the ``first'' positivity bound: $c^{2,0}\geq0$. 

To get an upper bound on $c^{2,0}$, we exploit the LP proposed in the last subsection, which in this case reads
\begin{align}
    &\mathbf{Decision~variables} \notag \\[-1mm]
    &\hspace{1.2em} \rho_{\ell,n}\nn\\
    &\mathbf{Maximize} \notag \\
    &\hspace{1.2em} c^{2,0} = \frac{1}{\Lambda^4} \sum\limits_{\ell=0,2,...,\ell_M;\ell_\infty} (2\ell+1) \sum\limits_{n=1}^N \frac{1}{N}\frac{n}{N} 32\rho_{\ell,n}\,, \label{eq:c1111} \\
    &\mathbf{Subject\, to} \notag  \\
    &\hspace{1.2em}  0\leq \rho_{\ell,n} \leq 2 \,,\\
    &\hspace{1.2em} \sum\limits_{\ell=0,2,...,\ell_M;\ell_\infty}(2\ell+1) \sum\limits_{n=1}^N 
    \frac{1}{N} \left(\frac{n}{N}\right)^{r+2}  C^{1111}_{r,i_r}(\ell) \rho_{\ell,n} = 0\,, \label{eq:nc_1field}
\end{align}
where $C^{1111}_{r,i_r}$ represents certain linear combinations of the corresponding $C(\ell),D(\ell),...,H(\ell)$ in Eq.~(\ref{eq:general nc}) when $i=j=k=l$, which are polynomials in $\ell$ of degree $\leq 2r+2$. In this case, we observe there to be $\lfloor (r-1)/3\rfloor+1$ null constraints at the $r$-th order, where $\lfloor x\rfloor$ denotes the floor integer function. 
Explicitly, at the first three orders, for example, we have
\begin{align}
    C^{1111}_{1,1} &= \ell^4 + 2\ell^3 - 7\ell^2 -8 \ell\,, \\
    C^{1111}_{2,1} &= \ell^6 + 3\ell^5 - \frac{37}{2}\ell^4 -42 \ell^3+\frac{107}{2}\ell^2 +75\ell\,, \\
    C^{1111}_{3,1} &= \ell^8 + 4\ell^7 - 38 \ell^6 -128\ell^5 + 457 \ell^4 + 1132 \ell^3 -1860\ell^2 -2448\ell\,. 
\end{align}

To evaluate this LP problem, we set a cutoff $\ell_M=100$ for the sum over the partial waves. However, we also generally include a very large partial wave term, $\ell_\infty$, to improve the numerical convergence; here we have chosen $\ell_\infty=10000$. The optimisation has ${\cal O}(\ell_M/2)\times N$ decision variables $\rho_{\ell,n}$ (recall that in this case $\rho_{\ell=\mathrm{odd},n}=0$), each subject to a unitarity bound and additionally constrained by the set of null constraints up to a given order, $r$.
\begin{figure}[htb!]
    \centering
    \includegraphics[height=0.35\textheight]{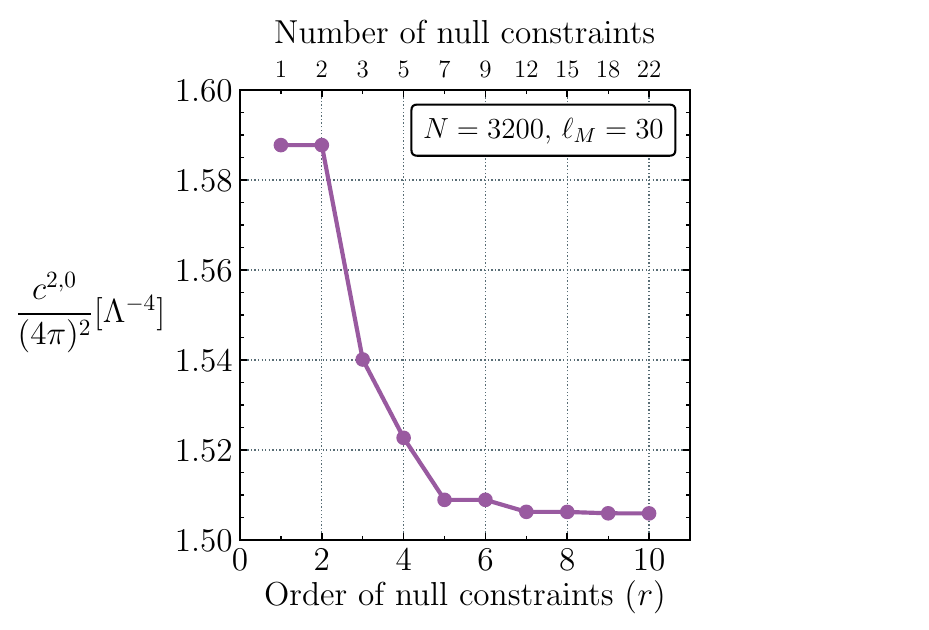}
    \caption{Upper bound on the amplitude coefficient for the single real scalar theory, $c^{2,0}$, in units of $(4\pi)^2$ and $\Lambda^{-4}$ resulting from our optimisation procedure as a function of the order of null constraints applied. The upper $x$-axis labels count the number of independent null constraints appearing up to each order. }\label{fig:singlez2}
\end{figure}
The result is shown in Figure \ref{fig:singlez2}, which plots the upper bound on $c^{2,0}$ in units of $(4\pi)^2$ and $\Lambda^{-4}$ as a function of the order of null constraints applied. $N=3200$ points are used for the discretized integral, which is observed to yield a solid convergence as $N$ increases (See Appendix~\ref{app:convergence}). 
We can see that the result converges quickly upon applying higher and higher orders of null constraints, and conclude that the $s^2$ coefficient $c^{2,0}$ is bounded both from above and below:
\begin{equation}
    0\leq \frac{c^{2,0}}{(4\pi)^2} \leq \frac{1.506}{\Lambda^4}.
\end{equation}
Note that the units of $(4\pi)^2$ are consistent with naive dimensional analysis \cite{Manohar:2018aog}. The upper bound on $c^{2,0}$ has been previously obtained semi-analytically using only the first null constraint in Ref.~\cite{Caron-Huot:2020cmc}\footnote{Note that our $c^{2,0}$ is twice the $g_2$ coefficient defined in Ref.~\cite{Caron-Huot:2020cmc}.}. We see that including the higher order null constraints leads to a 5\% improvement. We have also attempted to extend the method of \cite{Caron-Huot:2020cmc} to include higher order null constraints, but it becomes computationally costly and numerically unstable very quickly. However, for the low orders, the two methods do agree with each other.

\subsection{Example: bi-scalar theory} 
\label{sec:two_fields}

For a theory containing two scalar fields, $\phi_{1,2}$, we take $i,j,k,l$ to be permutations of 1 and 2. For simplicity, we focus on the case when $i,j,k,l$ includes two 1's and two 2's, \emph{i.e.}, we bound the coefficients $c_{1212}^{2,0}$ and $c_{1122}^{2,0}=c_{1221}^{2,0}$.\footnote{These coefficients correspond to the subset of bi-scalar amplitudes allowed when imposing a $\mathbb{Z}_2$ symmetry on each scalar field $\phi_i\leftrightarrow-\phi_i$.  For the bounds on
$c^{2,0}_{1112}$, $c^{2,0}_{1121}$,..., the derivation is similar to our example for $c^{2,0}_{1122}$.
}
The bounds on $c^{2,0}_{1111}$ and $c^{2,0}_{2222}$ are exactly the same as the previous single scalar example since the objective function, (linear) unitarity conditions and null constraints that make up each ``$iiii$'' system are independent and decoupled from all others.
For the null constraints, we set $\{i,j,k,l\}=\{1,1,2,2\},\{1,2,1,2\},\{1,2,2,1\}$ at each order, $r$, of Eq.(\ref{eq:discrete general nc}) to find the independent constraints for the LP, which results in $(r+1)$ null constraints at the $r$-th order. The null constraints can be classified into two groups, those that only involve $\rho_{\ell,n}^{1111}$ or $\rho_{\ell,n}^{2222}$, and those that only involve $\rho_{\ell,n}^{1122}$ and $\rho_{\ell,n}^{1212}$ (recall that $\rho_{\ell,n}^{1221}=(-1)^\ell \rho_{\ell,n}^{1212}$). They are connected by the inequality $\big|\rho_{\ell,n}^{1122}\big| \leq 1 - \big|1-(\rho_{\ell,n}^{1111}+\rho_{\ell,n}^{2222})/2 \big|$.
Now the LP problem can be expressed as
\begin{align} 
    &\mathbf{Decision~variables} \notag \\
    &\hspace{1.2em} \rho^{1111}_{\ell,n}, \ \rho^{2222}_{\ell,n}, \ \rho^{1122}_{\ell,n}, \ \rho^{1212}_{\ell,n}\,, \\
    &\mathbf{Maximize/Minimize} \notag \\
    &\hspace{1.2em} c_{1212}^{2,0} = \frac{1}{\Lambda^4} \smaller{\sum\limits_{\ell=0,...,\ell_M;\ell_\infty}}(2\ell+1) \smaller{\sum\limits_{n=1}^N \frac{1}{N}\frac{n}{N}}  32\rho_{\ell,n}^{1212}\,, \label{eq:c1212}  \\
    &\hspace{1.2em} c_{1122}^{2,0} = c_{1221}^{2,0} = \frac{1}{\Lambda^4} \smaller{\sum\limits_{\ell=0,...,\ell_M;\ell_\infty}}(2\ell+1) \smaller{\sum\limits_{n=1}^N \frac{1}{N}\frac{n}{N}}  16\Big(\rho_{\ell,n}^{1122}+(-1)^\ell \rho^{1212}_{\ell,n}\Big)\,, \label{eq:c1122} \\
    &\mathbf{Subject\, to} \notag  \\
    &\hspace{0.8em}\left\{ 
    \begin{array}{lr}
        0\leq \rho^{1111}_{\ell,n} \leq 2, \hspace{1.2em} 0\leq \rho^{2222}_{\ell,n} \leq 2\,, \\[1mm]
        \sum\limits_{\ell=0,...,\ell_M;\ell_\infty}(2\ell+1) \sum\limits_{n=1}^N \frac{1}{N}\left(\frac{n}{N}\right)^{r+2} C^{1111}_{r,i_r}(\ell) \rho^{1111}_{\ell,n} = 0\,, \\
        \sum\limits_{\ell=0,...,\ell_M;\ell_\infty}(2\ell+1) \sum\limits_{n=1}^N \frac{1}{N}(\frac{n}{N})^{r+2} C^{1111}_{r,i_r}(\ell) \rho^{2222}_{\ell,n} = 0\,,
    \end{array}
    \right. \label{eq:nc11112222} \\[3mm]
&\hspace{1.2em}\big|\rho_{\ell,n}^{1122}\big| \leq 1 - \Big|1-\frac{\rho_{\ell,n}^{1111}+\rho_{\ell,n}^{2222}}{2} \Big| \label{eq:ineq1122}\,, \\[3mm]
    &\hspace{0.8em}\left\{ 
    \begin{array}{lr}
    0\leq \rho_{\ell,n}^{1212} \leq \frac{1}{2}\,, \\[1mm]
    \sum\limits_{\ell=0,...,\ell_M;\ell_\infty}(2\ell+1) \sum\limits_{n=1}^N \frac{1}{N}\left(\frac{n}{N}\right)^{r+2}  \Big[C^{1122}_{r,i_r}(\ell)\rho_{\ell,n}^{1122}+C^{1212}_{r,i_r}(\ell) \rho_{\ell,n}^{1212} \Big] =0\,,
    \end{array}
    \right. \label{eq:nc1122}
\end{align}
where for the null constraints in the last line, $C^{1122}_{r,i_r}$ and $C^{1212}_{r,i_r}$, just like $C^{1111}_{r,i_r}$, are polynomials in $\ell$ of order $\leq 2r+2$. For example, the first order of them ($r=1$, $i_r=1,2$) are explicitly given by 
\begin{align}
    C^{1122}_{1,1} &= -\ell^4 - 2\ell^3 + \ell\,, \\
    C^{1212}_{1,1} &= \Big(\frac{5}{2}\ell^4+5 \ell^3-\frac{35}{2}\ell^2-20\ell+28\Big) + (-1)^\ell \Big( -\ell^4-2 \ell^3+14 \ell^2+15 \ell-28 \Big)\,, \\
    C^{1122}_{1,2} &= -\ell^4 - 2\ell^3 - \ell^2\,, \\
    C^{1212}_{1,2} &= \Big(3 \ell^4+6 \ell^3-21 \ell^2-24 \ell+32\Big) + (-1)^\ell \Big(-\ell^4-2 \ell^3+15 \ell^2+16 \ell-32\Big)\,.
\end{align}
It is worth noting that the upper bound on $|\rho^{1122}_{\ell,n}|$, which is acquired by first constraining $\rho^{1111}_{\ell,n}$ and $\rho^{2222}_{\ell,n}$, and then using the inequality on $|\rho^{1122}_{\ell,n}|$ (i.e., Eq.(\ref{eq:nc11112222}-\ref{eq:ineq1122})), is necessary to produce a convergent upper bound for $c_{1212}^{2,0}$ and $c_{1122}^{2,0}$ via Eq.(\ref{eq:c1212}-\ref{eq:c1122}). The constraints in Eq.(\ref{eq:nc1122}) alone are not sufficient, as explained in Appendix \ref{app:analytical}.

The formalism above can be further simplified as follows, in a way that will be crucial in reducing the number of variables in the case of SMEFT Higgs operators addressed in the next section. In this case it amounts to simply defining $R^{1111}_{\ell,n}\equiv (\rho^{1111}_{\ell,n}+\rho^{2222}_{\ell,n})/2$, which satisfies exactly the same bounds and null constraints as $\rho^{1111}_{\ell,n}$, leading to one fewer set of variables along with a reduction of constraints in the LP problem:
\begin{align} 
    &\mathbf{Decision~variables} \notag \\
    &\hspace{1.2em} R^{1111}_{\ell,n},\ \rho^{1122}_{\ell,n}, \ \rho^{1212}_{\ell,n}\,, \\
    &\mathbf{Maximize/Minimize} \notag \\
    &\hspace{1.2em} c_{1212}^{2,0} = \frac{1}{\Lambda^4} \smaller{\sum\limits_{\ell=0,1,2,...,l_M;l_\infty}}(2\ell+1) \smaller{\sum\limits_{n=1}^N \frac{1}{N}\frac{n}{N}} 32\rho_{\ell,n}^{1212}\,, \\
    &\hspace{1.2em} c_{1122}^{2,0} = c_{1221}^{2,0} = \frac{1}{\Lambda^4} \smaller{\sum\limits_{\ell=0,1,2,...,l_M;l_\infty}} (2\ell+1) \smaller {\sum\limits_{n=1}^N \frac{1}{N}\frac{n}{N}} 16\Big(\rho_{\ell,n}^{1122}+(-1)^l\rho_{\ell,n}^{1212}\Big)\,, \\
    &\mathbf{Subject\,to} \notag  \\
    &\hspace{0.8em}\left\{ 
    \begin{array}{lr}
        0\leq R^{1111}_{\ell,n} \leq 2\,, \\
        \sum\limits_{\ell=0,...,\ell_M;\ell_\infty}(2\ell+1) \sum\limits_{n=1}^N \frac{1}{N}(\frac{n}{N})^{r+2} C^{1111}_{r,i_r}(\ell) R^{1111}_{\ell,n} = 0 \,,
    \end{array}
    \right. \\[3mm]
    \label{eq:extra_inequality}
    &\hspace{1.2em}\big|\rho_{\ell,n}^{1122}\big| \leq 1 - \big|1- R^{1111}_{\ell,n} \big|\,, \\[3mm]
    &\hspace{0.8em}\left\{ 
    \begin{array}{lr}
        0\leq \rho_{\ell,n}^{1212} \leq \frac{1}{2}\,, \\[1mm]
        \sum\limits_{\ell=0,...,\ell_M;\ell_\infty}(2\ell+1) \sum\limits_{n=1}^N \frac{1}{N}(\frac{n}{N})^{r+2} \Big[C^{1122}_{r,i_r}(\ell)\rho_{\ell,n}^{1122}+C^{1212}_{r,i_r}(\ell) \rho_{\ell,n}^{1212} \Big]\,.
    \end{array}
    \right.
\end{align}
The problem has roughly four times as many variables as in the single scalar field case. Each variable is bounded by unitarity and there are a set of ${\cal O}(\ell_M\times N)$ additional inequalities from Eq.~\eqref{eq:extra_inequality}, relating the two field scattering amplitudes to the single field ones. 
The set of null constraints corresponds to the union of single and two-field constraints, of which we observe there are $\lfloor(r-1)/3\rfloor+r+2$ at the $r$-th order.

\begin{figure}[htbp]
  \begin{center}
     \includegraphics[width=1\textwidth]{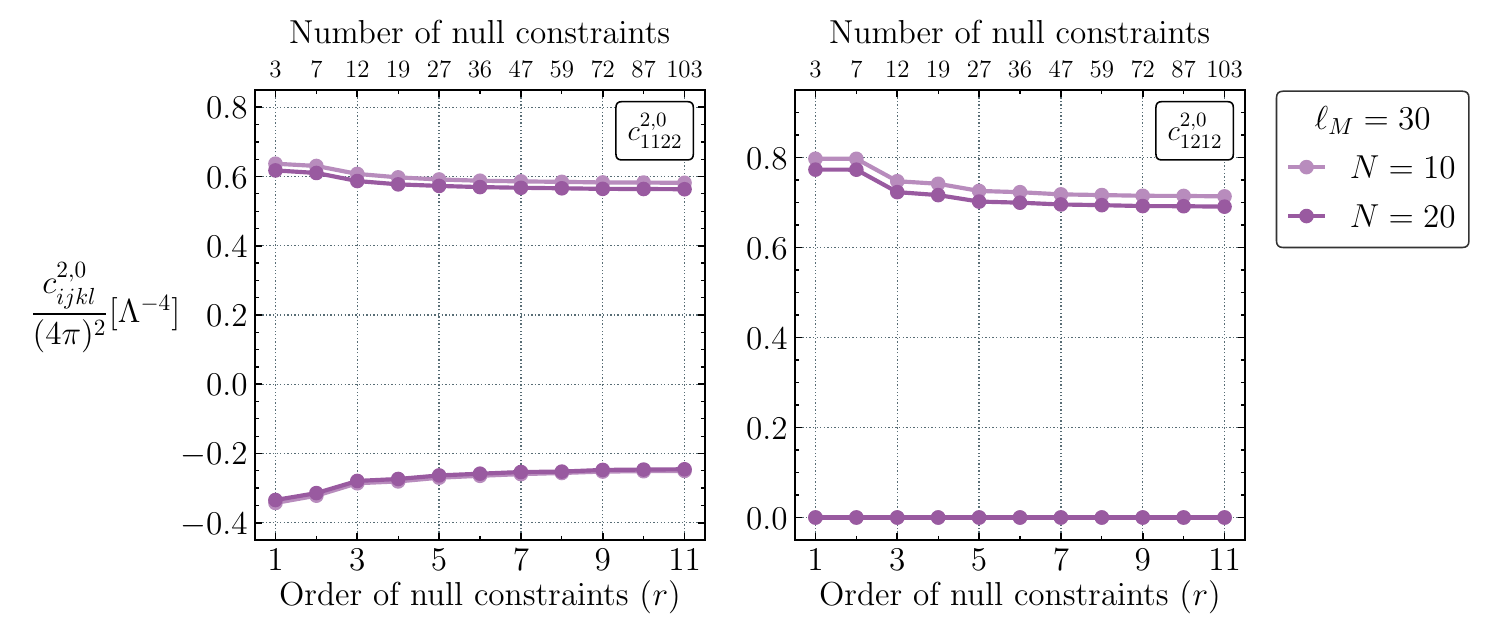}
  \end{center}
\caption{
Upper/lower bounds on the amplitude coefficients for the bi-scalar theory, $c^{2,0}_{1212}$ and $c^{2,0}_{1122}$, in units of $(4\pi)^2$ and $\Lambda^{-4}$ resulting from our optimisation procedure as a function of the order of null constraints applied. The upper axis label indicates the number of independent null constraints appearing up to each order.}
\label{fig:2field}
\end{figure}

\begin{figure}[htbp]
\centering
     \includegraphics[height=0.3\textheight]{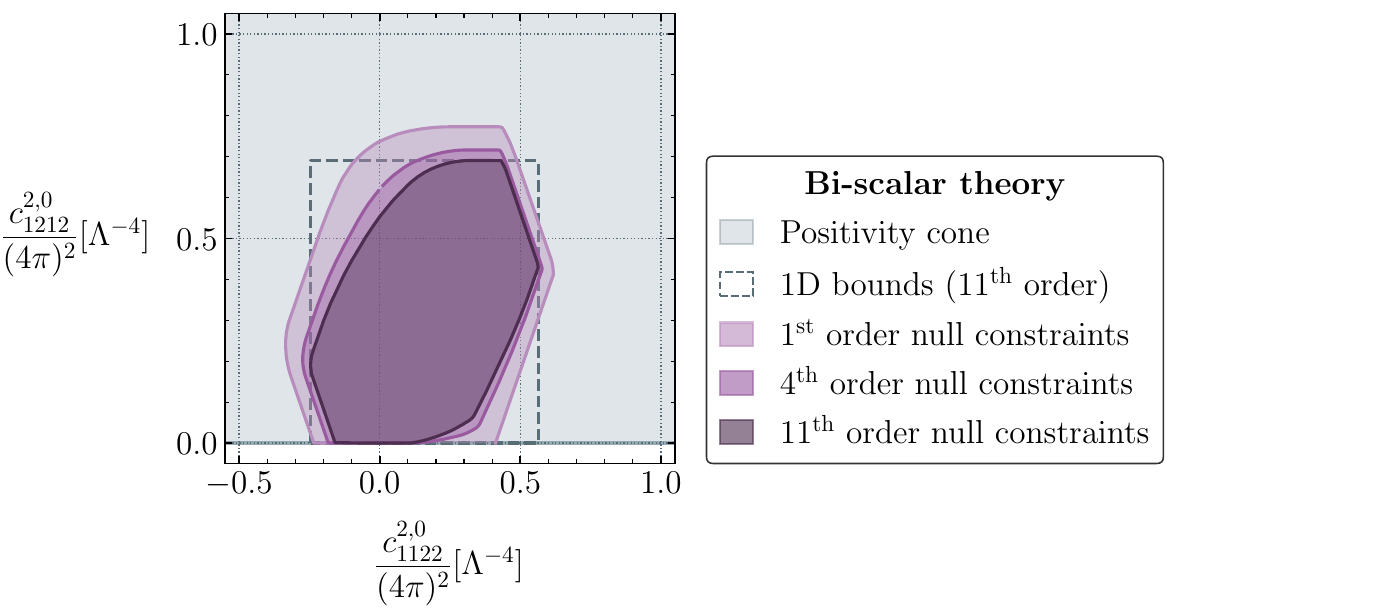}
\caption{
Allowed parameter space for the amplitude coefficients for the bi-scalar theory, $c^{2,0}_{1212}$ and $c^{2,0}_{1122}$, in units of $(4\pi)^2$ and $\Lambda^{-4}$ resulting from our optimisation procedure. Bounds obtained by employing null constraints at the 1$^\text{st}$, 4$^\text{th}$ and 11$^\text{th}$ order are shown in successive darker purple shaded regions. For comparison, the grey shaded region shows the allowed half-space projected down from the positivity cone of the bi-scalar theory. The dotted lines delimit the region enclosed by the corresponding one dimensional bounds obtained with 11$^{\text{th}}$ order null constraints shown in Figure~\ref{fig:2field}.} 
\label{fig:2field2D}
\end{figure}

To accomplish this LP computation economically, we choose $N=10$ and $20$, with $\ell_M=30$, $\ell_\infty=100$. Although these are not as large as for the single scalar case, we have carefully checked the convergence of the optimization (see  Appendix \ref{app:convergence}). The greatest inaccuracy arises from the coarser discretization parameter, $N$, which makes the resulting bound slightly ($\sim5 \%$) looser than its true value (approximated by the $N=100$ case).
The results of the optimisation are shown in Figure~\ref{fig:2field}. 
In this case, we go to the 11th order where there are 103 null constraints applied, and the result gradually levels off. We conclude that the bounds are
\begin{align}
    \label{eq:2field_1D_1}
    0 & \leq \frac{c_{1212}^{2,0}}{(4\pi)^2} \leq \frac{0.691}{\Lambda^4}\,, \\
     \label{eq:2field_1D_2}
   -\frac{0.246}{\Lambda^4} & \leq \frac{c_{1122}^{2,0}}{(4\pi)^2} = \frac{c_{1221}^{2,0}}{(4\pi)^2} \leq \frac{0.563}{\Lambda^4}\,.
\end{align}
We also determine the 2-D allowed region in the space of $c_{1122}^{2,0}$ and $c_{1212}^{2,0}$, by maximizing many linear combinations in the form of $\alpha c_{1122}^{2,0} +\beta c_{1212}^{2,0}$. The result is shown in Figure~\ref{fig:2field2D}, where successively darker shaded regions indicate the use of increasing orders of null constraints, where we have fixed $N=20$ and $\ell_M=30$.
 The allowed region converges as we increase the null constraint order, and we do not expect significant improvements beyond the 11$^\text{th}$ order. The new allowed region can be compared to the half-space that is allowed by traditional ``lower'' positivity bounds obtained from the positivity cone method~\cite{Li:2021lpe}, shaded in grey. The dashed lines also show the square region delimited by the 1D bounds in  Eqs.~\eqref{eq:2field_1D_1} and ~\eqref{eq:2field_1D_2}. We see that the 2D regions coincide along the line $c^{2,0}_{1212}=0$, where the traditional positivity bounds apply. The points where each coefficient is maximised also coincides with the 1D bounds that we obtained above, as expected. 
Overall, the true bounds in the 2D plane rule out a significant, additional portion of the parameter space with respect to the independent 1D case.
Since they make use of the most information available, these are the strongest bounds on the bi-scalar EFT Wilson coefficients that have been found to date.

\section{Upper bounds on Higgs scattering coefficients}
\label{sec: higgsupperbounds}
 Having established the formalism and gone through the warm-up examples, we now proceed to study the more realistic case of two Higgs boson scattering in the SMEFT. In this section, we identify relevant features of the theory and use them to develop the optimization procedure to bound the Wilson coefficients.
\subsection{Effective operators for Higgs scattering}
\label{sec: higgseft}
The Higgs field transforms as a complex doublet under the weak $\mathrm{SU}(2)$ gauge symmetry.
To make contact with the real scalar scatterings that we have previously analysed, 
we parametrize the Higgs doublet with four real components
\ba
H=\frac{1}{\sqrt{2}}\begin{pmatrix}
\phi_1+i\phi_2\\
\phi_3+i\phi_4
\end{pmatrix}\,.
\label{4scalarsH}
\ea
We consider the set of 2-to-2 scatterings $\phi_i \phi_j \to \phi_k \phi_l$, where $i,\,j,\,k,\,l$ runs from 1 to 4, and will compute the $s^2$ coefficients $c_{ijkl}^{2,0}$ defined in the last section from the effective Lagrangian. $SU(2)$ gauge invariance prevents the construction of an operator made out of three Higgs fields and derivatives, so there are no higher-derivative, three-point Higgs interactions in the massless limit that could contribute quadratically to the $s^2$ growth. This means that the only contributions to $c_{ijkl}^{2,0}$ come from single insertions of four-derivative dimension-8 operators~\cite{Eboli:2006wa, Almeida:2020ylr, Li:2020gnx, Murphy:2020rsh}
\begin{align}
\mathcal{O}^{(1)}_{H^4}&=\left(D_\mu H^\dagger D_\nu H\right)\left(D^\nu H^\dagger D^\mu H\right)\,,
\label{eq:dim8H1}\\
\mathcal{O}^{(2)}_{H^4}&=\left(D_\mu H^\dagger D_\nu H\right)\left(D^\mu H^\dagger D^\nu H\right)\,,
\label{eq:dim8H2}\\
\mathcal{O}^{(3)}_{H^4}&=\left(D^\mu H^\dagger D_\mu H\right)\left(D^\nu H^\dagger D_\nu H\right)\,.
\label{eq:dim8H3}
\end{align}

For convenience, we relabel the four real scalars in the Higgs doublet as
\ba
H=\frac{1}{\sqrt{2}}\begin{pmatrix}
H^1_1+i H^1_2\\[1mm]
H^2_1+i H^2_2
\end{pmatrix}
\,,
\ea
and consider the contribution of the dimension-8 operators in Eqs.\,(\ref{eq:dim8H1})--(\ref{eq:dim8H3}), to the four-Higgs scattering amplitude, $\mathcal{M}$. The $s^2$ coefficient of this amplitude can then be easily extracted:
\begin{align}
a^{2,0\,a b c d}_{\hspace{12pt}I J K L}&\equiv{1\over 2}\frac{{\rm d}^2}{{\rm d} s^2}\mathcal {M}\left(H_I^{a} H_J^{b}\to H_K^{c}H_L^{d}\right)
\nl
&=\frac{1}{4}\delta_{IJ}\delta_{KL}\left[\delta_{ab}\delta_{cd}\left(C_1+C_2+2C_3\right)+\delta_{ad}\delta_{bc}\left(C_1-C_2\right)\right]
\nl
&\quad+\frac{1}{4}\delta_{IK}\delta_{JL}\left[\delta_{ac}\delta_{bd}\left(2C_1+2C_2\right)+\left(\delta_{ad}\delta_{bc}+\delta_{ab}\delta_{cd}\right)\left(C_2-C_1\right)\right]
\nl
&\quad+\frac{1}{4}\delta_{IL}\delta_{JK}\left[\delta_{ad}\delta_{bc}\left(C_1+C_2+2C_3\right)+\delta_{ab}\delta_{cd}\left(C_1-C_2\right)\right]\,,
\label{realH}
\end{align}
where $C_1$, $C_2$ and $C_3$ are the Wilson coefficients for the operators $\mathcal{O}^{(1)}_{H^4}$, $\mathcal{O}^{(2)}_{H^4}$ and $\mathcal{O}^{(3)}_{H^4}$,  respectively, $a,\,b,\,c,\,d=1,2 $ are the weak $SU(2)$ indices and  $I,\,J,\,K,\,L=1,2 $ are the indices for the real and imaginary parts of a particular doublet component. We may convert the result in Eq.\,(\ref{realH}) into the $\left( \phi_1,\, \phi_2,\, \phi_3,\, \phi_4 \right)$ basis and obtain the corresponding coefficient, {\it i.e.}, the $s^2$ coefficient of the subtracted full amplitude of $\phi_i \phi_j \to \phi_k \phi_l$ (Eq.~(\ref{disampv}-\ref{Bexpansion})),
\begin{align}
c^{2,0}_{ijkl}={1\over 2}\frac{{\rm d}^2}{{\rm d} s^2}B_{ijkl}(s,0)
={1\over 2}\frac{{\rm d}^2}{{\rm d} s^2}\mathcal {M}\left(\phi_i \phi_j \to \phi_k \phi_l \right)=a^{2,0\,a b c d}_{\hspace{12pt} I J K L},
\label{Hc20}
\end{align}
where $i=I+2(a-1),\,j=J+2(b-1),\,k=K+2(c-1),\,l=L+2(d-1)$ are the indices of the four real scalars in Eq.\,(\ref{4scalarsH}). 
This defines the linear map between the Wilson coefficients $C_I$ and the amplitude coefficients $c^{2,0}_{ijkl}$:
 \begin{align}
c^{2,0}_{1111}&=C_1+C_2+C_3,
&
c^{2,0}_{1122}&=\frac{1}{2}\left(C_1+C_3\right),
\nl
c^{2,0}_{1212}&=C_2,
&
c^{2,0}_{1133}&=\frac{1}{4}\left(C_1+C_2+2C_3\right),
\nl
c^{2,0}_{1313}&=\frac{1}{2}\left(C_1+C_2\right),
&
c^{2,0}_{1234}&=\frac{1}{4}(C_2-C_1).
\label{eq:cijkl_Higgs}
\end{align}
In Appendix \ref{sec:symHiggs}, we show how Lorentz symmetries, the Higgs internal symmetries and crossing symmetries can be used to either relate the rest of the $c^{2,0}_{ijkl}$ coefficients to the above six $c^{2,0}$ ones or set some of them to zero, which is consistent with and more instructive than simply listing all the $c^{2,0}_{ijkl}$ through Eq.~(\ref{realH}-\ref{Hc20}). These six $c^{2,0}$ coefficients are of course not independent, as they can be expressed with just three Wilson coefficients, $C_i$.  

The LP allows us to compute bounds for arbitrary linear combinations of $c^{2,0}_{ijkl}$. However, obtaining accurate numerical bounds can be relatively costly, so we will only explicitly present 1D and 2D bounds of the three $C_I$'s below.

\subsection{Bounding the EFT coefficients of Higgs scattering}
\label{sec: higgsbounds}
The one-sided positivity bounds for this sector of operators were been calculated in Refs.~\cite{Zhang:2018shp, Remmen:2019cyz} using the scattering of electroweak states and read:
\begin{align}
    \label{eq:higgs_cone}
    C_2\geq0,\quad C_1+C_2\geq0,\quad C_1+C_2+C_3\geq0.
\end{align}
A subsequent analysis of the positivity cone in this sector was found to yield the same information~\cite{Zhang:2020jyn}, with the facets of the cone defined by the above inequalities\,\footnote{An enlarged Higgs cone has been identified in Ref \cite{Yang:2023ncf} by explicitly constructing partial UV completions with massive spin-2 particles. This is consistent with the positivity bounds, as these extra partial UV completions do not admit a standard Wilsonian UV completion because of the presence of the sharp massive spin-2 states \cite{Tolley:2020gtv, Bellazzini:2023nqj}.}. As such, these turn out to be be the optimal lower bounds that one can obtain from Higgs scattering.

In Section \ref{sec: mfbounds}, we have set up a linear program method to obtain the two-sided positivity bounds on the Wilson coefficients from the sum rules along with the null constraints and the unitarity conditions in a general multi-field EFT. In Section \ref{sec: higgseft}, we have extracted the model-specific ingredients needed to perform the linear program for the SMEFT Higgs scatterings. We are now ready to apply our linear program to obtain bounds for the Higgs operators. 

Instead of constraining the expansion coefficients of the amplitude such as $c^{2,0}_{1234}$, we want to directly bound the three dimension-8 Wilson coefficients $C_1,C_2,C_3$ in Eq.\,(\ref{eq:dim8H1}-\ref{eq:dim8H3}), using as much information as possible. The relations between the $c^{2,0}_{ijkl}$ and $C_1,C_2,C_3$ given in Eq.~\eqref{eq:cijkl_Higgs} and Appendix \ref{sec:symHiggs} can be reformulated as follows
\begin{align}
\begin{gathered}
    C_1=-c^{2,0}_{1212} + 2c^{2,0}_{1313},\quad
    C_2=c^{2,0}_{1212} ,\quad
    C_3=c^{2,0}_{1111}-2c^{2,0}_{1313}\,, 
    \label{eq:C123}
\end{gathered}\\
\begin{gathered}
    c^{2,0}_{1122}=\frac{c^{2,0}_{1111}}{2}-\frac{c^{2,0}_{1212}}{2},\quad
    c^{2,0}_{1133}=c^{2,0}_{1144}=\frac{c^{2,0}_{1111}}{2}-\frac{c^{2,0}_{1313}}{2}\,, \\
    c^{2,0}_{1414}=c^{2,0}_{1313},\quad
    c^{2,0}_{1234}=\frac{c^{2,0}_{1212}}{2}-\frac{c^{2,0}_{1313}}{2},\quad
    c^{2,0}_{1234}+c^{2,0}_{1243}=0\,, \\
    c^{2,0}_{1324}=0,\quad
    c^{2,0}_{ijkl}=c^{2,0}_{ijkl}\Big|_{1\leftrightarrow3,\ 2\leftrightarrow4},\quad
    c^{2,0}_{ijkl}=c^{2,0}_{ijkl}\Big|_{1\leftrightarrow2,\ 3\leftrightarrow4}\,,
    \label{eq:addconstraint}
\end{gathered}
\end{align}
where the last two equations denote that $c^{2,0}_{ijkl}$ is symmetric under the simultaneous interchange of $1\leftrightarrow3$ and $2\leftrightarrow4$, and the simultaneous interchange of $1\leftrightarrow2$ and $3\leftrightarrow4$. After expanding $c^{2,0}_{ijkl}$ into sums of $\rho^{ijkl}_{\ell,n}$ according to Eq.~(\ref{eq:cijkl}), the expressions for $C_I$ in Eq.~\eqref{eq:C123}, or linear combinations of them, are regarded as the objective functions to optimize, and the remaining relations in Eq.~\eqref{eq:addconstraint} become additional constraints in the LP problem. 

Compared to the bi-scalar case, in addition to the null constraints on $\rho^{1111}_{\ell,n}$, $\rho^{1212}_{\ell,n}$,..., we now also include those containing 4 different scalar fields, {\it i.e.}, the null constraints on $\rho^{1234}_{\ell,n}$, $\rho^{1324}_{\ell,n}$, $\rho^{1423}_{\ell,n}$,..., by the standard procedure of first choosing $i,j,k,l$ to be permutations of $1,2,3,4$ in Eq.~(\ref{eq:general nc}) and then finding the linearly independent ones. Similar to the bi-scalar case, the first group of constraints on $\rho^{iiii}_{\ell,n}$ are connected to the constraints on $\rho^{iijj}_{\ell,n}$ and $\rho^{ijij}_{\ell,n}$ by the inequality $\big|\rho_{\ell}^{iijj}\big| \leq 1 - \big|1-(\rho_{\ell}^{iiii}+\rho_{\ell}^{jjjj})/2 \big|$.  In turn, these are connected to the constraints for $\rho^{ijkl}_{\ell,n}$ by the inequality $\big|\rho_{\ell}^{ijkl}\big| \leq 1/4 - \big|1/4-(\rho_{\ell}^{ijij}+\rho_{\ell}^{klkl})/2 \big|$ (cf.~Eq.~(\ref{eq:sb1})). Note that every link in this chain is indispensable for producing convergent and finite upper bounds of the objective functions, for the reason explained in Appendix \ref{app:analytical}. The chain of unitarity conditions and null constraints are explicitly given by:
\begin{align}
    &\left\{ 
    \begin{array}{lr}
        0\leq \rho^{iiii}_{\ell,n} \leq 2\,, \\
        \sum\limits_{\ell=0,...,\ell_M;\ell_\infty}(2\ell+1) \sum\limits_{n=1}^N \frac{1}{N}\left(\frac{n}{N}\right)^{r+2} C^{1111}_{r,i_r}(\ell) \rho^{iiii}_{\ell,n} = 0\,,
    \end{array}
    \right. \hspace{1.2em} i\in \{1,2,3,4 \}
    \\[2mm]
    &\hspace{0.6em} \big|\rho_{\ell}^{iijj}\big| \leq 1 - \Big|1-\frac{\rho_{\ell}^{iiii}+\rho_{\ell}^{jjjj}}{2} \Big|\,,  \hspace{2em} i,j\in \{1,2,3,4 \} \label{eq:rhoiijj} \\[3mm] 
    &\left\{ 
    \begin{array}{lr}
    0\leq \rho_{\ell,n}^{ijij} \leq \frac{1}{2}\,,\\[2mm]
    \big|(\rho^{iijj}_{\ell,n} + \rho^{kkll}_{\ell,n}) \pm (\rho^{iikk}_{\ell,n} + \rho^{jjll}_{\ell,n})\big| \leq 2\,,
    \\[1mm]
    \sum\limits_{\ell=0,...,\ell_M;\ell_\infty} \!\! (2\ell+1) \sum\limits_{n=1}^N \frac{1}{N}\left(\frac{n}{N}\right)^{r+2}  \Big[C^{1122}_{r,i_r}(\ell)\rho_{\ell,n}^{iijj}+C^{1212}_{r,i_r}(\ell) \rho_{\ell,n}^{ijij}\Big] =0\,,
    \end{array}
    \right. \hspace{1em} i,j,k,l\in \{1,2,3,4 \} \label{eq:rhoiijjkkll}
    \\
    &\hspace{0.6em}
    \big|\rho_{\ell}^{1jkl}\big| \leq \frac{1}{4} - \Big|\frac{1}{4}-\frac{\rho_{\ell}^{1jij}+\rho_{\ell}^{klkl}}{2} \Big|\,,  \hspace{2em} j,k,l\in \{2,3,4 \} \label{eq:rho1ijk}
    \\[3mm]
    &\left\{ 
    \begin{array}{lr}
    \sum\limits_{\ell=0,...,\ell_M;\ell_\infty}(2\ell+1) \sum\limits_{n=1}^N \frac{1}{N}\left(\frac{n}{N}\right)^{r+2}  \Big[C^{1234}_{r,i_r}(\ell)\rho_{\ell,n}^{1234}+C^{1324}_{r,i_r}(\ell) \rho_{\ell,n}^{1342} +C^{1423}_{r,i_r}(\ell)\rho_{\ell,n}^{1432} \Big]=0\,.
    \end{array} \label{eq:nc1234}
    \right. 
\end{align}
In the null constraints for 4 different fields (Eq.~(\ref{eq:rho1ijk}), we have set the first index of $\rho^{ijkl}_{\ell,n}$ to be $i=1$, since those are the only relevant variables that are related to our objective functions through Eq.~(\ref{eq:C123}) and Eq.~(\ref{eq:addconstraint}). In the first two groups of null constraints, on the other hand, we need to include $i\neq 1$ since they are related to $\rho^{1jkl}_{\ell,n}$ through Eq.~(\ref{eq:rho1ijk}). $C^{1111}_{r,i_r}(\ell)$, $C^{1122}_{r,i_r}(\ell)$ and $C^{1212}_{r,i_r}(\ell)$ are the same as those appeared in Section \ref{sec:onephi} and \ref{sec:two_fields}. Similarly, $C^{1234}_{r,i_r}(\ell)$, $C^{1324}_{r,i_r}(\ell)$, $C^{1423}_{r,i_r}(\ell)$ are $\ell$ polynomials of degree less than or equal to $2r+2$, $r=1,2,...$ being the order of null constraints. In the case of 4 fields, the label for independent ones at each order is $i_r=1,2,...,2r+2$. For example, for $r=1$, $i_r=1$, we have
\begin{align}
     C^{1234}_{1,1} &= -C^{1324}_{1,1}= -\ell^4-2 \ell^3+\ell^2+2 \ell\,,  \\
     C^{1423}_{1,1} &=\big(1+ (-1)^\ell \big)\Big( \ell^4+2 \ell^3-13 \ell^2-14 \ell+24 \Big) \,,
\end{align}
where the $(-1)^\ell$ originates from using the general symmetry of $\rho^{ijkl}_{\ell,n} = (-1)^\ell \rho^{ijlk}_{\ell,n}$ (Eq.~(\ref{eq:rhoellRel})) to combine related coefficients.

We can further simplify the LP problem in an analogous way to what was done in Section~\ref{sec:two_fields}, reducing the number of variables and constraints by using new variables $R^{ijkl}_{\ell,n}$ as certain linear combinations of $\rho^{ijkl}_{\ell,n}$. We show this process in Appendix~\ref{app:COV}. This is very useful practically because, despite the efficiency of the modern linear program solvers, it becomes time and resource consuming to extract these bounds as the number of fields increases in the model. It is also worth noting that this change of variables we propose is completely general and does not rely on any symmetry of the scalar fields, such as the symmetry of Higgs. Now, the LP problem can be reformulated as 
\begin{align} 
    &\textbf{Variables} \notag \\
    &\hspace{1.2em} R^{1111}_{\ell,n},\ R^{11ii}_{\ell,n},\ R^{1i1i}_{\ell,n},\  \rho^{1jkl}_{\ell,n}\,, \hspace{2em} i,j,k,l \in \{ 2,3,4\}\\
    &\textbf{Maximize} \notag \\
    &\hspace{1.2em} \sum_I \alpha_I C_I\,, ~~~~ \text{where}\\
    &\hspace{1.2em} C_1=  \frac{1}{\Lambda^4} \smaller{ \sum\limits_{\ell=0,...,\ell_M;\ell_\infty}}(2\ell+1) \smaller{\sum\limits_{n=1}^N \frac{1}{N}\frac{n}{N} } 32(-R_{\ell,n}^{1212}+ 2R_{\ell,n}^{1313})\,,  \\
    &\hspace{1.2em} C_2= \frac{1}{\Lambda^4} \smaller{ \sum\limits_{\ell=0,...,\ell_M;\ell_\infty}}(2\ell+1) \smaller{\sum\limits_{n=1}^N \frac{1}{N}\frac{n}{N} } 32R_{\ell,n}^{1212}\,, \\
    &\hspace{1.2em} C_3= \frac{1}{\Lambda^4} \smaller{ \sum\limits_{\ell=0,...,\ell_M;\ell_\infty}}(2\ell+1) \smaller{\sum\limits_{n=1}^N \frac{1}{N}\frac{n}{N} } 32(R_{\ell,n}^{1111}- 2R_{\ell,n}^{1313})\,, \\
    &\hspace{1.2em} \text{and}\, \alpha_I \text{ specify directions to optimze in }(C_1,C_2,C_3)\text{ space}\,, \notag \\
    &\textbf{Subject to} \notag  \\
    &\hspace{1.2em} \mbox{Unitarity conditions and null constraints in Eq.(\ref{eq:higgsnc1}-\ref{eq:higgsnc2})}\,,\notag  \\
    &\hspace{1.2em} \mbox{Additional constraints in Eq.(\ref{eq:addconstraint})}\,. \notag 
\end{align}

For example, to find the two-sided bounds on $C_1$, we choose $\alpha_I=(1,0,0)$ and $\alpha_I=(-1,0,0)$ respectively. In general, the $\alpha_I$ vector specifies which direction to optimize in the LP, and running $\alpha_I$ over different directions, we can plot the positivity bounds in 2D or 3D.

\begin{figure}[htbp]
     \centering
\includegraphics[width=\linewidth]{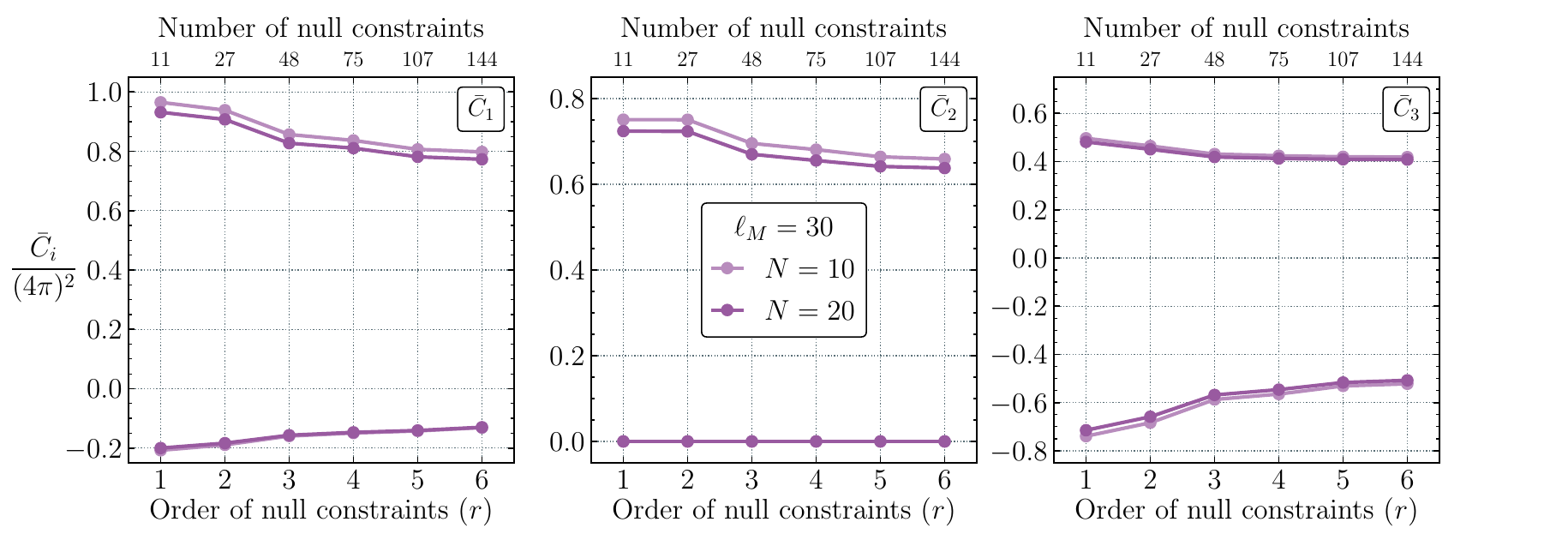}
        \caption{
        \label{fig:C123_1D}
        Upper/lower bounds on the dimension-8 Higgs scattering Wilson coefficients, $\bar{C}_1$, $\bar{C}_2$ and $\bar{C}_3$, in units of $(4\pi)^2$ resulting from our optimisation procedure as a function of the order of null constraints applied. The dimensionless coefficients are defined as $\bar{C}_I=C_I \Lambda^4$. The upper axis label indicates the total number of independent null constraints imposed at each order. }
\end{figure}

Fig~\ref{fig:C123_1D} shows the individual results for the upper and lower bounds on $C_{1-3}$. The pairs of lines correspond to using $N=10$ and $20$, with fixed $\ell_M=30$, $\ell_\infty=100$. Rather coarse discretization values for the $\mu$ integral were used in order to have a reasonable computing time, and in Appendix \ref{app:convergence} we discuss some convergence checks that were performed. The inaccuracy due to the discretisation results in looser bounds than the continuum limit $N\rightarrow \infty$, so the bounds we derive can be interpreted as conservative.

\begin{align}
    \label{eq:C1_bound}
    -\frac{0.130}{\Lambda^4} &\leq \frac{C_1}{(4\pi)^2} \leq \frac{0.774}{\Lambda^4}\,, \\
    \label{eq:C2_bound}
    0 &\leq \frac{C_2}{(4\pi)^2} \leq \frac{0.638}{\Lambda^4}\,, \\
    \label{eq:C3_bound}
    -\frac{0.508}{\Lambda^4} &\leq \frac{C_3}{(4\pi)^2} \leq \frac{0.408}{\Lambda^4}\,.
\end{align}

\begin{figure}[htbp]
     \centering
         \includegraphics[width=0.8\textwidth]{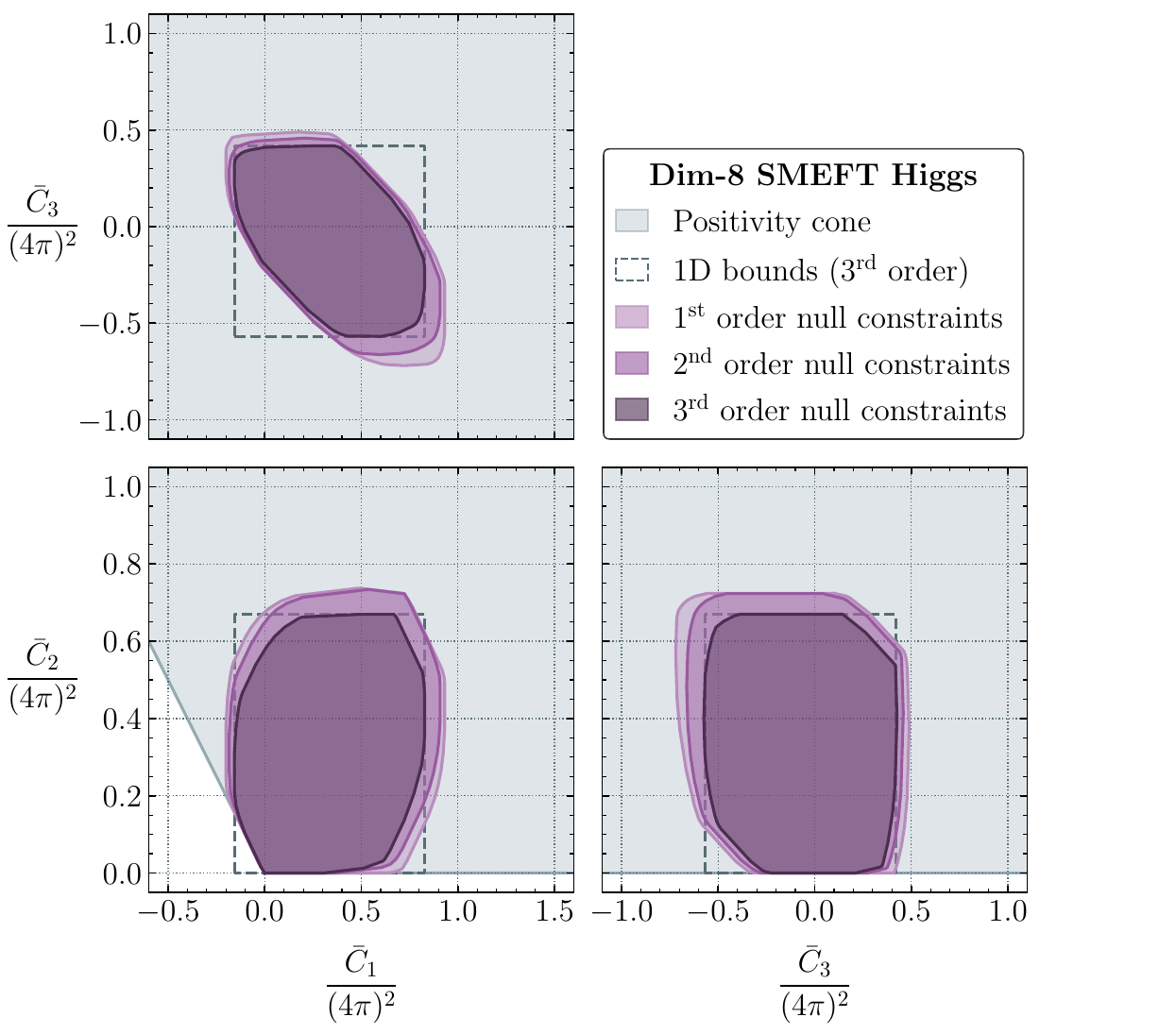}
         \label{fig:C123_2D}
        \caption{Allowed region in the two dimensional subspaces of $C_1$, $C_2$ and $C_3$, when applying 1$^{\text{st}}$, 2$^{\text{nd}}$ and 3$^{\text{rd}}$ order null constraints (purple shaded) compared to the allowed regions by the positivity bounds on elastic scattering~\cite{Zhang:2020jyn} (grey shaded). The dimensionless coefficients are defined as $\bar{C}_I=C_I \Lambda^4$.}
        \label{fig:4field_2D}
\end{figure}

In Figure~\ref{fig:4field_2D}, we show the 2D projections of the positivity bounds onto the planes spanned by two of $C_1$, $C_2$ and $C_3$. Each 2D plot is obtained by an angular optimization along 50 different directions. That is, one chooses 50 different combinations of $C_i$ and $C_j$, each of which picks out one direction, and for each combination or direction, one optimizes to get a upper bound. The numerical parameters chosen are $N=20$, $\ell_M=30$ and $\ell_\infty=100$. For the lower bounds, although we have not used the full positivity of the UV unitarity conditions, they are identical to those obtained via the convex cone approach, which does use the full UV positivity \cite{Zhang:2020jyn}. However, now, using the non-positivity part of the UV unitarity conditions along with the null constraints, we have also derived the upper bounds for all of the Wilson coefficients $C_i$ for Higgs scattering, capping the Higgs positivity cone.

\section{Comparison with experimental bounds and perturbative unitarity\label{sec: pheno}}

Having successfully bounded the space of dimension-8 operators for Higgs scattering from first principles, we would like to compare the allowed space with the existing and future experimental limits on the operators in question. We will also compare the bounds arising from partial wave unitarity in the low energy EFT. To help us with the discussion, we will work in terms of the dimensionless Wilson coefficients, $\bar{C}_I$ since the bounds from experiment and partial wave unitarity depend on the cutoff $\Lambda_{\rm EFT}$. From here on, we identify $\Lambda_{\mathrm{_{EFT}}}$ and $\Lambda$, as per the discussion at the end of Section \ref{sec:constWilson}.

The Higgs scattering operators that we are interested in are subject to experimental constraints from LHC data, specifically from measurements of vector boson scattering (VBS). In this process a pair of weak bosons ($W$, $Z$ or photon) are emitted, one from each initial state quark, and undergo a hard, $2\to2$ scattering that results in a distinctive signature of two weak bosons accompanied by two forward jets. This process is seen as a key probe of the gauge structure of the Standard Model that can test the unitarising properties of the Higgs boson exchange in weak boson scattering. It was measured for the first time in 2017 by the ATLAS and CMS collaborations~\cite{CMS:2017fhs,ATLAS:2019cbr}, in the same-sign $WW$ channel and many other channels have since been observed and used to set bounds on anomalous quartic gauge couplings (aQGC). This process served as one of the primary motivations to experimentally probe dimension-8 operators in the SMEFT~\cite{Yang:2020rjt,Yang:2021pcf}. This is because there are only a few operators at dimension-6 that can generate aQGC and they all lead to correlated effects in anomalous triple gauge couplings, electroweak precision observables or Higgs coupling modifications. Dimension-8 operators, on the other hand, can generate independent aQGCs. Therefore, although dimension-6 operators can impact VBS (See~\cite{Gomez-Ambrosio:2018pnl,Ethier:2021ydt} for recent studies), it is expected that these operators will be better constrained elsewhere, hence opening up the opportunity to use this process as a probe of dimension-8 coefficients. The only exception to this argument applies to neutral triple gauge couplings, which are absent in both the SM and the SMEFT at dimension-6, hence also constitute a well motivated scenario in which to study dimension-8 operators~\cite{Degrande:2013kka,Senol:2018cks,Rahaman:2018ujg,Senol:2019swu,Ellis:2019zex,Ellis:2020ljj,Fu:2021mub,Lombardi:2021wug,Jahedi:2022duc,Senol:2022snc,Ellis:2022zdw,Spor:2022zob,Degrande:2023iob}.

In the unitary gauge, the three operators for which we obtain theoretical bounds specifically lead to four-derivative Higgs boson self-interactions, two-derivative interactions between pairs of Higgs and vector bosons, and a modification of the SM quartic gauge interactions. Although they do not explicitly contain extra derivatives, it is the latter that lead to unitarity-violating behaviour in 4-point, longitudinally polarised vector boson amplitudes, by spoiling the cancellations enforced by the SM gauge symmetries~\cite{LlewellynSmith:1973yud,Lee:1977eg,Lee:1977yc}. One can also understand this energy growth more intuitively by considering, \emph{e.g.}, the Feynman gauge, where the Goldstone modes are kept explicit. There, our operators also induce four-derivative contact interactions between the Goldstone boson components of the Higgs field. By the Goldstone boson equivalence theorem~\cite{Cornwall:1974km}, these can be identified with the longitudinal polarisations of the massive vector bosons of the SM. The strong energy growth with respect to the SM induced in fully longitudinal $2\to2$ scattering amplitudes between $W$ and $Z$ bosons can be probed by measuring the high invariant mass region of the vector boson pair in VBS. This growth is the very same feature that we use to impose theoretical bounds on the coefficients, so it is interesting to compare the bounds that we have obtained with the current experimental sensitivity. 

The LHC experiments use a basis of dimension-8 operators~\cite{Eboli:2006wa,Eboli:2016kko} that predates even the determination of a complete non-redundant dimension-6 SMEFT basis. Our operators of interest, $\mathcal{O}^{(1)}_{H^4}$, $\mathcal{O}^{(2)}_{H^4}$ and $\mathcal{O}^{(3)}_{H^4}$ are identified with the operators $\mathcal{O}_{S,2}$, $\mathcal{O}_{S,0}$ and $\mathcal{O}_{S,1}$ of that basis, respectively. The LHC experiments have delivered a significant number of VBS analyses, deriving constraints on the coefficients of these operators\footnote{An up-to-date summary of these results can be found at \url{https://twiki.cern.ch/twiki/bin/view/CMSPublic/PhysicsResultsSMPaTGC}}. Among these, the most stringent bounds on the four-Higgs operators are reported by the CMS experiment, in a search for anomalous VBS production in semi-leptonic $WW$, $WZ$ and $ZZ$ final states~\cite{CMS:2019qfk} that we summarise in Table~\ref{tab:exp_bounds}, for $\Lambda=1$ TeV. 

The original set of dimension-8 operators was proposed only identified $\mathcal{O}_{S,0}$ and $\mathcal{O}_{S,1}$, with $\mathcal{O}_{S,2}$ only being added in follow up works some years later~\cite{Eboli:2016kko}. To our knowledge, the experimental collaborations have so far continued to only report results for $\mathcal{O}_{S,0}$ and $\mathcal{O}_{S,1}$. A recent phenomenological study~\cite{Cappati:2022skp} undertook a validation exercise reproducing the limits obtained by the CMS analysis, where a similar sensitivity to $\mathcal{O}_{S,1}$ was found for $\mathcal{O}_{S,2}$. We therefore assume that the experimental bound on $\mathcal{O}_{S,1}$ applies equally to $\mathcal{O}_{S,2}$ for our purposes, denoting the assumed bound by `$\ast$'. 
\begin{table}[!h]
  \centering
  \renewcommand{\arraystretch}{1.4}
  \begin{tabular}{|c|c|c|c|}
    \hline
    \multirow{2}{*}{\shortstack[c]{Coefficient \\[0.5ex] ($\Lambda=1$ TeV)}} & 
    \multirow{2}{*}{\shortstack[c]{ Observed \\[0.5ex] \phantom{kkk} Bound~\cite{CMS:2019qfk}} } &
    \multicolumn{2}{c|}{HL-LHC projection~\cite{Cappati:2022skp}}\\
    \cline{3-4}
    & & w/o unitarity & w unitarity\\
    \hline
    $\bar{C}_1$ & \phantom{k}$[-3.4,\,3.4]^\ast$ & $[-2.3,\,2.4]$& $[-4.8,\,5.2]$ \\ 
    $\bar{C}_2$ & $[-2.7,\, 2.7]$                  & $[-1.8,\,2.0]$ & $[-2.6,\,3.3]$ \\ 
    $\bar{C}_3$ & $[-3.4,\, 3.4] $                 & $[-2.4,\,2.4]$ & $[-5.8,\,6.1]$\\ 
    \hline
  \end{tabular}
  \caption{ \label{tab:exp_bounds}
  Summary of current and projected experimental bounds on four-Higgs dimension-8 operator coefficients. Current experimental limits are taken from Ref.~\cite{CMS:2019qfk} and future projections from a phenomenological analysis of Ref.~\cite{Cappati:2022skp}. The bounds on $\bar{C}_1$ denoted with an `$\ast$' are not reported by the CMS analysis. Based on the results of Ref.~\cite{Cappati:2022skp}, we assume they are identical to those of $\bar{C}_3$.
  }
  \renewcommand{\arraystretch}{1.}
\end{table}
The current experimental sensitivity lies around 2-4 TeV$^{-4}$ for the absolute value of the coefficients, and is dominated by the $WV$ channel, with one leptonically decaying $W$ boson and a hadronically decaying $W$ or $Z$ boson. The limits are symmetric around zero, which indicates that they are dominated by the quadratic, $\mathcal{O}(\Lambda^{-8})$ contribution of the coefficients to the VBS cross-section.

One of the challenges faced by the experimental collaborations when interpreting VBS data in terms of aQGC is that of unitarity violation. On one hand, sensitivity to aQGC is optimized by probing the high invariant-mass tails of VBS, which is already a rare process. On the other hand, the massive ($E^4$) energy growth of the underlying weak boson scattering amplitudes will eventually violate unitarity at some scale. Partial wave unitarity bounds on the dimension-8 aQGC Wilson coefficients were computed in Ref.~\cite{Almeida:2020ylr}, based on an analysis of $2\to2$ Higgs and vector boson scattering amplitudes of various total charge and helicity. The bounds on our four-Higgs operators arise from the eigenvalues of the $J=0$ scattering matrices, and are summarised by
\begin{align}
\begin{split}
\label{eq:unitarity_bounds}
\big|\bar{C}_1 + 3\bar{C}_2 + \bar{C}_3\big|
 &< 96\pi \frac{\Lambda^4}{s^2},\\
  \big|3\bar{C}_1+ \bar{C}_2 + \bar{C}_3\big|
 &< 96\pi \frac{\Lambda^4}{s^2},\\
  \big|5\bar{C}_1 + 3\bar{C}_2 + 7\bar{C}_3\big|
 &< 96\pi \frac{\Lambda^4}{s^2}.
\end{split}
\end{align}
These imply that, for bounds of around 3 TeV$^{-4}$ given in Table~\ref{tab:exp_bounds}, unitarity is violated at a scale of 1.9-2.4 TeV. In VBS, the diboson invariant mass, $m_{VV}$ can be used as a proxy for the centre of mass energy, $\sqrt{s}$, of the $2\to2$ sub-amplitude, and used to apply unitarity violation constraints on this process.  Given that the analyses measure the $m_{VV}$ distribution out to around 2.5 TeV, the current sensitivity lies in a region that risks violating unitarity. Several procedures for mitigating this fact have been studied over the years, ranging from ad-hoc damping factors, physically motivated unitarization methods to simply forbidding the signal amplitudes from violating unitarity. None of these methods that modify the signal amplitudes are truly consistent with the model-independent SMEFT approach and the bounds reported in Table~\ref{tab:exp_bounds} are obtained without any unitarization method.

Another option is to consider limiting the energy reach of the data to a region where unitarity is not violated. Ref.~\cite{Cappati:2022skp} also studied the impact of unitarity bounds on the experimental constraints from VBS by computing them as a function of an upper cut on $m_{VV}$. Limiting the scattering energy simultaneously weakens the experimental and unitarity constraints on a given coefficients. New limits are then derived at the point where the experimentally derived bound with the upper cut on $m_{VV}$ is stronger than the unitarity bound obtained from identifying $\sqrt{s}=m_{VV}$ in Eq.~\eqref{eq:unitarity_bounds}. If there is no such point, then no limit can be derived. These bounds are summarised in the second and third columns of Table~\ref{tab:exp_bounds} and correspond to High-Luminosity LHC (HL-LHC) projected sensitivities. These show that unitarity considerations lead to order one effects on the experimental sensitivity on the Wilson coefficients.

\begin{figure}[htbp]
     \centering
\centerline{\includegraphics[width=1.15\linewidth]{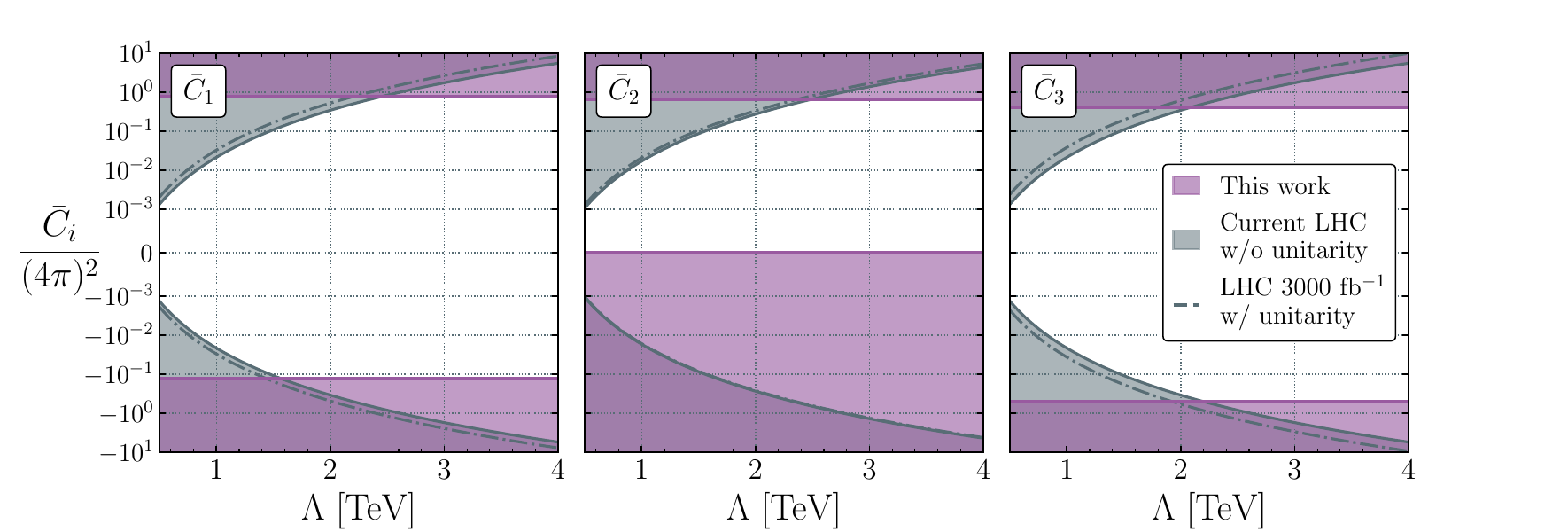}}
        \caption{Upper and lower positivity bounds on the Higgs scattering Wilson coefficients obtained in this work (purple shaded) compared to the current~\cite{CMS:2019qfk} (grey shaded) and projected~\cite{Cappati:2022skp} (grey dashed) LHC exclusion limits from VBS measurements as a function of the EFT cutoff, $\Lambda$. }
        \label{fig:compare_exp}
\end{figure}

We now turn to a comparison of our positivity bounds summarised in Eqs.~\eqref{eq:C1_bound}--\eqref{eq:C3_bound} with the experimental sensitivity from current and projected VBS measurements as well as those arising from perturbative unitarity in Eq.~\eqref{eq:unitarity_bounds}. Our LP procedure produces bounds on individual coefficients, allowing the remaining coefficients to freely vary within their allowed range. The bounds are therefore the maximally conservative ones obtained from considering four-Higgs scattering amplitudes in the dimension-8 SMEFT. In contrast, the experimental constraints are obtained by setting all other coefficients to zero. Allowing all coefficients to vary would significantly reduce the sensitivity, and it may even be that a closed bound is not possible in this case. Furthermore, as discussed in Section~\ref{sec:constWilson}, the experimental analyses bound the dimensionful combination, $\bar{C}_I/\Lambda^4$, such that the bound on the value of $\bar{C}_I$ varies depending on the assumed value of the EFT cutoff, $\Lambda$. This is reflected in Figure~\ref{fig:compare_exp}, which plots the experimental bounds as a function of $\Lambda$, as well as the $\Lambda$-independent bounds obtained from our numerical analysis. The solid grey line represents the current experimental limits that were obtained without mitigating possible unitarity violation in the signal modelling, while the dot-dashed grey lines show the projected HL-LHC bounds from Ref.~\cite{Cappati:2022skp} obtained with the unitarity mitigation prescription described above. 
The fact that the HL-LHC projections are weaker than the current bounds underlines the significant effect that unitarity violation has on the prospects of probing these dimension-8 operators through VBS at the LHC. 

One can see how the two-sided positivity bounds provide complementary information to the experimental limits. For instance, the lower bounds on $\bar{C}_2$ from the positivity cone immediately rules out half of the available parameter space. The new bounds obtained from our analysis cut out a further, significant portion of the parameter space and become stronger than the experimental limits for $\Lambda$ between 1.5 and 2.5 TeV, depending on the operator and the sign of its coefficient. Given that the experimental sensitivity -- despite having assumed all other coefficients to be zero -- implies unitarity violation around the same scale, our bounds appear highly competitive and even dominant over the experimental sensitivity. For example, if evidence for the presence of one of these operators were observed during the LHC lifetime, it would either imply an upper bound on the scale of new physics of order 1.5-2.5 TeV or a violation of unitarity, causality and/or locality in the UV. These processes are also therefore an interesting testbed for testing whether these fundamental principle hold true in physics beyond the SM.

\begin{figure}[htbp]
     \centering
\centerline{\includegraphics[width=1.15\linewidth]{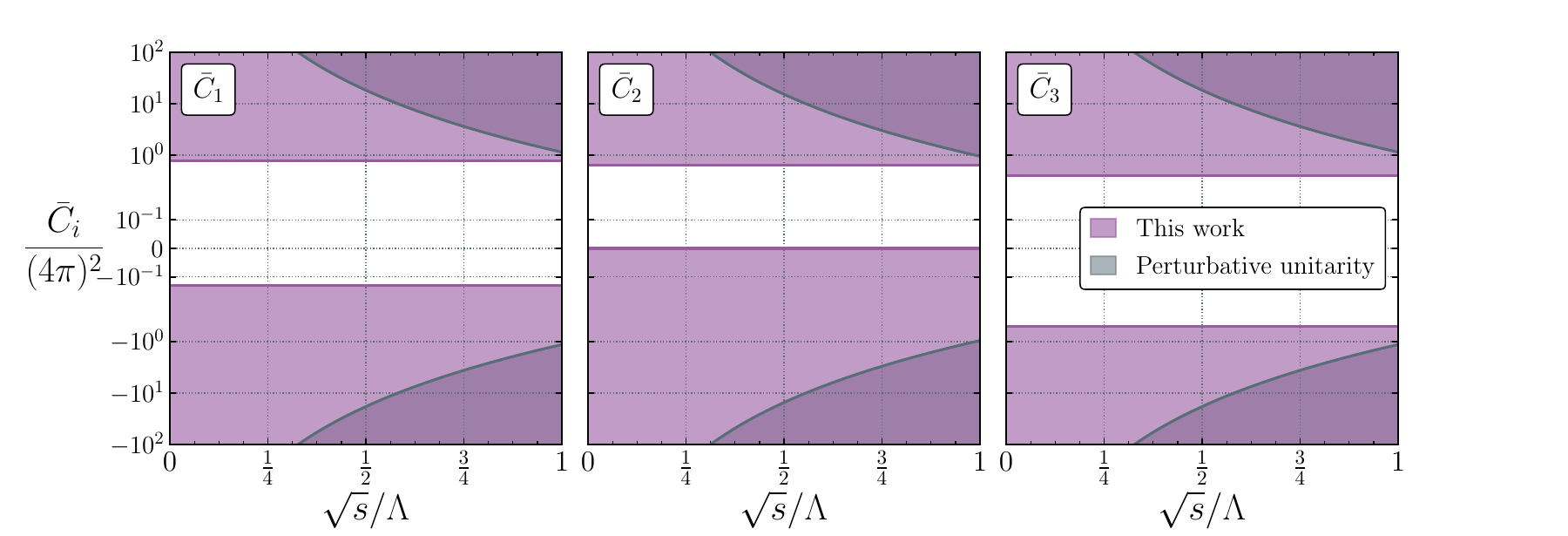}}
        \caption{Upper and lower positivity bounds on the Higgs scattering Wilson coefficients obtained in this work (purple shaded) compared to the (tradiational) perturbative unitarity bounds obtained in the EFT (grey shaded) as a function of the ratio of centre of mass energy of the $2\to2$ scattering, $\sqrt{s}$ relative to the EFT cutoff, $\Lambda$. The current LHC bounds do not impose unitarity on the EFT amplitudes while the projections take into account the unitarity bounds, providing the maximal limits that can be obtained while remaining consistent with perturbative unitarity. }
        \label{fig:compare_unit}
\end{figure}

In order to fairly compare our positivity bounds with those from perturbative unitarity, we infer `profiled' unitarity bounds on each coefficient, in analogy with the fact that our bounds on a given coefficient allow the values of the others to float. Extremizing the value of each coefficient separately within the volume defined by Eq.~\eqref{eq:unitarity_bounds} yields:
\begin{align}
\label{eq:unitarity_bounds_profiled}
\big|\bar{C}_1\big|,\big|\bar{C}_3\big|
 < 0.6\times96\pi \frac{\Lambda^4}{s^2},\quad\big|\bar{C}_2\big|< 0.5\times96\pi \frac{\Lambda^4}{s^2},
\end{align}
which we compare with our positivity bounds in Figure~\ref{fig:compare_unit}. The values of the Wilson coefficients are plotted in units of $(4\pi)^2$ and the unitarity bounds are shown as a function of $\sqrt{s}/\Lambda$, where $\sqrt{s}$ denotes the scale up to which one is willing to trust the EFT approximation for the tree-level scattering amplitudes that go into deriving them. The most aggressive unitarity bounds are then obtained by taking $\sqrt{s}=\Lambda$. 
From the conservative perspective of remaining agnostic about the possible values of the Wilson coefficients, our positivity bounds are stronger than the aggressive perturbative unitarity bounds by up to an order of magnitude. This is not to say that positivity is always more constraining than perturbative unitarity. Indeed, in certain directions of the parameter space, perturbative unitarity can yield a more powerful constraint. For example, by setting up our LP to maximise the specific linear combinations constrained in Eq.~\eqref{eq:unitarity_bounds}, we obtain the following positivity bounds, using the 6$^{\text{th}}$ order null constraints and fixing $N=20$ and $\ell_M=30$:
\begin{align}
\begin{split}
0<\bar{C}_1 + 3\bar{C}_2 + \bar{C}_3
 &< (4\pi)^2\times 2.42,\\
 -(4\pi)^2\times 0.18< 3\bar{C}_1+ \bar{C}_2 + \bar{C}_3
 &< (4\pi)^2\times 2.66,\\
 -(4\pi)^2\times 1.07< 5\bar{C}_1 + 3\bar{C}_2 + 7\bar{C}_3
 &< (4\pi)^2\times 5.89.
\end{split}
\end{align}
Comparing the numbers to the most optimistic unitarity bound of $96\pi=1.91\times(4\pi)^2$, we see that for these particular directions, our upper bound from positivity is slightly weaker for the first two  and about a factor of 3 weaker for the third. However, the lower bounds are considerably stronger. The first direction is bounded to be positive by the Higgs positivity cone, since it is a positive linear combination of $\bar{C}_2$ and $\bar{C}_1+\bar{C}_2+\bar{C}_3$, both of which are positive (See Eq.~\eqref{eq:higgs_cone}). The others are not bounded by the positivity cone and hence can only be constrained by our two-sided approach. 

Overall, it is clear that our positivity bounds provide highly complementary information on the viable space for dimension-8 operators that mediate Higgs scattering amplitudes to experimental and unitarity-based probes. Since they have a completely independent origin, they can be used as part of a data interpretation, \emph{i.e.}, by including them as part of the statistical prior to obtain tighter bounds on the Wilson coefficient space, adding only the relatively weak assumptions of unitarity, causailty and locality in the UV. Since the constraints can be stronger than perturbative unitarity, they may help experimental collaborations to avoid the recurrent issues of unitarity violation in the EFT interpretations of VBS measurements.

\acknowledgments

We would like to thank Dong-Yu Hong, Yue-Zhou Li, Andrew Tolley, Shi-Lin Wan and Zhuo-Hui Wang for helpful discussions. QC acknowledges postdoctorate fellowship supported by University of Science and Technology of China and Peng Huanwu Center for Fundamental Theory (PCFT), Hefei.
PCFT is supported by National Natural Science Foundation of China under grant No.\,12247103. SYZ acknowledges support from the National Natural Science Foundation of China under grant No.~12075233 and 12247103, from the Fundamental Research Funds for the Central Universities under grant No.~WK2030000036, and from the National Key R\&D Program of China under grant No. 2022YFC220010.

\appendix

\section{Partial wave unitarity}
\label{app:partial wave unitarity}

Partial wave unitarity comes from the fact that in a scattering angular momentum is conserved, so the S-matrix can be block-diagonalized into different angular momenta: $S_\ell S^\dagger_\ell=I_\ell$. In this appendix, we first review an explicit derivation of the full partial wave unitarity conditions starting from the full scattering amplitudes, based largely on Ref~\cite{Soldate:1986mk}. Then, we derive a set of linear, two-sided unitarity conditions from full partial wave unitarity that can be easily implemented in our linear programming to constrain the Wilson coefficients using the dispersion relations.   

\subsection{Review}

We work in 4 dimensions and evaluate amplitudes in the center of mass frame. The amplitude for a scalar $2\to2$ scattering process, $ij\to kl$, can be decomposed into partial waves,
\begin{align}
A_{ij kl}(\theta) \equiv \bra{\vb*{i}} T\ket{\vb*{f}} = 16\pi \sum_{\ell=0}^\infty a_{\ell}^{ijkl}\qty(2\ell+1)P_\ell(\cos{\theta})\,,
\label{amp}
\end{align}
where $\vb*{i}$ and $\vb*{f}$ denote the two particle states $ij$ and $kl$, respectively, both of which consist of a pair of states in a back-to-back momentum configuration. $\theta$ is the polar angle between the   $\hat{i}$ and $\hat{f}$ directions and we use the standard convention for Legendre polynomials where $P_\ell(1)=1$. 
From the unitarity of the $S$-matrix, we have
\begin{equation}
    \bra{\vb*{i}} T\ket{\vb*{f}} - \bra{\vb*{f}} T\ket{\vb*{i}}^* = i\sum_{\vb*{x}} \bra{\vb*{i}} T\ket{\vb*{x}} \bra{\vb*{f}} T\ket{\vb*{x}}^* (2\pi)^4 \delta^4 (p_{\vb*{x}}-p_{\vb*{i}})\,,
    \label{eq:unitarity}
\end{equation}
where the $\vb*{x}$ summation includes both the sum over all possible intermediate states and the integration over their phase space. If the theory is time-reversal (CP) invariant, $A^{kl ij}=A^{ij kl}$, then the LHS of Eq.~\eqref{eq:unitarity} becomes
\begin{equation}
    \mathrm{LHS}=2i\, \mathrm{Im}A_{ij kl} = 32i\,\pi \sum_{\ell=0}^\infty  \mathrm{Im} a_{\ell}^{ijkl}\qty(2\ell+1)P_\ell(\cos{\theta})\,. \label{eq:RHS1}
\end{equation}
For every phase space configuration of an intermediate state, $\vb*{x}$, there exists a set of configurations that are related to it by an overall spatial rotation, $R(\Omega_n)$, which can be characterised by its action on $\hat{i}$, rotating it to a new direction, $\hat{n}$.
Denoting $\{X\}$ as the set of intermediate states that are not related by an overall rotation, each state $\vb*{x}$ can be specified by $\{X,\Omega_n\}$. Now the RHS of Eq.~\eqref{eq:unitarity} can be written as
\begin{equation}
    \mathrm{RHS} = i\sum_{X} \int \dd \Omega_n \bra{\vb*{i}} T\ket{X,\Omega_n} \bra{\vb*{f}} T\ket{X,\Omega_n}^* (2\pi)^4 \delta^4 (p_X-p_{\vb*{i}}), \label{eq:LHS1}
\end{equation}
where the integration over the solid angle of $\hat{n}$ has been factored out of the overall phase space integral.
We can then expand each intermediate amplitude into partial waves
\begin{equation}
    \bra{\vb*{i}} T\ket{X,\Omega_n} = 16\pi \sum_{\ell=0}^\infty a_{\ell}^{ij\rightarrow X}\qty(2\ell+1) P_\ell(\cos{\theta_{in}})\,,
\end{equation}
where $\cos \theta_{in}=\hat{i}\cdot \hat{n}$, and do the same for $\bra{\vb*{f}} T\ket{X,\Omega_n}$. Using geometric relations and the addition formula for Legendre polynomials
\begin{align}
    \cos \theta_{fn} &= \cos \theta \cos\theta_{in} + \sin \theta \sin \theta_{in} \cos \phi_{fn},
    \\
    P_\ell (\cos \theta_{fn}) &= P_\ell (\cos \theta) P_\ell (\cos \theta_{in}) + 2\!\sum_{m=1}^{\ell}\!\frac{(l-m)!}{(l+m)!} P_\ell^m (\cos \theta) P_\ell^m (\cos \theta_{in}) \cos(m \phi_{fn}),
\end{align}
and evaluating the $\Omega_n$ integral in Eq.~(\ref{eq:LHS1}), we get 
\begin{equation}
    \mathrm{RHS} = i(16\pi)^2 (4\pi) \sum_{\ell=0}^\infty \sum_X a_{\ell}^{ij\rightarrow X} \qty(a_{\ell}^{kl\rightarrow X})^* \qty(2\ell+1) P_\ell(\cos{\theta}) (2\pi)^4\delta^4 (p_X-p_{\vb*{i}})\,.\label{eq:LHS2}
\end{equation}
Comparing Eq.\,(\ref{eq:RHS1}) and Eq.\,(\ref{eq:LHS2}), we obtain the partial wave unitarity condition
\begin{equation}
    \mathrm{Im} a_{\ell}^{ijkl} = 32\pi^2 \sum_X a_{\ell}^{ij\rightarrow X} \qty(a_{\ell}^{kl\rightarrow X})^*  (2\pi)^4 \delta^4 (p_X-p_{\vb*{i}})\,.\label{eq:uni1}
\end{equation}

For the special case of $X=\{mn\}$, i.e., when the intermediate state is also a two scalar field state, 
$\sum_X$ reduces to a phase space integral over the magnitude of the momentum
\begin{align}
\label{aijmnEta}
    & 32\pi^2 \,  \eta_{mn} \int \frac{p_f^2 \dd p_f}{(2\pi)^6}\frac{1}{(2E_{p_f})^2} a_{\ell}^{ijmn} \qty(a_{\ell}^{klmn})^* (2\pi)^4 \delta(2E_{p_f}-E_i) \\ 
    =\, &\eta_{mn} \,  a_{\ell}^{ijmn} \qty(a_{\ell}^{klmn})^*\,, \hspace{2em} \mbox{(massless limit)}
\end{align}
where $\eta_{mn}$ is a prefactor that accounts for identical final-state particles: $\eta_{mn}=1$ for $m\neq n$ and $\eta_{mn}=1/2$ for $m=n$. (For massive particles, we can again get Eq.~(\ref{aijmnEta}) if the partial wave expansion is defined with an extra kinematic factor.) For the rest of the positive weighted sum in Eq.~(\ref{eq:uni1}), we can define the normalized $a_\ell^{ij,kl\rightarrow X}$ to have unit weight for our purposes. We can now write the unitarity condition in a more compact form
\begin{equation}
    \mathrm{Im} a_{\ell}^{ijkl} = \sum_{mn} \, \eta_{mn} a_{\ell}^{ijmn} \qty(a_{\ell}^{klmn})^* + \sum_{X\neq mn} a_{\ell}^{ij\rightarrow X} \qty(a_{\ell}^{kl\rightarrow X})^*\,.\label{eq:uni2}
\end{equation}
In the text, sometimes we omit the prefactor $\eta_{mn}$ and refer to this condition simply as $\mathrm{Im} a_{\ell}^{ijkl} = \sum'_{X} a_{\ell}^{ij\rightarrow X} \qty(a_{\ell}^{kl\rightarrow X})^*$.

\subsection{Linear unitarity conditions}

From partial wave unitarity (\ref{eq:uni2}), we can derive some linear unitarity conditions on the imaginary part of partial wave amplitude $\mathrm{Im}a_{\ell}^{ijkl}$, which can be easily fed into the linear programs to constrain the EFT coefficients. While the linear conditions presented below are already very constraining, it is likely that we have not exhausted all possible linear unitarity conditions.
\\[1mm]
\noindent \textbullet\ \textbf{Individual partial amplitudes} 
\\[1mm]
First, we derive simple two-sided constraints on individual $\mathrm{Im}a^{ijkl}$, making use of the positive nature of the sum on the right hand side of (\ref{eq:uni2}) for elastic scatterings. 

For the process of $ii\to ii$, picking out only $X=ii$ for the intermediate states gives
\begin{align}
&\mathrm{Im}a_{\ell}^{ii ii}\geq \frac{1}{2}|a_{\ell}^{ii ii}|^2  = \frac{1}{2}\left(\mathrm{Im}a_{\ell}^{ii ii}\right)^2+ \frac{1}{2}\left(\mathrm{Re}a_{\ell}^{ii ii}\right)^2 \notag
\\[2mm]
& \hspace{4em} \Rightarrow ~~~0\leq \mathrm{Im}a_{\ell}^{ii ii} \leq 2, 
\label{Imaiiii2}
\end{align}
picking out $X=ii,jj$ gives
\begin{align}
&\mathrm{Im}a_{\ell}^{ii ii}\geq \frac{|a_{\ell}^{ii ii}|^2}{2}+\frac{|a_{\ell}^{ii jj}|^2}{2}
 = \frac{\left(\mathrm{Im}a_{\ell}^{ii ii}\right)^2}{2}+ \frac{\left(\mathrm{Re}a_{\ell}^{ii ii}\right)^2}{2}
+ \frac{\big(\mathrm{Im}a_{\ell}^{ii jj}\big)^2}{2}+ \frac{\big(\mathrm{Re}a_{\ell}^{ii jj}\big)^2}{2}\,,
\notag
\\[2mm]
& \hspace{11.2em} \Rightarrow ~~~ -1\leq \mathrm{Im}a_{\ell}^{ii jj} \leq 1\,, \label{Imaiijjweak}
\end{align}
For the process of $ij\to ij$, picking out $X=ij,ji$ leads to
\begin{align}
&\mathrm{Im}a_{\ell}^{ij ij} \geq |a_{\ell}^{ij ij}|^2+|a_{\ell}^{ij ji}|^2
=\big(\mathrm{Im}a_{\ell}^{ij ij}\big)^2+\big(\mathrm{Re}a_{\ell}^{ij ij}\big)^2
+ \big(\mathrm{Im}a_{\ell}^{ij ji}\big)^2+\big(\mathrm{Re}a_{\ell}^{ij ji}\big)^2 
\geq 2 \big(\mathrm{Im}a_{\ell}^{ij ij}\big)^2 \notag
\\[2mm]
& \hspace{13.6em} \Rightarrow ~~~ 0 \leq \mathrm{Im}a_{\ell}^{ij ij} \leq \frac{1}{2}\,,
\label{Imaijij2}
\end{align}
where we have used the fact that $a^{ijji}_\ell = (-1)^\ell\, a^{ijij}_\ell$. Picking out multiple distinct intermediate states $X=ij,ji,kl,lk$ ($kl,lk\neq ij$) gives
\begin{align}
\mathrm{Im}a_{\ell}^{ij ij} &\geq \big|a_{\ell}^{ij ij}\big|^2 + \big|a_{\ell}^{ij ji}\big|^2 + \big|a_{\ell}^{ij kl}\big|^2 + \big|a_{\ell}^{ij lk}\big|^2 = 2\big|a_{\ell}^{ij ij}\big|^2 + 2\big|a_{\ell}^{ij kl}\big|^2 \notag
\\[1mm]
 &\geq 2 \big(\mathrm{Im}a_{\ell}^{ij ij}\big)^2 + 2 \big(\mathrm{Im}a_{\ell}^{ij kl}\big)^2 ~~~~
\Rightarrow ~~~ -\frac{1}{4}\leq \mathrm{Im}a_{\ell}^{ij kl} \leq \frac{1}{4}\,,   \label{Imaijkl14}
\end{align}
where we have again used $a^{ijkl}_\ell = (-1)^\ell\, a^{ijlk}_\ell$. In the numerical LP, we will use conditions (\ref{Imaiiii2}) and (\ref{Imaijij2}), but will not use conditions (\ref{Imaiijjweak}) and (\ref{Imaijkl14}), as we can obtain stronger conditions than (\ref{Imaiijjweak}) and (\ref{Imaijkl14}), as shown below.
\\
\\
\noindent \textbullet\ \textbf{Mixing different partial amplitudes} 
\\[1mm]
Given the above simple bounds, now we show that some of them can be further enhanced with a more careful treatment of unitaritiy conditions (\ref{eq:uni2}).

First, note that a simple re-arrangement of Eq.~(\ref{eq:uni2}) for $iiii$ gives
\begin{equation}
    \qty(2-\mathrm{Im}a_{\ell}^{ii ii}) = \frac{1}{2}\qty(2-\mathrm{Im}a_{\ell}^{ii ii})^2 +\frac{1}{2}\qty(\mathrm{Re}a_{\ell}^{ii ii})^2+ \sum_{X \neq ii}\!\!{}' \big|a_{\ell}^{ii\rightarrow X}\big|^2\,. \label{eq:2-rho1111}
\end{equation}
On the other hand, Eq.~(\ref{eq:uni2}) for $iijj$ can be explicitly written as
\begin{align}
\label{ImaiijjReal}
    \mathrm{Im}a_{\ell}^{ii jj} &= \frac{1}{2} a_{\ell}^{ii ii} \big(a_{\ell}^{jj ii}\big)^* + \frac{1}{2} a_{\ell}^{ii jj} \big(a_{\ell}^{jj jj}\big)^* + \sum_{X \neq ii,jj}\!\!\!\!{}'~ a_{\ell}^{ii\rightarrow X} \big(a_{\ell}^{jj\rightarrow X}\big)^* \\
    &= \frac{1}{2} \mathrm{Im}a_{\ell}^{ii jj} \big(\mathrm{Im}a_{\ell}^{ii ii}+\mathrm{Im}a_{\ell}^{jj jj}\big) + \frac{1}{2} \big( \mathrm{Re}a_{\ell}^{ii ii} \mathrm{Re}a_{\ell}^{jj ii} +\mathrm{Re}a_{\ell}^{ii jj} \mathrm{Re}a_{\ell}^{jj jj} \big)
    \notag\\[0.5mm]
    & \hspace{2em} +  \sum_{X \neq ii,jj}\!\!\!\!{}'~ \Big(\mathrm{Im}\,a_{\ell}^{ii\rightarrow X} \mathrm{Im}\,a_{\ell}^{jj\rightarrow X} + \mathrm{Re}\,a_{\ell}^{ii\rightarrow X} \mathrm{Re}\,a_{\ell}^{jj\rightarrow X} \Big)\,,
    \label{ImaiijjReal2}
\end{align}
where we have used the time-reversal invariance $a_{\ell}^{ii jj}=a_{\ell}^{jj ii}$ for the first term on the second line, and the imaginary terms on the right hand side of Eq.~(\ref{ImaiijjReal}) must cancel because $\mathrm{Im}a_{\ell}^{ii jj}$ is real. A simple rearrangement of Eq.~(\ref{ImaiijjReal2}) then gives
\begin{align}
    -\mathrm{Im}a_{\ell}^{ii jj} &= -\frac{1}{2} \mathrm{Im}a_{\ell}^{ii jj} \qty(\qty(2-\mathrm{Im}a_{\ell}^{ii ii})+\big(2-\mathrm{Im}a_{\ell}^{jj jj}\big)) + \frac{1}{2} \big( \mathrm{Re}a_{\ell}^{ii ii} \mathrm{Re}a_{\ell}^{jj ii} +\mathrm{Re}a_{\ell}^{ii jj} \mathrm{Re}a_{\ell}^{jj jj} \big)
    \notag\\[0.5mm]
    & \hspace{2em} +  \sum_{X \neq ii,jj}\!\!\!\!{}'~ \Big(\mathrm{Im}\,a_{\ell}^{ii\rightarrow X} \mathrm{Im}\,a_{\ell}^{jj\rightarrow X} + \mathrm{Re}\,a_{\ell}^{ii\rightarrow X} \mathrm{Re}\,a_{\ell}^{jj\rightarrow X} \Big)\,.
    \label{ImaiijjRealFF}
\end{align}
Notice that conditions (\ref{ImaiijjReal2}) and (\ref{ImaiijjRealFF}) are structurely very similar if we swap $\mathrm{Im}a^{iiii}_\ell\leftrightarrow (2-\mathrm{Im}a^{iiii}_\ell)$. This allows us to obtain an inequality using a technique similar to that leads to Eq.~(\ref{eq:proofaijkl}) and the Cauchy-Schwarz inequality. More explicitly,
\begin{align}
       \big|\mathrm{Im}a_{\ell}^{ii jj}\big|^2 &\leq \Bigg(\frac{(2-\mathrm{Im}a^{iiii}_\ell)^2}{2} +  \frac{(\mathrm{Im}a^{iijj}_\ell )^2}{2}  + \frac{(\mathrm{Re}a_{\ell}^{ii ii})^2}{2} + \frac{(\mathrm{Re}a_{\ell}^{ii jj})^2}{2} + \sum_{X \neq ii,jj}\!\!\!\!{}' \; \big|a_{\ell}^{ii\rightarrow X} \big|^2\Bigg) \notag
    \\
    & \times \Bigg( \frac{(\mathrm{Im}a^{jjii}_\ell )^2 }{2} + \frac{(2-\mathrm{Im}a^{jjjj}_\ell)^2}{2} +   \frac{ (\mathrm{Re}a_{\ell}^{jjii})^2}{2} + \frac{(\mathrm{Re}a_{\ell}^{jjjj})^2}{2}  + \sum_{X \neq ii,jj}\!\!\!\!{}'\; \big|a_{\ell}^{jj\rightarrow X} \big|^2\Bigg)  \notag
    \\
    & = \big(2-\mathrm{Im}a^{iiii}_\ell\big) \big(2-\mathrm{Im}a^{jjjj}_\ell\big)
    \leq \Big(2-\frac{\mathrm{Im}a^{iiii}_\ell+\mathrm{Im}a^{jjjj}_\ell}{2}\Big)^2\,,
    \label{ImaiijjInequ}
\end{align}
where in the last line we have used Eq.~(\ref{eq:2-rho1111}). Taking the square root, we get a linear inequality
\be
\label{ImaiijjBoundFin}
 \big|\mathrm{Im}a_{\ell}^{ii jj}\big|\leq  2-\frac{\mathrm{Im}a^{iiii}_\ell+\mathrm{Im}a^{jjjj}_\ell}{2}\,.
\ee

For unitarity conditions with $a_{\ell}^{ijij}$ and $a_{\ell}^{i\neq j\neq k\neq l}$, we can analogously obtain
\be
\label{Imaijkl1414}
\big| \mathrm{Im}a_{\ell}^{ij kl} \big| \leq \frac{1}{2}-\frac{\mathrm{Im}a_{\ell}^{ij ij} + \mathrm{Im}a_{\ell}^{kl kl}}{2}\,.
\ee
Recall that, in the main text, we have established a similar set of inequalities (\ref{ImijklUpperB}) for $\mathrm{Im}a^{iijj}_\ell$ and $\mathrm{Im}a^{ijkl}_\ell$:
\begin{equation}
\label{ImijklUpperBagain}
    \big|\mathrm{Im}a^{iijj}_\ell\big|  \leq  \frac{\mathrm{Im}a^{iiii}_\ell+\mathrm{Im}a^{jjjj}_\ell}{2}, ~~~~~~ \big|\mathrm{Im}a^{ijkl}_\ell\big|  \leq  \frac{\mathrm{Im}a^{ijij}_\ell+\mathrm{Im}a^{klkl}_\ell}{2}\,.
\end{equation}
Combining Eqs.~(\ref{ImaiijjBoundFin}-\ref{ImijklUpperBagain}), we conclude
\begin{align}
\big|\mathrm{Im}a_{\ell}^{iijj}\big| &\leq 1 - \Big|1-\frac{\mathrm{Im}a_{\ell}^{iiii}+\mathrm{Im}a_{\ell}^{jjjj}}{2} \Big|\,,
\\
\big| \mathrm{Im}a_{\ell}^{ij kl} \big| &\leq \frac14- \Big|\frac14 -\frac{\mathrm{Im}a_{\ell}^{ij ij} + \mathrm{Im}a_{\ell}^{kl kl}}{2}\Big| ,
\end{align}
which are stronger than Eqs.~(\ref{Imaiijjweak}) and (\ref{Imaijkl14}).
\\
\\
\noindent \textbullet\ \textbf{Special inequalities with four fields} 
\\[1mm]
For the case of 4 fields\,\footnote{For the case with more than 4 fields, we can pick up any 4 of them, the same derivation also applies.}, we can derive some extra linear constraints. To highlight the variables whose constraints we are after, let us denote $\mathrm{Im}a^{1122}_\ell$, $\mathrm{Im}a^{1133}_\ell$, $\mathrm{Im}a^{2244}_\ell$ and $\mathrm{Im}a^{3344}_\ell$ as ${\cal A}$, ${\cal B}$, ${\cal C}$ and ${\cal D}$ respectively. After some simple re-arrangements, the unitarity condition (Eq.~(\ref{eq:uni2})) for the case of $ijkl=1111,4444,1144$ can be written as, respectively,
\begin{align}
    \big(1-\mathrm{Im}a^{1111}_\ell \big)^2 + \big(\mathrm{Im}a^{1144}_\ell \big)^2 &= 1- {\cal A}^2 - {\cal 
    B}^2 -2 {\cal S}^{1111}\,, \label{eq:A1111} \\
    \big(1-\mathrm{Im}a^{4444}_\ell \big)^2 + \big(\mathrm{Im}a^{1144}_\ell \big)^2 &= 1- {\cal C}^2 - {\cal D}^2 -2 {\cal S}^{4444}\,, \label{eq:A4444} \\
    \mathrm{Im}a^{1144}_\ell \big( (1-\mathrm{Im}a^{1111}_\ell) + (1-\mathrm{Im}a^{4444}_\ell) \big) &= {\cal A}{\cal C} + {\cal B}{\cal D} + 2{\cal S}^{1144}\,, \label{eq:A1144}
\end{align}
where we have introduced the shorthand 
\be
{\cal S}^{iijj}=\sum_X{}{}'\; \mathrm{Re}\,a_{\ell}^{ii\rightarrow X} \mathrm{Re}\,a_{\ell}^{jj\rightarrow X} + \sum_{X \neq 11,22,33,44}\hspace{-18pt}{}'~~~~\mathrm{Im}\,a_{\ell}^{ii\rightarrow X} \mathrm{Im}\,a_{\ell}^{jj\rightarrow X}\,.
\ee
Combining the three equations as (\ref{eq:A1111})$+$(\ref{eq:A4444})$-2$(\ref{eq:A1144}) and using the fact that ${\cal S}^{1144}\geq -\sqrt{{\cal S}^{1111} {\cal S}^{4444}}$, we can obtain
\begin{align}
    &\, 2-({\cal A}+{\cal C})^2-({\cal B}+{\cal D})^2  \notag \\
    = &\, \big(1-\mathrm{Im}a^{1111}_\ell -\mathrm{Im}a^{1144}_\ell \big)^2 + \big(1-\mathrm{Im}a^{4444}_\ell -\mathrm{Im}a^{1144}_\ell \big)^2 +2\big({\cal S}^{1111}+{\cal S}^{4444}+2{\cal S}^{1144}\big) \notag \\
    \geq &\, 2\big(\sqrt{{\cal S}^{1111}}-\sqrt{{\cal S}^{4444}}\big)^2  \geq  0\,,
    \label{ABCD1}
\end{align}
Similarly, combining the three equations as (\ref{eq:A1111})$+$(\ref{eq:A4444})$+2$(\ref{eq:A1144}) and using the fact that ${\cal S}^{1144}\leq \sqrt{{\cal S}^{1111} {\cal S}^{4444}}$, we have
\be
({\cal A}-{\cal C})^2+({\cal B}-{\cal D})^2\leq 2\,,
\label{ABCD2}
\ee
In the numerical LP, we will use the linear conditions that can be inferred from conditions (\ref{ABCD1}) and (\ref{ABCD2}) respectively:
\begin{align}
\label{ImaS1}
    |\mathrm{Im}a^{1122}_\ell + \mathrm{Im}a^{1133}_\ell + \mathrm{Im}a^{2244}_\ell + \mathrm{Im}a^{3344}_\ell| &\leq 2\,,
    \\
\label{ImaS2}
    |\mathrm{Im}a^{1122}_\ell - \mathrm{Im}a^{1133}_\ell - \mathrm{Im}a^{2244}_\ell + \mathrm{Im}a^{3344}_\ell| &\leq 2\,.
\end{align}
The above two conditions are the strongest in the 2D subspace furnished by $({\cal A}+{\cal D})$ and  $({\cal B}+{\cal C})$ from the unitarity conditions in the form of (\ref{eq:A1111}) to (\ref{eq:A1144}). Obviously, for general $i\neq j\neq k\neq l$, the same result can be formulated as
\begin{equation}
    \big|(\mathrm{Im}a^{iijj}_\ell + \mathrm{Im}a^{kkll}_\ell) \pm (\mathrm{Im}a^{iikk}_\ell + \mathrm{Im}a^{jjll}_\ell)\big| \leq 2\,.
\end{equation}

\section{Existence of upper bounds: simple analytical example}
\label{app:analytical}

In this appendix, we shall use a simple example to illustrate why the LP problem (\ref{LPgeneral1}-\ref{eq:discrete general nc}) leads to upper bounds on the amplitude coefficients. We will not attempt to use all available null constraints and unitarity conditions, but utilize just a few conditions to establish the existence of an upper bound, the upshot being that we can show this analytically. Furthermore, in the numerical LP, we have to truncate the partial waves at a finite order $\ell_M$, while in the analytical example below, we can sum over all partial waves, and find that the result does not diverge as $\ell$ goes to infinity. This, together with our numerical convergence study in Appendix~\ref{app:convergence} justifies the finite $\ell$ truncation in our numerical results.

We shall take the $c_{1212}^{2,0}$ coefficient as an example, analytically bounding it from the above. The sum rule for this coefficient is given by
\begin{equation}
    c_{1212}^{2,0} = \frac{1}{\Lambda^4}\, 32\sum_{\ell\geq 0}(2\ell+1)\int_{0}^1 \dd z\, z\, \rho_\ell^{1212}(z)\,, \label{eq:c121220}
\end{equation}
where we have changed the integration variable from $\mu$ to $z={\Lambda^2}/{\mu}$ in Eq.~(\ref{eq:cijkl20}). Likewise, the first order null constraint for $\rho^{1111}_\ell$ can be written as
\begin{equation}
    \sum_{\ell\geq 0} (2\ell +1) \int_0^1 \dd z\, z^3  (\ell^4 +2\ell^3 -7\ell^2 -8\ell) \rho^{1111}_\ell (z)  = 0\,.
\end{equation}
By $\rho_{\ell}^{i j k l}=(-1)^{\ell} \rho_{\ell}^{j i k l}$, we can infer that $\rho^{1111}_{\ell={\rm odd}}=0$. 
Moving the negative $\ell=2$ term to the right hand side ($0\leq \rho^{1111}_\ell\leq 2$) and using the unitarity condition of $\rho^{1111}_2\leq 2$, we get
\begin{align}
    &\sum_{\ell\geq 4} (2\ell +1) \int_0^1 \dd z\, z^3  (\ell^4 +2\ell^3 -7\ell^2 -8\ell) \rho^{1111}_\ell (z) 
    =\int_0^1 \dd z\, z^3  60 \rho^{1111}_2(z)
    \leq  {30}\,. \label{eq:sumrho1111}
\end{align}
That is, an infinite positive sum converges to a finite number, which implies that $\int_0^1\! \dd z\, z^3\rho^{1111}_\ell$ must be sufficiently small at large $\ell$, sometimes known as low spin dominance. We stress that this is a result of null constraints. Since $\rho^{1111}_\ell$ and $\rho^{2222}_\ell$ satisfy exactly the same constraints, we can replace $\rho^{1111}_\ell$ with $(\rho^{1111}_\ell+\rho^{2222}_\ell)/2$ in the inequality above, and further use it to constrain a similar positive sum on $\rho^{1122}_\ell$ through $|\rho^{1122}_\ell|\leq (\rho^{1111}_\ell+\rho^{2222}_\ell)/2$,
\begin{align}
    & \sum_{\ell\geq 4} (2\ell +1) \int_0^1 \dd z\, z^3  (\ell^4 +2\ell^3 +\ell^2) \, \qty|\rho^{1122}_\ell (z)| 
    \label{Psum1122}\\
    \leq  & \,\frac{5}{3} \sum_{\ell\geq 4} (2\ell +1) \int_0^1 \dd z\, z^3  (\ell^4 +2\ell^3 -7\ell^2 -8\ell) \, \qty|\rho^{1122}_\ell (z)|
    \\
    \leq  & \,\frac{5}{3} \sum_{\ell\geq 4} (2\ell +1) \int_0^1 \dd z\, z^3  (\ell^4 +2\ell^3 -7\ell^2 -8\ell) \, \frac{\rho^{1111}_\ell (z)+\rho^{2222}_\ell(z)}{2}
    \leq  \,\frac{5}{3}\times {30} = {50}\,. 
    \label{Psum1122Final}
\end{align}
The sum (\ref{Psum1122}) appears in one of the first order null constraints for $\rho^{1122}_\ell$ and $\rho^{1212}_\ell$
\begin{align}
    \sum_{\ell\geq 0} (2\ell +1) \int_0^1 \dd z\, z^3  \Big[-(\ell^4 +2\ell^3 +\ell^2) \rho^{1122}_\ell (z)\, + 
    Q(\ell)\, \rho^{1212}_\ell (z) \Big] = 0\,,
\end{align}
where $Q(\ell) = (3\ell^4 +6\ell^3 -21\ell^2 -24\ell +32) + (-1)^\ell (-\ell^4 -2\ell^2 +15\ell^2 +16\ell -32)$. We can then constrain the sum of $\rho^{1212}_\ell$
\begin{align}
    \sum_{\ell\geq 0} (2\ell +1) \int_0^1 \dd z\, z^3 Q(\ell)\, \rho^{1212}_\ell (z)
     &= \sum_{\ell\geq 0} (2\ell +1) \int_0^1 \dd z\, z^3  (\ell^4 +2\ell^3 +\ell^2) \rho^{1122}_\ell (z)
    \\
     & \leq  \int_0^1 \dd z\, z^3 180 \rho^{1122}_2 (z) + {50}
     \leq  {95}\,. \label{eq:constrainrho1212}
\end{align}
where in the first inequality we have used Eq.~(\ref{Psum1122Final}) and $\rho^{1122}_{1}=\rho^{1122}_{3}=0$, and in the second inequality $|\rho^{1122}_2| \leq 1$ are used. 

Now, we are ready to constrain $c^{2,0}_{1212}$ (Eq.~(\ref{eq:c121220})) by subtracting Eq.~(\ref{eq:constrainrho1212}) multiplied by an arbitrary constant $k$.
\begin{align}
    c^{2,0}_{1212} &\leq \frac{1}{\Lambda^4} \bigg[\sum_{\ell\geq 0}(2\ell +1)\int_{0}^1 \dd z\, z\, \qty(32-k z^2 Q(\ell))\rho_\ell^{1212}(z) + 95 k \bigg]
    \\
    & \leq \frac{1}{\Lambda^4} \bigg[ \int_{0}^1 \dd z\, z\, \qty(32\rho_0^{1212} + 96\rho_1^{1212})
     + \sum_{\ell\geq 2}(2\ell+1)\int_{0}^1 \dd z\, z\, \Big(32-\frac{3}{2} k z^2 \ell^4 \Big)\rho_\ell^{1212} +{95} k \bigg]\notag
    \\
    & \leq \frac{1}{\Lambda^4} \bigg[ 32 + \sum_{\ell\geq 2}(2\ell+1)\int_{0}^{\mathrm{Min}(1,\sqrt{\frac{64}{3k}} \frac{1}{\ell^2})} \dd z\, z\, \Big(32-\frac{3}{2} k z^2 \ell^4 \Big)\Big(\frac{1}{2}\Big) +{95} k \bigg]\,,
\end{align}
where in the second line we have used $Q(0)=Q(1)=0$ and $Q(\ell)\geq \frac{3}{2}\ell^4$ for $ \geq2$, and in the third line we have used $\rho^{1212}_\ell \leq \frac{1}{2}$, and also changed the upper limit of the integral to drop the negative contribution. Since $k$ is arbitrary, we can choose $k=0.419$ to minimize the result, which concludes an upper bound of
\begin{equation}
    \frac{c^{2,0}_{1212}}{(4\pi)^2} \leq \frac{0.89}{\Lambda^4}\,.
\end{equation}
This is actually very close to the numerical upper bound of 0.77 obtained by performing linear programs with all first order null constraints and more unitarity conditions.

As emphasized in the main text, this example also highlights the necessity of the unitarity condition $|\rho^{1122}_\ell|\leq (\rho^{1111}_\ell+\rho^{2222}_\ell)/2$ for establishing the upper bound (see Eq.~(\ref{Psum1122Final})). Without it, we would not be able to bound $\rho^{1122}_\ell$ neither from above nor from below, due to the fact that the sign of $\rho^{1122}_\ell$ is undetermined. We can also analytically find the upper bound for any other $c_{ijkl}^{2,0}$ analogously, which again crucially relies on the unitarity relation of $|\rho^{ijkl}_\ell|\leq (\rho^{ijij}_\ell+\rho^{klkl}_\ell)/2$.

\section{Symmetries of Higgs amplitudes}
\label{sec:symHiggs}

It is instructive to see how the symmetries of the Higgs amplitudes reduce the number of independent $s^2$ coefficients $c^{2,0}_{ijkl}$. To this end, first, note that the $\phi_i \phi_j \to \phi_k \phi_l$ amplitude is invariant under $SU(2)_L\times U(1)_Y$. Considering all possible time-reversal (CP) and gauge invariant operators with an arbitrary derivative structure, 
the following amplitudes must vanish
\begin{align}
&A_{iiij}=A_{iiji}=A_{ijii}=A_{jiii}=A_{iijk}=A_{ijki}=A_{ijik}=A_{jkii}=0\quad(i\neq j\text{ and } j\neq k)\,.
\end{align}
The only non-vanishing amplitudes are those involving four identical states, two pairs of identical states, or four different states. The only other general relations among these non-vanishing amplitudes that can be obtained from the operator structure are for the scattering of all identical particles, $\phi_i \phi_ i \to \phi_i \phi_i$, where the internal symmetry alone dictates that
\begin{align}
A_{1111}=A_{2222}=A_{3333}=A_{4444}\,. \label{eq:symmetryaiiii}
\end{align}
On the other hand, time-reversal symmetry implies
\begin{align}
A_{ijkl}=A_{klij}\,,
\end{align}
for the non-vanishing amplitudes.
For scalar scattering amplitudes, we also have a symmetry when simultaneously exchanging the initial state particles and the two final state particles, (which corresponds to a rotation by $\pi$ in the scattering plane or parity conservation), {\it i.e.}, 
\begin{align}
A_{ijkl}=A_{jilk}\,,
\end{align}
which can also be inferred from Eq.~\eqref{eq:rhoellRel}.
We also observe that under either of the interchanges $\phi_1\leftrightarrow \phi_2$ or $\phi_3\leftrightarrow \phi_4$, the interactions involving all four fields flip sign, while the remaining terms are unchanged. This implies that the operators are symmetric under the simultaneous exchange $\phi_1\leftrightarrow \phi_2$ and $\phi_3\leftrightarrow \phi_4$. We also observe a symmetry under simultaneous $\phi_1\leftrightarrow \phi_3$ and $\phi_2\leftrightarrow \phi_4$ interchange (which in turn implies a simultaneous $\phi_1\leftrightarrow \phi_4$ and $\phi_2\leftrightarrow \phi_3$ symmetry).
This further reduces the number of independent amplitudes with two pairs of identical particles and halves the number of amplitudes with all different particles.
\begin{align}
\begin{matrix}
    A_{1122}&=A_{3344},\quad &
    A_{1133}&=A_{2244}=
    A_{1144}=A_{2233},\\
    A_{1212}&=A_{3434},\quad &
    A_{1313}&=A_{2424}=
    A_{1414}=A_{2323},\\
    A_{1221}&=A_{3443},\quad &
    A_{1331}&=A_{2442}=
    A_{1441}=A_{2332},
\end{matrix}
\end{align}
\begin{align}
\label{eq:A1234}
\begin{matrix}
A_{1234}&=-A_{1243},\quad &
    A_{1324}&=-A_{1423},\quad &
    A_{1342}=-A_{1432}.
\end{matrix}
\end{align}

Considering all of the above symmetries, we arrive at the following set of 10 independent amplitudes for a generic $2\to2$ scattering of Higgs doublet components,
\begin{itemize}
\item 1 of $\{i,\,i,\,i,\,i\}$ type: $A_{1111}$
\item $2\times 3$ of $\{i,\,i,\,j,\,j\}$ type: $A_{11jj},\, A_{1j1j},\, A_{1jj1}$ $(j\neq1)$
\item $4!/(2\times2\times2)=3$ of $\{1,\,2,\,3,\,4\}$ type:
$A_{1234},\,A_{1324},\,A_{1432}$.
\end{itemize}
Full crossing symmetry implies that the $c_{ijkl}^{2,0}$ coefficients fully characterise the coefficients of $A_{ijkl}$ amplitudes at the quadratic order in $(s,t)$.
It also requires them to be $su$-symmetric ($j \leftrightarrow l$), leading to further relations, $c^{2,0}_{11jj}=c^{2,0}_{1jj1}$, $c^{2,0}_{1234}=c^{2,0}_{1432}$, which together with Eq.~\eqref{eq:A1234} implies $c^{2,0}_{1324}=0$. Finally, we are left with $1+4+1=6$ independent $c_{ijkl}^{2,0}$ for Higgs scattering:  $c_{1111}$, $c_{1122}$, $c_{1212}$, $c_{1133}$, $c_{1313}$ and $c_{1234}$.
Matching the amplitudes to the three dimension-8 coefficients $C_1,\,C_2$ and $C_3$, we find the 6 non-vanishing coefficients given in Eq.~\eqref{eq:cijkl_Higgs}.

\section{Reducing decision variables for the Higgs case}
\label{app:COV}

In Section~\ref{sec: higgsbounds}, when performing LP for Higgs operators, we define the following new variables
\begin{align}
    \begin{split}
    & R^{1111}_{\ell,n}\equiv\frac{\rho^{1111}_{\ell,n}+\rho^{2222}_{\ell,n}+\rho^{3333}_{\ell,n}+\rho^{4444}_{\ell,n}}{4}, \\
    & R^{1122}_{\ell,n}\equiv \frac{\rho^{1122}_{\ell,n}+\rho^{3344}_{\ell,n}}{2}, \hspace{1.2em} R^{1212}_{\ell,n}\equiv \frac{\rho^{1212}_{\ell,n}+\rho^{3434}_{\ell,n}}{2}, \\
    & R^{1133}_{\ell,n}\equiv \frac{\rho^{1133}_{\ell,n}+\rho^{2244}_{\ell,n}}{2}, \hspace{1.2em} R^{1313}_{\ell,n}\equiv \frac{\rho^{1313}_{\ell,n}+\rho^{2424}_{\ell,n}}{2}, \\
    & R^{1144}_{\ell,n}\equiv \frac{\rho^{1144}_{\ell,n}+\rho^{2233}_{\ell,n}}{2}, \hspace{1.2em} R^{1414}_{\ell,n}\equiv \frac{\rho^{1414}_{\ell,n}+\rho^{2323}_{\ell,n}}{2}.        
    \end{split} \label{eq:higgs_changevariable}
\end{align}
It is easy to see that $R^{1111}_{\ell,n}$ and $\rho^{iiii}_{\ell,n}$ satisfy the same upper/lower bounds and null constraints, as do {$R^{11jj}_{\ell,n},R^{1j1j}_{\ell,n}$} and {$\rho^{iijj}_{\ell,n},\rho^{ijij}_{\ell,n}$}, due to the fact that their formalism is independent of the explicit choice of $i$ and $j$. The unitarity inequalities that relate different $\rho^{ijkl}_{\ell,n}$ in Eq.~(\ref{eq:rhoiijj}-\ref{eq:rho1ijk}), now become
\begin{align}
    &\big|R^{11jj}_{\ell,n}\big|\leq 1- \big|1-R^{1111}_{\ell,n}\big|, \\[2mm]
    &\big|R^{11jj}_\ell \pm R^{11kk}_\ell \big| \leq 1, \\[2mm]
    &\big|\rho^{1jkl}_{\ell,n}\big| \leq \frac{1}{4} - \Big|\frac{1}{4} - R^{1j1j}_{\ell,n}\Big|,
\end{align}
where the last inequality relies on the symmetry of $\rho^{ijij}_{\ell,n}=\rho^{jiji}_{\ell,n}$ which is implied by Eq.~\eqref{eq:rhoellRel}. With this, the LP constraints in terms of the new variables $R$ and $\rho^{1jkl}_{\ell,n}|_{j\neq k\neq l\neq 1}$ read
\begin{align}
    &\left\{ 
    \begin{array}{lr}
        0\leq R^{1111}_{\ell,n} \leq 2\,, \\
        \sum\limits_{\ell=0,...,\ell_M;\ell_\infty}(2\ell+1) \sum\limits_{n=1}^N \frac{1}{N}(\frac{n}{N})^{r+2} C^{1111}_{r,i_r}(\ell) R^{1111}_{\ell,n} = 0\,, 
    \end{array}
    \right. \label{eq:higgsnc1}
    \\[2mm]
    &\hspace{0.6em}\big|R^{1122}_{\ell,n}\big|\leq 1- \big|1-R^{1111}_{\ell,n}\big|, \hspace{0.9em} \notag \\ &\hspace{0.6em}\big|R^{1133}_{\ell,n}\big|\leq 1- \big|1-R^{1111}_{\ell,n}\big|, \hspace{0.9em} \\ &\hspace{0.6em}\big|R^{1144}_{\ell,n}\big|\leq 1- \big|1-R^{1111}_{\ell,n}\big|, \notag \\[2mm]
    &\left\{ 
    \begin{array}{lr}
     0\leq R_{\ell,n}^{1212} \leq \frac{1}{2}, \hspace{1em} 0\leq R_{\ell,n}^{1313} \leq \frac{1}{2}, \hspace{1em} 0\leq R_{\ell,n}^{1414} \leq \frac{1}{2},\\[2mm]
    \big| R^{1122}_{\ell,n} \pm R^{1133}_{\ell,n} \big| \leq 1, \hspace{1.2em} \big|R^{1122}_{\ell,n} \pm R^{1133}_{\ell,n} \big| \leq 1, \hspace{1.2em} \big|R^{1122}_{\ell,n} \pm R^{1133}_{\ell,n} \big| \leq 1\,,\\[2mm]
    \sum\limits_{\ell=0,...,\ell_M;\ell_\infty}(2\ell+1) \sum\limits_{n=1}^N \frac{1}{N}(\frac{n}{N})^{r+2}  \Big[C^{1122}_{r,i_r}(\ell) R_{\ell,n}^{1122}+C^{1212}_{r,i_r}(\ell) R_{\ell,n}^{1212} \Big] =0\,, \\
     \sum\limits_{\ell=0,...,\ell_M;\ell_\infty}(2\ell+1) \sum\limits_{n=1}^N \frac{1}{N}(\frac{n}{N})^{r+2}  \Big[C^{1122}_{r,i_r}(\ell) R_{\ell,n}^{1133}+C^{1212}_{r,i_r}(\ell) R_{\ell,n}^{1313} \Big] =0\,, \\
     \sum\limits_{\ell=0,...,\ell_M;\ell_\infty}(2\ell+1) \sum\limits_{n=1}^N \frac{1}{N}(\frac{n}{N})^{r+2}  \Big[C^{1122}_{r,i_r}(\ell) R_{\ell,n}^{1144}+C^{1212}_{r,i_r}(\ell) R_{\ell,n}^{1414} \Big] =0\,,
    \end{array}
    \right. 
    \\[2mm]
    &\hspace{0.6em}\big|\rho^{1234}_{\ell,n}\big| \leq \text{$\tfrac{1}{4}$} - \big|\tfrac{1}{4} - R^{1212}_{\ell,n}\big|, \notag \\
    &\hspace{0.6em}\big|\rho^{1324}_{\ell,n}\big| \leq \tfrac{1}{4} - \big|\tfrac{1}{4} - R^{1313}_{\ell,n}\big|,  \\
    &\hspace{0.6em}\big|\rho^{1423}_{\ell,n}\big| \leq \tfrac{1}{4} - \big|\tfrac{1}{4} - R^{1414}_{\ell,n}\big|, \notag
    \\[1mm]
    &\left\{ 
    \begin{array}{lr}
    \sum\limits_{\ell=0,...,\ell_M;\ell_\infty}(2\ell+1) \sum\limits_{n=1}^N \frac{1}{N}(\frac{n}{N})^{r+2}  \Big[C^{1234}_{r,i_r}(\ell)\rho_{\ell,n}^{1234}+C^{1324}_{r,i_r}(\ell) \rho_{\ell,n}^{1342} +C^{1423}_{r,i_r}(\ell)\rho_{\ell,n}^{1432} \Big]=0\,.
    \end{array}
    \right.  \label{eq:higgsnc2}
\end{align}
This reduces the number of variables from $14\times \ell \times N$ to $8\times \ell \times N$, and there are now $\lfloor(r-1)/3\rfloor+5r+6$ null constraints at the $r^\text{th}$ order, which is $3(r+1)$ fewer than the previous LP formulation.

We note that, to get the sum rules $c_{ijkl}^{2,0}$ in terms of $R$ and $\rho^{ijkl}_{\ell,n}|_{i\neq j\neq k\neq l}$, we simply change $\rho$ to $R$ on the right hand sides of the corresponding sum rules, thanks to the symmetries of $c_{ijkl}^{2,0}$. For example, since $c_{1111}^{2,0}=c_{2222}^{2,0}=c_{3333}^{2,0}=c_{4444}^{2,0}$, we can express $c_{1111}^{2,0}$ as $(c_{1111}^{2,0}+c_{2222}^{2,0}+c_{3333}^{2,0}+c_{4444}^{2,0})/{4}$, so we can write the $c_{1111}^{2,0}$ sum rule as follows
\begin{equation}
    c_{1111}^{2,0} = \frac{1}{\Lambda^4} \sum\limits_{\ell=0,...,\ell_M;\ell_\infty}(2\ell+1) \sum\limits_{n=1}^N \frac{1}{N}\frac{n}{N} 32R^{1111}_{\ell,n}\,.
\end{equation}
Similarly, since $c_{1122}^{2,0}=c_{3344}^{2,0}$, $c_{1122}^{2,0} = ({c_{1122}^{2,0}+c_{3344}^{2,0}})/{2}$, we can rewrite the $c_{1122}^{2,0}$ sum rule as 
\begin{equation}
    c_{1122}^{2,0} = \frac{1}{\Lambda^4} \sum\limits_{\ell=0,...,\ell_M;\ell_\infty}(2\ell+1) \sum\limits_{n=1}^N \frac{1}{N}\frac{n}{N} 16(R_{\ell,n}^{1122}+(-1)^\ell R_{\ell,n}^{1212})\,.
\end{equation}
In this way, we can reformulate all $c_{ijkl}^{2,0}$ in Eq.(\ref{eq:C123},~\ref{eq:addconstraint}) in terms of the corresponding $R$ variables, and therefore reformulate the target functions $C_1,C_2,C_3$. Now the LP problem becomes well-defined with fewer variables $R^{1111}_{\ell,n}$, $R^{11ii}_{\ell,n}$, $R^{1i1i}_{\ell,n}$, $\rho^{1jkl}_{\ell,n}$, as shown in Section~\ref{sec: higgsbounds}.

\begin{figure}[htbp]
\centering
\begin{center}  
\hspace{-3.2em}
\includegraphics[width=0.395\textwidth] 
{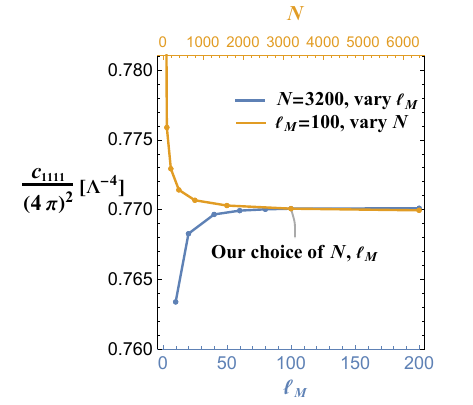}
\hspace{-1.7em}
\includegraphics[width=0.357\textwidth] {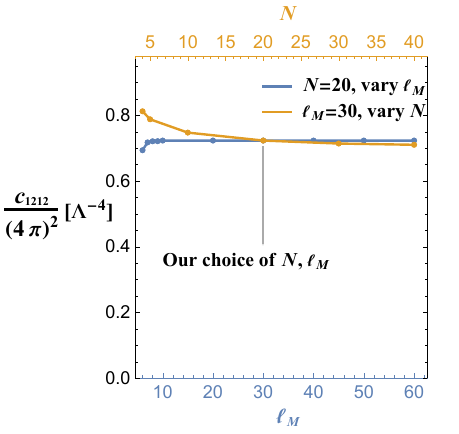}
\hspace{-1em}
\includegraphics[width=0.355\textwidth] {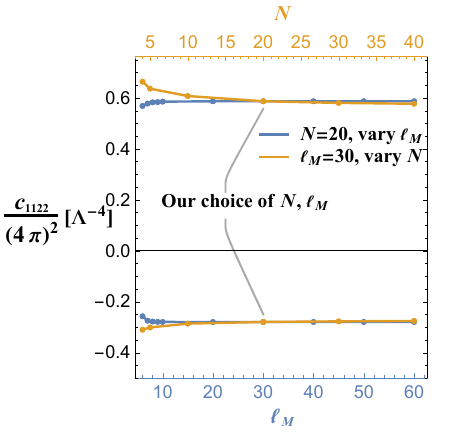}
\\[1mm]
\hspace{-1.2em}
\includegraphics[width=0.345\textwidth] {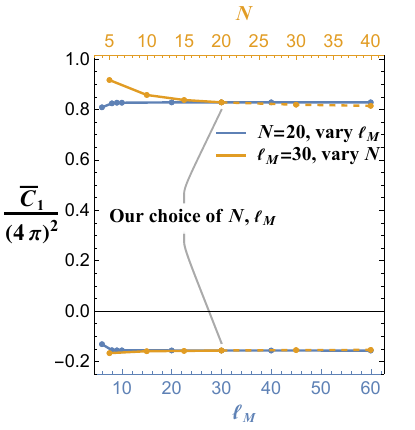}
\hspace{-0.8em}
\includegraphics[width=0.345\textwidth] {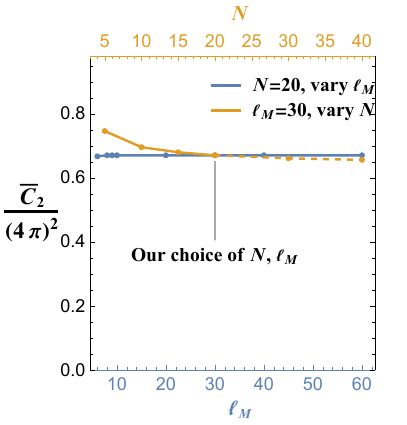}
\hspace{-0.9em}
\includegraphics[width=0.345\textwidth] {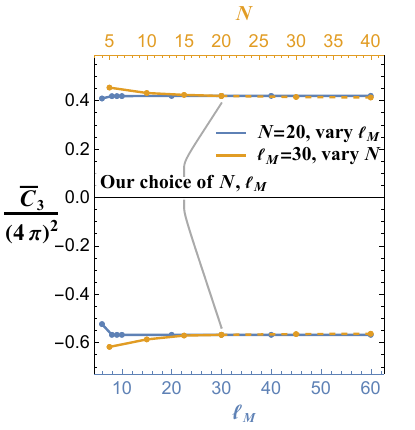}

\caption{Convergence of the upper/lower bounds with respect to $N$ and $\ell_M$. Null constraints are used up to 3rd-order. The lines in the upper and lower half plane are the values of the upper and lower bounds, respectively, except for $c_{1111}$, $c_{1212}$ and $C_2$ where only the upper bounds are presented since their lower bounds are exactly zero. }
\label{fig:covergence}
\end{center}
\end{figure}

\section{Numerical implementation and convergence}
\label{app:convergence}
\noindent \textbullet\ \textbf{Some numerical treatments} \\
When performing the LP numerically, we rotate and rescale the decision variables to speed up the runs. For example, consider the unitarity condition $\big|\rho^{1234}_{\ell,n}\big| \leq \text{$\tfrac{1}{4}$} - \big|\tfrac{1}{4} - R^{1212}_{\ell,n}\big|$, which actually defines a square geometrically. So if we can redefine
\begin{equation}
    \mathcal{R}_{\ell,n}=\frac{R^{1212}_{\ell,n}+\rho^{1234}_{\ell,n}}{2},\quad \mathcal{S}_{\ell,n}=\frac{R^{1212}_{\ell,n}-\rho^{1234}_{\ell,n}}{2} ,
\end{equation}
this condition takes the much simpler form
\begin{equation}
    0\leq \mathcal{R}_{\ell,n} \leq \frac{1}{4},\quad 0\leq \mathcal{S}_{\ell,n} \leq \frac{1}{4} \,.
\end{equation}

After discretization, a null constraint schematically takes the form of $\sum_{\ell,n} (n/N)^r Q(\ell) \rho^{ijkl}_{\ell,n}=0$, where $r\geq3$ is an integer and $Q(\ell)$ is some polynomial in $\ell$. When summing to a large $\ell$ and $n$, we see that the coefficients in the sum differ significantly in values, which slows down the numerical evaluation. To mitigate this problem, we can absorb certain powers of $\ell$ and $n$ into the spectral densities $\rho^{ijkl}_{\ell,n}\rightarrow \ell^{\hspace{0.05em}\alpha} n^\beta \rho^{ijkl}_{\ell,n}$, $\alpha,\beta$ being integers. We have observed that this enhances the speed of the LP by several times. 

\noindent \textbullet\ \textbf{Checking convergence} \\
Figure~\ref{fig:covergence} shows how the upper/lower positivity bounds change with respect to $N$ and $\ell_M$, using up to 3rd-order null constraints. We see that despite the ideal case being $N, \ell_M \rightarrow \infty$, our choice of finite $N$ and $\ell_M$ is a very good approximation. The difference caused by finite $\ell_M$ is found to be negligible, and further increasing $N$ will improve the bounds by a few percent, which however becomes time-consuming. For $C_i$, the cases of $N>20$ are found to be too time-consuming, so we use extrapolation to get the bounds for larger $N$'s (shown as dashed lines in Figure~\ref{fig:covergence}). Specifically, we assume that the deviation from the true value goes like $\delta \propto 1/N$, which is true for $c_{1111}$, $c_{1212}$ and $c_{1122}$ and is consistent with the fact that the objective function is a discretized integration in the form of Eq.~(\ref{eq:c20_objective}). This convergence study justifies our discretization scheme used in Section \ref{sec:constWilson}.

\bibliographystyle{JHEP}
\bibliography{refs}

\providecommand{\href}[2]{#2}\begingroup\raggedright\begin{thebibliography}{100}

\bibitem{Adams:2006sv}
A.~Adams, N.~Arkani-Hamed, S.~Dubovsky, A.~Nicolis and R.~Rattazzi,
  \emph{{Causality, analyticity and an IR obstruction to UV completion}},
  \href{http://dx.doi.org/10.1088/1126-6708/2006/10/014}{\emph{JHEP} {\bfseries
  10} (2006) 014}, [\href{https://arxiv.org/abs/hep-th/0602178}{{\ttfamily
  hep-th/0602178}}].

\bibitem{Distler:2007}
J.~Distler, B.~Grinstein, R.~A. Porto and I.~Z. Rothstein, \emph{Falsifying
  models of new physics via $ww$ scattering},
  \href{http://dx.doi.org/10.1103/PhysRevLett.98.041601}{\emph{Phys. Rev.
  Lett.} {\bfseries 98} (Jan, 2007) 041601}.

\bibitem{Ian:2010}
I.~Low, R.~Rattazzi and A.~Vichi, \emph{Theoretical constraints on the higgs
  effective couplings},
  \href{http://dx.doi.org/10.1007/JHEP04(2010)126}{\emph{Journal of High Energy
  Physics} {\bfseries 2010} (2010) }.

\bibitem{deRham:2017avq}
C.~de~Rham, S.~Melville, A.~J. Tolley and S.-Y. Zhou, \emph{{Positivity bounds
  for scalar field theories}},
  \href{http://dx.doi.org/10.1103/PhysRevD.96.081702}{\emph{Phys. Rev. D}
  {\bfseries 96} (2017) 081702},
  [\href{https://arxiv.org/abs/1702.06134}{{\ttfamily 1702.06134}}].

\bibitem{deRham:2017zjm}
C.~de~Rham, S.~Melville, A.~J. Tolley and S.-Y. Zhou, \emph{{UV complete me:
  Positivity Bounds for Particles with Spin}},
  \href{http://dx.doi.org/10.1007/JHEP03(2018)011}{\emph{JHEP} {\bfseries 03}
  (2018) 011}, [\href{https://arxiv.org/abs/1706.02712}{{\ttfamily
  1706.02712}}].

\bibitem{Bellazzini:2017bkb}
B.~Bellazzini, F.~Riva, J.~Serra and F.~Sgarlata, \emph{{The other effective
  fermion compositeness}},
  \href{http://dx.doi.org/10.1007/JHEP11(2017)020}{\emph{JHEP} {\bfseries 11}
  (2017) 020}, [\href{https://arxiv.org/abs/1706.03070}{{\ttfamily
  1706.03070}}].

\bibitem{Arkani-Hamed:2020blm}
N.~Arkani-Hamed, T.-C. Huang and Y.-t. Huang, \emph{{The EFT-Hedron}},
  \href{http://dx.doi.org/10.1007/JHEP05(2021)259}{\emph{JHEP} {\bfseries 05}
  (2021) 259}, [\href{https://arxiv.org/abs/2012.15849}{{\ttfamily
  2012.15849}}].

\bibitem{Bellazzini:2020cot}
B.~Bellazzini, J.~Elias~Mir\'o, R.~Rattazzi, M.~Riembau and F.~Riva,
  \emph{{Positive moments for scattering amplitudes}},
  \href{http://dx.doi.org/10.1103/PhysRevD.104.036006}{\emph{Phys. Rev. D}
  {\bfseries 104} (2021) 036006},
  [\href{https://arxiv.org/abs/2011.00037}{{\ttfamily 2011.00037}}].

\bibitem{Tolley:2020gtv}
A.~J. Tolley, Z.-Y. Wang and S.-Y. Zhou, \emph{{New positivity bounds from full
  crossing symmetry}},
  \href{http://dx.doi.org/10.1007/JHEP05(2021)255}{\emph{JHEP} {\bfseries 05}
  (2021) 255}, [\href{https://arxiv.org/abs/2011.02400}{{\ttfamily
  2011.02400}}].

\bibitem{Caron-Huot:2020cmc}
S.~Caron-Huot and V.~Van~Duong, \emph{{Extremal Effective Field Theories}},
  \href{http://dx.doi.org/10.1007/JHEP05(2021)280}{\emph{JHEP} {\bfseries 05}
  (2021) 280}, [\href{https://arxiv.org/abs/2011.02957}{{\ttfamily
  2011.02957}}].

\bibitem{Chiang:2021ziz}
L.-Y. Chiang, Y.-t. Huang, W.~Li, L.~Rodina and H.-C. Weng, \emph{{Into the
  EFThedron and UV constraints from IR consistency}},
  \href{http://dx.doi.org/10.1007/JHEP03(2022)063}{\emph{JHEP} {\bfseries 03}
  (2022) 063}, [\href{https://arxiv.org/abs/2105.02862}{{\ttfamily
  2105.02862}}].

\bibitem{Sinha:2020win}
A.~Sinha and A.~Zahed, \emph{{Crossing Symmetric Dispersion Relations in
  Quantum Field Theories}},
  \href{http://dx.doi.org/10.1103/PhysRevLett.126.181601}{\emph{Phys. Rev.
  Lett.} {\bfseries 126} (2021) 181601},
  [\href{https://arxiv.org/abs/2012.04877}{{\ttfamily 2012.04877}}].

\bibitem{Zhang:2020jyn}
C.~Zhang and S.-Y. Zhou, \emph{{Convex Geometry Perspective on the (Standard
  Model) Effective Field Theory Space}},
  \href{http://dx.doi.org/10.1103/PhysRevLett.125.201601}{\emph{Phys. Rev.
  Lett.} {\bfseries 125} (2020) 201601},
  [\href{https://arxiv.org/abs/2005.03047}{{\ttfamily 2005.03047}}].

\bibitem{Li:2021lpe}
X.~Li, H.~Xu, C.~Yang, C.~Zhang and S.-Y. Zhou, \emph{{Positivity in Multifield
  Effective Field Theories}},
  \href{http://dx.doi.org/10.1103/PhysRevLett.127.121601}{\emph{Phys. Rev.
  Lett.} {\bfseries 127} (2021) 121601},
  [\href{https://arxiv.org/abs/2101.01191}{{\ttfamily 2101.01191}}].

\bibitem{Bellazzini:2014waa}
B.~Bellazzini, L.~Martucci and R.~Torre, \emph{{Symmetries, Sum Rules and
  Constraints on Effective Field Theories}},
  \href{http://dx.doi.org/10.1007/JHEP09(2014)100}{\emph{JHEP} {\bfseries 09}
  (2014) 100}, [\href{https://arxiv.org/abs/1405.2960}{{\ttfamily 1405.2960}}].

\bibitem{Bellazzini:2016xrt}
B.~Bellazzini, \emph{{Softness and amplitudes\textquoteright{} positivity for
  spinning particles}},
  \href{http://dx.doi.org/10.1007/JHEP02(2017)034}{\emph{JHEP} {\bfseries 02}
  (2017) 034}, [\href{https://arxiv.org/abs/1605.06111}{{\ttfamily
  1605.06111}}].

\bibitem{Bern:2021ppb}
Z.~Bern, D.~Kosmopoulos and A.~Zhiboedov, \emph{{Gravitational effective field
  theory islands, low-spin dominance, and the four-graviton amplitude}},
  \href{http://dx.doi.org/10.1088/1751-8121/ac0e51}{\emph{J. Phys. A}
  {\bfseries 54} (2021) 344002},
  [\href{https://arxiv.org/abs/2103.12728}{{\ttfamily 2103.12728}}].

\bibitem{Alberte:2020jsk}
L.~Alberte, C.~de~Rham, S.~Jaitly and A.~J. Tolley, \emph{{Positivity Bounds
  and the Massless Spin-2 Pole}},
  \href{http://dx.doi.org/10.1103/PhysRevD.102.125023}{\emph{Phys. Rev. D}
  {\bfseries 102} (2020) 125023},
  [\href{https://arxiv.org/abs/2007.12667}{{\ttfamily 2007.12667}}].

\bibitem{Tokuda:2020mlf}
J.~Tokuda, K.~Aoki and S.~Hirano, \emph{{Gravitational positivity bounds}},
  \href{http://dx.doi.org/10.1007/JHEP11(2020)054}{\emph{JHEP} {\bfseries 11}
  (2020) 054}, [\href{https://arxiv.org/abs/2007.15009}{{\ttfamily
  2007.15009}}].

\bibitem{Caron-Huot:2021rmr}
S.~Caron-Huot, D.~Mazac, L.~Rastelli and D.~Simmons-Duffin, \emph{{Sharp
  boundaries for the swampland}},
  \href{http://dx.doi.org/10.1007/JHEP07(2021)110}{\emph{JHEP} {\bfseries 07}
  (2021) 110}, [\href{https://arxiv.org/abs/2102.08951}{{\ttfamily
  2102.08951}}].

\bibitem{Guerrieri:2021tak}
A.~Guerrieri and A.~Sever, \emph{{Rigorous Bounds on the Analytic S Matrix}},
  \href{http://dx.doi.org/10.1103/PhysRevLett.127.251601}{\emph{Phys. Rev.
  Lett.} {\bfseries 127} (2021) 251601},
  [\href{https://arxiv.org/abs/2106.10257}{{\ttfamily 2106.10257}}].

\bibitem{Du:2021byy}
Z.-Z. Du, C.~Zhang and S.-Y. Zhou, \emph{{Triple crossing positivity bounds for
  multi-field theories}},
  \href{http://dx.doi.org/10.1007/JHEP12(2021)115}{\emph{JHEP} {\bfseries 12}
  (2021) 115}, [\href{https://arxiv.org/abs/2111.01169}{{\ttfamily
  2111.01169}}].

\bibitem{Alberte:2021dnj}
L.~Alberte, C.~de~Rham, S.~Jaitly and A.~J. Tolley, \emph{{Reverse
  Bootstrapping: IR Lessons for UV Physics}},
  \href{http://dx.doi.org/10.1103/PhysRevLett.128.051602}{\emph{Phys. Rev.
  Lett.} {\bfseries 128} (2022) 051602},
  [\href{https://arxiv.org/abs/2111.09226}{{\ttfamily 2111.09226}}].

\bibitem{Bellazzini:2021oaj}
B.~Bellazzini, M.~Riembau and F.~Riva, \emph{{IR side of positivity bounds}},
  \href{http://dx.doi.org/10.1103/PhysRevD.106.105008}{\emph{Phys. Rev. D}
  {\bfseries 106} (2022) 105008},
  [\href{https://arxiv.org/abs/2112.12561}{{\ttfamily 2112.12561}}].

\bibitem{Chiang:2022ltp}
L.-Y. Chiang, Y.-t. Huang, L.~Rodina and H.-C. Weng, \emph{{De-projecting the
  EFThedron}},  \href{https://arxiv.org/abs/2204.07140}{{\ttfamily
  2204.07140}}.

\bibitem{Caron-Huot:2022ugt}
S.~Caron-Huot, Y.-Z. Li, J.~Parra-Martinez and D.~Simmons-Duffin,
  \emph{{Causality constraints on corrections to Einstein gravity}},
  \href{http://dx.doi.org/10.1007/JHEP05(2023)122}{\emph{JHEP} {\bfseries 05}
  (2023) 122}, [\href{https://arxiv.org/abs/2201.06602}{{\ttfamily
  2201.06602}}].

\bibitem{Chiang:2022jep}
L.-Y. Chiang, Y.-t. Huang, W.~Li, L.~Rodina and H.-C. Weng,
  \emph{{(Non)-projective bounds on gravitational EFT}},
  \href{https://arxiv.org/abs/2201.07177}{{\ttfamily 2201.07177}}.

\bibitem{CarrilloGonzalez:2023cbf}
M.~Carrillo~Gonz\'alez, C.~de~Rham, S.~Jaitly, V.~Pozsgay and A.~Tokareva,
  \emph{{Positivity-causality competition: a road to ultimate EFT consistency
  constraints}},  \href{https://arxiv.org/abs/2307.04784}{{\ttfamily
  2307.04784}}.

\bibitem{Hong:2023zgm}
D.-Y. Hong, Z.-H. Wang and S.-Y. Zhou, \emph{{Causality bounds on scalar-tensor
  EFTs}},  \href{https://arxiv.org/abs/2304.01259}{{\ttfamily 2304.01259}}.

\bibitem{deRham:2022hpx}
C.~de~Rham, S.~Kundu, M.~Reece, A.~J. Tolley and S.-Y. Zhou, \emph{{Snowmass
  White Paper: UV Constraints on IR Physics}},  in \emph{{Snowmass 2021}}, 3,
  2022.
\newblock \href{https://arxiv.org/abs/2203.06805}{{\ttfamily 2203.06805}}.

\bibitem{Zhang:2018shp}
C.~Zhang and S.-Y. Zhou, \emph{{Positivity bounds on vector boson scattering at
  the LHC}}, \href{http://dx.doi.org/10.1103/PhysRevD.100.095003}{\emph{Phys.
  Rev. D} {\bfseries 100} (2019) 095003},
  [\href{https://arxiv.org/abs/1808.00010}{{\ttfamily 1808.00010}}].

\bibitem{Bi:2019phv}
Q.~Bi, C.~Zhang and S.-Y. Zhou, \emph{{Positivity constraints on aQGC: carving
  out the physical parameter space}},
  \href{http://dx.doi.org/10.1007/JHEP06(2019)137}{\emph{JHEP} {\bfseries 06}
  (2019) 137}, [\href{https://arxiv.org/abs/1902.08977}{{\ttfamily
  1902.08977}}].

\bibitem{Bellazzini:2018paj}
B.~Bellazzini and F.~Riva, \emph{{New phenomenological and theoretical
  perspective on anomalous ZZ and Z\ensuremath{\gamma} processes}},
  \href{http://dx.doi.org/10.1103/PhysRevD.98.095021}{\emph{Phys. Rev. D}
  {\bfseries 98} (2018) 095021},
  [\href{https://arxiv.org/abs/1806.09640}{{\ttfamily 1806.09640}}].

\bibitem{Remmen:2019cyz}
G.~N. Remmen and N.~L. Rodd, \emph{{Consistency of the Standard Model Effective
  Field Theory}}, \href{http://dx.doi.org/10.1007/JHEP12(2019)032}{\emph{JHEP}
  {\bfseries 12} (2019) 032},
  [\href{https://arxiv.org/abs/1908.09845}{{\ttfamily 1908.09845}}].

\bibitem{Yamashita:2020gtt}
K.~Yamashita, C.~Zhang and S.-Y. Zhou, \emph{{Elastic positivity vs extremal
  positivity bounds in SMEFT: a case study in transversal electroweak
  gauge-boson scatterings}},
  \href{http://dx.doi.org/10.1007/JHEP01(2021)095}{\emph{JHEP} {\bfseries 01}
  (2021) 095}, [\href{https://arxiv.org/abs/2009.04490}{{\ttfamily
  2009.04490}}].

\bibitem{Trott:2020ebl}
T.~Trott, \emph{{Causality, unitarity and symmetry in effective field theory}},
  \href{http://dx.doi.org/10.1007/JHEP07(2021)143}{\emph{JHEP} {\bfseries 07}
  (2021) 143}, [\href{https://arxiv.org/abs/2011.10058}{{\ttfamily
  2011.10058}}].

\bibitem{Remmen:2020vts}
G.~N. Remmen and N.~L. Rodd, \emph{{Flavor Constraints from Unitarity and
  Analyticity}},
  \href{http://dx.doi.org/10.1103/PhysRevLett.127.149901}{\emph{Phys. Rev.
  Lett.} {\bfseries 125} (2020) 081601},
  [\href{https://arxiv.org/abs/2004.02885}{{\ttfamily 2004.02885}}].

\bibitem{Remmen:2020uze}
G.~N. Remmen and N.~L. Rodd, \emph{{Signs, spin, SMEFT: Sum rules at dimension
  six}}, \href{http://dx.doi.org/10.1103/PhysRevD.105.036006}{\emph{Phys. Rev.
  D} {\bfseries 105} (2022) 036006},
  [\href{https://arxiv.org/abs/2010.04723}{{\ttfamily 2010.04723}}].

\bibitem{Gu:2020thj}
J.~Gu and L.-T. Wang, \emph{{Sum Rules in the Standard Model Effective Field
  Theory from Helicity Amplitudes}},
  \href{http://dx.doi.org/10.1007/JHEP03(2021)149}{\emph{JHEP} {\bfseries 03}
  (2021) 149}, [\href{https://arxiv.org/abs/2008.07551}{{\ttfamily
  2008.07551}}].

\bibitem{Fuks:2020ujk}
B.~Fuks, Y.~Liu, C.~Zhang and S.-Y. Zhou, \emph{{Positivity in
  electron-positron scattering: testing the axiomatic quantum field theory
  principles and probing the existence of UV states}},
  \href{http://dx.doi.org/10.1088/1674-1137/abcd8c}{\emph{Chin. Phys. C}
  {\bfseries 45} (2021) 023108},
  [\href{https://arxiv.org/abs/2009.02212}{{\ttfamily 2009.02212}}].

\bibitem{Gu:2020ldn}
J.~Gu, L.-T. Wang and C.~Zhang, \emph{{Unambiguously Testing Positivity at
  Lepton Colliders}},
  \href{http://dx.doi.org/10.1103/PhysRevLett.129.011805}{\emph{Phys. Rev.
  Lett.} {\bfseries 129} (2022) 011805},
  [\href{https://arxiv.org/abs/2011.03055}{{\ttfamily 2011.03055}}].

\bibitem{Bonnefoy:2020yee}
Q.~Bonnefoy, E.~Gendy and C.~Grojean, \emph{{Positivity bounds on Minimal
  Flavor Violation}},
  \href{http://dx.doi.org/10.1007/JHEP04(2021)115}{\emph{JHEP} {\bfseries 04}
  (2021) 115}, [\href{https://arxiv.org/abs/2011.12855}{{\ttfamily
  2011.12855}}].

\bibitem{Davighi:2021osh}
J.~Davighi, S.~Melville and T.~You, \emph{{Natural selection rules: new
  positivity bounds for massive spinning particles}},
  \href{http://dx.doi.org/10.1007/JHEP02(2022)167}{\emph{JHEP} {\bfseries 02}
  (2022) 167}, [\href{https://arxiv.org/abs/2108.06334}{{\ttfamily
  2108.06334}}].

\bibitem{Chala:2021wpj}
M.~Chala and J.~Santiago, \emph{{Positivity bounds in the standard model
  effective field theory beyond tree level}},
  \href{http://dx.doi.org/10.1103/PhysRevD.105.L111901}{\emph{Phys. Rev. D}
  {\bfseries 105} (2022) L111901},
  [\href{https://arxiv.org/abs/2110.01624}{{\ttfamily 2110.01624}}].

\bibitem{Zhang:2021eeo}
C.~Zhang, \emph{{SMEFTs living on the edge: determining the UV theories from
  positivity and extremality}},
  \href{http://dx.doi.org/10.1007/JHEP12(2022)096}{\emph{JHEP} {\bfseries 12}
  (2022) 096}, [\href{https://arxiv.org/abs/2112.11665}{{\ttfamily
  2112.11665}}].

\bibitem{Ghosh:2022qqq}
D.~Ghosh, R.~Sharma and F.~Ullah, \emph{{Amplitude\textquoteright{}s positivity
  vs. subluminality: causality and unitarity constraints on dimension 6 \& 8
  gluonic operators in the SMEFT}},
  \href{http://dx.doi.org/10.1007/JHEP02(2023)199}{\emph{JHEP} {\bfseries 02}
  (2023) 199}, [\href{https://arxiv.org/abs/2211.01322}{{\ttfamily
  2211.01322}}].

\bibitem{Remmen:2022orj}
G.~N. Remmen and N.~L. Rodd, \emph{{Spinning sum rules for the dimension-six
  SMEFT}}, \href{http://dx.doi.org/10.1007/JHEP09(2022)030}{\emph{JHEP}
  {\bfseries 09} (2022) 030},
  [\href{https://arxiv.org/abs/2206.13524}{{\ttfamily 2206.13524}}].

\bibitem{Li:2022tcz}
X.~Li and S.~Zhou, \emph{{Origin of neutrino masses on the convex cone of
  positivity bounds}},
  \href{http://dx.doi.org/10.1103/PhysRevD.107.L031902}{\emph{Phys. Rev. D}
  {\bfseries 107} (2023) L031902},
  [\href{https://arxiv.org/abs/2202.12907}{{\ttfamily 2202.12907}}].

\bibitem{Li:2022rag}
X.~Li, K.~Mimasu, K.~Yamashita, C.~Yang, C.~Zhang and S.-Y. Zhou,
  \emph{{Moments for positivity: using Drell-Yan data to test positivity bounds
  and reverse-engineer new physics}},
  \href{http://dx.doi.org/10.1007/JHEP10(2022)107}{\emph{JHEP} {\bfseries 10}
  (2022) 107}, [\href{https://arxiv.org/abs/2204.13121}{{\ttfamily
  2204.13121}}].

\bibitem{Li:2022aby}
X.~Li, \emph{{Positivity bounds at one-loop level: the Higgs sector}},
  \href{http://dx.doi.org/10.1007/JHEP05(2023)230}{\emph{JHEP} {\bfseries 05}
  (2023) 230}, [\href{https://arxiv.org/abs/2212.12227}{{\ttfamily
  2212.12227}}].

\bibitem{Altmannshofer:2023bfk}
W.~Altmannshofer, S.~Gori, B.~V. Lehmann and J.~Zuo, \emph{{UV physics from IR
  features: New prospects from top flavor violation}},
  \href{http://dx.doi.org/10.1103/PhysRevD.107.095025}{\emph{Phys. Rev. D}
  {\bfseries 107} (2023) 095025},
  [\href{https://arxiv.org/abs/2303.00781}{{\ttfamily 2303.00781}}].

\bibitem{Davighi:2023acq}
J.~Davighi, S.~Melville, K.~Mimasu and T.~You, \emph{{Positivity and the
  Electroweak Hierarchy}},  \href{https://arxiv.org/abs/2308.06226}{{\ttfamily
  2308.06226}}.

\bibitem{Lehman:2015coa}
L.~Lehman and A.~Martin, \emph{{Low-derivative operators of the Standard Model
  effective field theory via Hilbert series methods}},
  \href{http://dx.doi.org/10.1007/JHEP02(2016)081}{\emph{JHEP} {\bfseries 02}
  (2016) 081}, [\href{https://arxiv.org/abs/1510.00372}{{\ttfamily
  1510.00372}}].

\bibitem{Henning:2015alf}
B.~Henning, X.~Lu, T.~Melia and H.~Murayama, \emph{{2, 84, 30, 993, 560, 15456,
  11962, 261485, ...: Higher dimension operators in the SM EFT}},
  \href{http://dx.doi.org/10.1007/JHEP08(2017)016}{\emph{JHEP} {\bfseries 08}
  (2017) 016}, [\href{https://arxiv.org/abs/1512.03433}{{\ttfamily
  1512.03433}}].

\bibitem{Vecchi:2007na}
L.~Vecchi, \emph{{Causal versus analytic constraints on anomalous quartic gauge
  couplings}},
  \href{http://dx.doi.org/10.1088/1126-6708/2007/11/054}{\emph{JHEP} {\bfseries
  11} (2007) 054}, [\href{https://arxiv.org/abs/0704.1900}{{\ttfamily
  0704.1900}}].

\bibitem{Paulos:2017fhb}
M.~F. Paulos, J.~Penedones, J.~Toledo, B.~C. van Rees and P.~Vieira, \emph{{The
  S-matrix bootstrap. Part III: higher dimensional amplitudes}},
  \href{http://dx.doi.org/10.1007/JHEP12(2019)040}{\emph{JHEP} {\bfseries 12}
  (2019) 040}, [\href{https://arxiv.org/abs/1708.06765}{{\ttfamily
  1708.06765}}].

\bibitem{Guerrieri:2020bto}
A.~L. Guerrieri, J.~Penedones and P.~Vieira, \emph{{S-matrix bootstrap for
  effective field theories: massless pions}},
  \href{http://dx.doi.org/10.1007/JHEP06(2021)088}{\emph{JHEP} {\bfseries 06}
  (2021) 088}, [\href{https://arxiv.org/abs/2011.02802}{{\ttfamily
  2011.02802}}].

\bibitem{Guerrieri:2021ivu}
A.~Guerrieri, J.~Penedones and P.~Vieira, \emph{{Where Is String Theory in the
  Space of Scattering Amplitudes?}},
  \href{http://dx.doi.org/10.1103/PhysRevLett.127.081601}{\emph{Phys. Rev.
  Lett.} {\bfseries 127} (2021) 081601},
  [\href{https://arxiv.org/abs/2102.02847}{{\ttfamily 2102.02847}}].

\bibitem{EliasMiro:2022xaa}
J.~Elias~Miro, A.~Guerrieri and M.~A. Gumus, \emph{{Bridging positivity and
  S-matrix bootstrap bounds}},
  \href{http://dx.doi.org/10.1007/JHEP05(2023)001}{\emph{JHEP} {\bfseries 05}
  (2023) 001}, [\href{https://arxiv.org/abs/2210.01502}{{\ttfamily
  2210.01502}}].

\bibitem{Haring:2022sdp}
K.~H\"aring, A.~Hebbar, D.~Karateev, M.~Meineri and J.~a. Penedones,
  \emph{{Bounds on photon scattering}},
  \href{https://arxiv.org/abs/2211.05795}{{\ttfamily 2211.05795}}.

\bibitem{Kruczenski:2022lot}
M.~Kruczenski, J.~Penedones and B.~C. van Rees, \emph{{Snowmass White Paper:
  S-matrix Bootstrap}},  \href{https://arxiv.org/abs/2203.02421}{{\ttfamily
  2203.02421}}.

\bibitem{Xu:2023lpq}
H.~Xu and S.-Y. Zhou, \emph{{Triple crossing positivity bounds, mass dependence
  and cosmological scalars: Horndeski theory and DHOST}},
  \href{https://arxiv.org/abs/2306.06639}{{\ttfamily 2306.06639}}.

\bibitem{Poland:2018epd}
D.~Poland, S.~Rychkov and A.~Vichi, \emph{{The Conformal Bootstrap: Theory,
  Numerical Techniques, and Applications}},
  \href{http://dx.doi.org/10.1103/RevModPhys.91.015002}{\emph{Rev. Mod. Phys.}
  {\bfseries 91} (2019) 015002},
  [\href{https://arxiv.org/abs/1805.04405}{{\ttfamily 1805.04405}}].

\bibitem{Froissart:1961ux}
M.~Froissart, \emph{{Asymptotic behavior and subtractions in the Mandelstam
  representation}},
  \href{http://dx.doi.org/10.1103/PhysRev.123.1053}{\emph{Phys. Rev.}
  {\bfseries 123} (1961) 1053--1057}.

\bibitem{Martin:1962rt}
A.~Martin, \emph{{Unitarity and high-energy behavior of scattering
  amplitudes}}, \href{http://dx.doi.org/10.1103/PhysRev.129.1432}{\emph{Phys.
  Rev.} {\bfseries 129} (1963) 1432--1436}.

\bibitem{deRham:2017imi}
C.~de~Rham, S.~Melville, A.~J. Tolley and S.-Y. Zhou, \emph{{Massive Galileon
  Positivity Bounds}},
  \href{http://dx.doi.org/10.1007/JHEP09(2017)072}{\emph{JHEP} {\bfseries 09}
  (2017) 072}, [\href{https://arxiv.org/abs/1702.08577}{{\ttfamily
  1702.08577}}].

\bibitem{EliasMiro:2023fqi}
J.~Elias~Miro, A.~Guerrieri and M.~A. Gumus, \emph{{Extremal Higgs couplings}},
   \href{https://arxiv.org/abs/2311.09283}{{\ttfamily 2311.09283}}.

\bibitem{Manohar:2018aog}
A.~V. Manohar, \emph{{Introduction to Effective Field Theories}},
  \href{https://arxiv.org/abs/1804.05863}{{\ttfamily 1804.05863}}.

\bibitem{Eboli:2006wa}
O.~J.~P. Eboli, M.~C. Gonzalez-Garcia and J.~K. Mizukoshi, \emph{{p p
  ---\ensuremath{>} j j e+- mu+- nu nu and j j e+- mu-+ nu nu at O(
  alpha(em)**6) and O(alpha(em)**4 alpha(s)**2) for the study of the quartic
  electroweak gauge boson vertex at CERN LHC}},
  \href{http://dx.doi.org/10.1103/PhysRevD.74.073005}{\emph{Phys. Rev. D}
  {\bfseries 74} (2006) 073005},
  [\href{https://arxiv.org/abs/hep-ph/0606118}{{\ttfamily hep-ph/0606118}}].

\bibitem{Almeida:2020ylr}
E.~d.~S. Almeida, O.~J.~P. \'Eboli and M.~C. Gonzalez\textendash{}Garcia,
  \emph{{Unitarity constraints on anomalous quartic couplings}},
  \href{http://dx.doi.org/10.1103/PhysRevD.101.113003}{\emph{Phys. Rev. D}
  {\bfseries 101} (2020) 113003},
  [\href{https://arxiv.org/abs/2004.05174}{{\ttfamily 2004.05174}}].

\bibitem{Li:2020gnx}
H.-L. Li, Z.~Ren, J.~Shu, M.-L. Xiao, J.-H. Yu and Y.-H. Zheng, \emph{{Complete
  set of dimension-eight operators in the standard model effective field
  theory}}, \href{http://dx.doi.org/10.1103/PhysRevD.104.015026}{\emph{Phys.
  Rev. D} {\bfseries 104} (2021) 015026},
  [\href{https://arxiv.org/abs/2005.00008}{{\ttfamily 2005.00008}}].

\bibitem{Murphy:2020rsh}
C.~W. Murphy, \emph{{Dimension-8 operators in the Standard Model Eective Field
  Theory}}, \href{http://dx.doi.org/10.1007/JHEP10(2020)174}{\emph{JHEP}
  {\bfseries 10} (2020) 174},
  [\href{https://arxiv.org/abs/2005.00059}{{\ttfamily 2005.00059}}].

\bibitem{Yang:2023ncf}
C.~Yang, Z.~Ren and J.-H. Yu, \emph{{Positivity from J-Basis Operators in the
  Standard Model Effective Field Theory}},
  \href{https://arxiv.org/abs/2312.04663}{{\ttfamily 2312.04663}}.

\bibitem{Bellazzini:2023nqj}
B.~Bellazzini, G.~Isabella, S.~Ricossa and F.~Riva, \emph{{Massive Gravity is
  not Positive}},  \href{https://arxiv.org/abs/2304.02550}{{\ttfamily
  2304.02550}}.

\bibitem{CMS:2017fhs}
{\scshape CMS} collaboration, A.~M. Sirunyan et~al., \emph{{Observation of
  electroweak production of same-sign W boson pairs in the two jet and two
  same-sign lepton final state in proton-proton collisions at $\sqrt{s} = $ 13
  TeV}}, \href{http://dx.doi.org/10.1103/PhysRevLett.120.081801}{\emph{Phys.
  Rev. Lett.} {\bfseries 120} (2018) 081801},
  [\href{https://arxiv.org/abs/1709.05822}{{\ttfamily 1709.05822}}].

\bibitem{ATLAS:2019cbr}
{\scshape ATLAS} collaboration, M.~Aaboud et~al., \emph{{Observation of
  electroweak production of a same-sign $W$ boson pair in association with two
  jets in $pp$ collisions at $\sqrt{s}=13$ TeV with the ATLAS detector}},
  \href{http://dx.doi.org/10.1103/PhysRevLett.123.161801}{\emph{Phys. Rev.
  Lett.} {\bfseries 123} (2019) 161801},
  [\href{https://arxiv.org/abs/1906.03203}{{\ttfamily 1906.03203}}].

\bibitem{Yang:2020rjt}
J.-C. Yang, Z.-B. Qing, X.-Y. Han, Y.-C. Guo and T.~Li, \emph{{Tri-photon at
  muon collider: a new process to probe the anomalous quartic gauge
  couplings}}, \href{http://dx.doi.org/10.1007/JHEP07(2022)053}{\emph{JHEP}
  {\bfseries 22} (2020) 053},
  [\href{https://arxiv.org/abs/2204.08195}{{\ttfamily 2204.08195}}].

\bibitem{Yang:2021pcf}
J.-C. Yang, Y.-C. Guo, C.-X. Yue and Q.~Fu, \emph{{Constraints on anomalous
  quartic gauge couplings via Z\ensuremath{\gamma}jj production at the LHC}},
  \href{http://dx.doi.org/10.1103/PhysRevD.104.035015}{\emph{Phys. Rev. D}
  {\bfseries 104} (2021) 035015},
  [\href{https://arxiv.org/abs/2107.01123}{{\ttfamily 2107.01123}}].

\bibitem{Gomez-Ambrosio:2018pnl}
R.~Gomez-Ambrosio, \emph{{Studies of Dimension-Six EFT effects in Vector Boson
  Scattering}},
  \href{http://dx.doi.org/10.1140/epjc/s10052-019-6893-2}{\emph{Eur. Phys. J.
  C} {\bfseries 79} (2019) 389},
  [\href{https://arxiv.org/abs/1809.04189}{{\ttfamily 1809.04189}}].

\bibitem{Ethier:2021ydt}
J.~J. Ethier, R.~Gomez-Ambrosio, G.~Magni and J.~Rojo, \emph{{SMEFT analysis of
  vector boson scattering and diboson data from the LHC Run II}},
  \href{http://dx.doi.org/10.1140/epjc/s10052-021-09347-7}{\emph{Eur. Phys. J.
  C} {\bfseries 81} (2021) 560},
  [\href{https://arxiv.org/abs/2101.03180}{{\ttfamily 2101.03180}}].

\bibitem{Degrande:2013kka}
C.~Degrande, \emph{{A basis of dimension-eight operators for anomalous neutral
  triple gauge boson interactions}},
  \href{http://dx.doi.org/10.1007/JHEP02(2014)101}{\emph{JHEP} {\bfseries 02}
  (2014) 101}, [\href{https://arxiv.org/abs/1308.6323}{{\ttfamily 1308.6323}}].

\bibitem{Senol:2018cks}
A.~Senol, H.~Denizli, A.~Yilmaz, I.~Turk~Cakir, K.~Y. Oyulmaz, O.~Karadeniz
  et~al., \emph{{Probing the Effects of Dimension-eight Operators Describing
  Anomalous Neutral Triple Gauge Boson Interactions at FCC-hh}},
  \href{http://dx.doi.org/10.1016/j.nuclphysb.2018.08.018}{\emph{Nucl. Phys. B}
  {\bfseries 935} (2018) 365--376},
  [\href{https://arxiv.org/abs/1805.03475}{{\ttfamily 1805.03475}}].

\bibitem{Rahaman:2018ujg}
R.~Rahaman and R.~K. Singh, \emph{{Anomalous triple gauge boson couplings in
  $ZZ$ production at the LHC and the role of $Z$ boson polarizations}},
  \href{http://dx.doi.org/10.1016/j.nuclphysb.2019.114754}{\emph{Nucl. Phys. B}
  {\bfseries 948} (2019) 114754},
  [\href{https://arxiv.org/abs/1810.11657}{{\ttfamily 1810.11657}}].

\bibitem{Senol:2019swu}
A.~Senol, H.~Denizli, A.~Yilmaz, I.~Turk~Cakir and O.~Cakir, \emph{{Study on
  Anomalous Neutral Triple Gauge Boson Couplings from Dimension-eight Operators
  at the HL-LHC}},  \href{https://arxiv.org/abs/1906.04589}{{\ttfamily
  1906.04589}}.

\bibitem{Ellis:2019zex}
J.~Ellis, S.-F. Ge, H.-J. He and R.-Q. Xiao, \emph{{Probing the scale of new
  physics in the $ZZ\gamma$ coupling at $e^+e^-$ colliders}},
  \href{http://dx.doi.org/10.1088/1674-1137/44/6/063106}{\emph{Chin. Phys. C}
  {\bfseries 44} (2020) 063106},
  [\href{https://arxiv.org/abs/1902.06631}{{\ttfamily 1902.06631}}].

\bibitem{Ellis:2020ljj}
J.~Ellis, H.-J. He and R.-Q. Xiao, \emph{{Probing new physics in dimension-8
  neutral gauge couplings at e+e- colliders}},
  \href{http://dx.doi.org/10.1007/s11433-020-1617-3}{\emph{Sci. China Phys.
  Mech. Astron.} {\bfseries 64} (2021) 221062},
  [\href{https://arxiv.org/abs/2008.04298}{{\ttfamily 2008.04298}}].

\bibitem{Fu:2021mub}
Q.~Fu, J.-C. Yang, C.-X. Yue and Y.-C. Guo, \emph{{The study of neutral triple
  gauge couplings in the process
  e+e\ensuremath{-}\textrightarrow{}Z\ensuremath{\gamma} including unitarity
  bounds}},
  \href{http://dx.doi.org/10.1016/j.nuclphysb.2021.115543}{\emph{Nucl. Phys. B}
  {\bfseries 972} (2021) 115543},
  [\href{https://arxiv.org/abs/2102.03623}{{\ttfamily 2102.03623}}].

\bibitem{Lombardi:2021wug}
D.~Lombardi, M.~Wiesemann and G.~Zanderighi, \emph{{Anomalous couplings in
  Z\ensuremath{\gamma} events at NNLO+PS and improving
  \ensuremath{\nu}\ensuremath{\nu}\textasciimacron{}\ensuremath{\gamma}
  backgrounds in dark-matter searches}},
  \href{http://dx.doi.org/10.1016/j.physletb.2021.136846}{\emph{Phys. Lett. B}
  {\bfseries 824} (2022) 136846},
  [\href{https://arxiv.org/abs/2108.11315}{{\ttfamily 2108.11315}}].

\bibitem{Jahedi:2022duc}
S.~Jahedi and J.~Lahiri, \emph{{Probing anomalous ZZ\ensuremath{\gamma} and
  Z\ensuremath{\gamma}\ensuremath{\gamma} couplings at the e+e- colliders using
  optimal observable technique}},
  \href{http://dx.doi.org/10.1007/JHEP04(2023)085}{\emph{JHEP} {\bfseries 04}
  (2023) 085}, [\href{https://arxiv.org/abs/2212.05121}{{\ttfamily
  2212.05121}}].

\bibitem{Senol:2022snc}
A.~Senol, S.~Spor, E.~Gurkanli, V.~Cetinkaya, H.~Denizli and M.~K\"oksal,
  \emph{{Model-independent study on the anomalous $ZZ\gamma $ and $Z\gamma
  \gamma $ couplings at the future muon collider}},
  \href{http://dx.doi.org/10.1140/epjp/s13360-022-03569-8}{\emph{Eur. Phys. J.
  Plus} {\bfseries 137} (2022) 1354},
  [\href{https://arxiv.org/abs/2205.02912}{{\ttfamily 2205.02912}}].

\bibitem{Ellis:2022zdw}
J.~Ellis, H.-J. He and R.-Q. Xiao, \emph{{Probing neutral triple gauge
  couplings at the LHC and future hadron colliders}},
  \href{http://dx.doi.org/10.1103/PhysRevD.107.035005}{\emph{Phys. Rev. D}
  {\bfseries 107} (2023) 035005},
  [\href{https://arxiv.org/abs/2206.11676}{{\ttfamily 2206.11676}}].

\bibitem{Spor:2022zob}
S.~Spor, E.~Gurkanli and M.~K\"oksal, \emph{{Search for the anomalous
  ZZ\ensuremath{\gamma} and Z\ensuremath{\gamma}\ensuremath{\gamma} couplings
  via \ensuremath{\nu}\ensuremath{\nu}\ensuremath{\gamma} production at the
  CLIC}}, \href{http://dx.doi.org/10.1016/j.nuclphysb.2022.115785}{\emph{Nucl.
  Phys. B} {\bfseries 979} (2022) 115785},
  [\href{https://arxiv.org/abs/2203.02352}{{\ttfamily 2203.02352}}].

\bibitem{Degrande:2023iob}
C.~Degrande and H.-L. Li, \emph{{Impact of dimension-8 SMEFT operators on
  diboson productions}},
  \href{http://dx.doi.org/10.1007/JHEP06(2023)149}{\emph{JHEP} {\bfseries 06}
  (2023) 149}, [\href{https://arxiv.org/abs/2303.10493}{{\ttfamily
  2303.10493}}].

\bibitem{LlewellynSmith:1973yud}
C.~H. Llewellyn~Smith, \emph{{High-Energy Behavior and Gauge Symmetry}},
  \href{http://dx.doi.org/10.1016/0370-2693(73)90692-8}{\emph{Phys. Lett. B}
  {\bfseries 46} (1973) 233--236}.

\bibitem{Lee:1977eg}
B.~W. Lee, C.~Quigg and H.~B. Thacker, \emph{{Weak Interactions at Very
  High-Energies: The Role of the Higgs Boson Mass}},
  \href{http://dx.doi.org/10.1103/PhysRevD.16.1519}{\emph{Phys. Rev. D}
  {\bfseries 16} (1977) 1519}.

\bibitem{Lee:1977yc}
B.~W. Lee, C.~Quigg and H.~B. Thacker, \emph{{The Strength of Weak Interactions
  at Very High-Energies and the Higgs Boson Mass}},
  \href{http://dx.doi.org/10.1103/PhysRevLett.38.883}{\emph{Phys. Rev. Lett.}
  {\bfseries 38} (1977) 883--885}.

\bibitem{Cornwall:1974km}
J.~M. Cornwall, D.~N. Levin and G.~Tiktopoulos, \emph{{Derivation of Gauge
  Invariance from High-Energy Unitarity Bounds on the s Matrix}},
  \href{http://dx.doi.org/10.1103/PhysRevD.10.1145}{\emph{Phys. Rev. D}
  {\bfseries 10} (1974) 1145}.

\bibitem{Eboli:2016kko}
O.~J.~P. \'Eboli and M.~C. Gonzalez-Garcia, \emph{{Classifying the bosonic
  quartic couplings}},
  \href{http://dx.doi.org/10.1103/PhysRevD.93.093013}{\emph{Phys. Rev. D}
  {\bfseries 93} (2016) 093013},
  [\href{https://arxiv.org/abs/1604.03555}{{\ttfamily 1604.03555}}].

\bibitem{CMS:2019qfk}
{\scshape CMS} collaboration, A.~M. Sirunyan et~al., \emph{{Search for
  anomalous electroweak production of vector boson pairs in association with
  two jets in proton-proton collisions at 13 TeV}},
  \href{http://dx.doi.org/10.1016/j.physletb.2019.134985}{\emph{Phys. Lett. B}
  {\bfseries 798} (2019) 134985},
  [\href{https://arxiv.org/abs/1905.07445}{{\ttfamily 1905.07445}}].

\bibitem{Cappati:2022skp}
A.~Cappati, R.~Covarelli, P.~Torrielli and M.~Zaro, \emph{{Sensitivity to new
  physics in final states with multiple gauge and Higgs bosons}},
  \href{http://dx.doi.org/10.1007/JHEP09(2022)038}{\emph{JHEP} {\bfseries 09}
  (2022) 038}, [\href{https://arxiv.org/abs/2205.15959}{{\ttfamily
  2205.15959}}].

\bibitem{Soldate:1986mk}
M.~Soldate, \emph{{Partial Wave Unitarity and Closed String Amplitudes}},
  \href{http://dx.doi.org/10.1016/0370-2693(87)90302-9}{\emph{Phys. Lett. B}
  {\bfseries 186} (1987) 321--327}.

\end{thebibliography}\endgroup

\end{document}